\documentclass[
  journal=pasa,
  manuscript=research-paper,
  year=2024,
  volume=37,
]{cup-journal}

\usepackage{amsmath}
\usepackage[nopatch]{microtype}
\usepackage{booktabs}
\usepackage{graphicx}
\usepackage{subcaption}
\usepackage{makecell}
\usepackage{hyperref}

\newcommand{\XY}[2]{\left[\textrm{#1/#2}\right]}
\newcommand{\FeH}{\XY{Fe}{H}}
\newcommand{\XFe}[1]{\XY{#1}{Fe}}

\newcommand{\kms}{km\,s$^{-1}$}
\newcommand{\Teff}{T_{\textrm{eff}}}
\newcommand{\logg}{\log g}

\title{The Rise of the Milky Way Disk through EMP Stars}

\author{B.\,Lowe}
\affiliation{Research School of Astronomy and Astrophysics, Australian National University, Canberra, ACT 2611, Australia}
\alsoaffiliation{ARC Centre of Excellence for All Sky Astrophysics in 3 Dimensions (ASTRO 3D), Australia}
\email[B. Lowe]{ben.lowe@anu.edu.au}

\author{T.\,Nordlander}
\affiliation{Research School of Astronomy and Astrophysics, Australian National University, Canberra, ACT 2611, Australia}
\alsoaffiliation{ARC Centre of Excellence for All Sky Astrophysics in 3 Dimensions (ASTRO 3D), Australia}
\alsoaffiliation{Theoretical Astrophysics, Department of Physics and Astronomy, Uppsala University, Box 516, 751 20 Uppsala, Sweden}

\author{L.\,Casagrande}
\affiliation{Research School of Astronomy and Astrophysics, Australian National University, Canberra, ACT 2611, Australia}
\alsoaffiliation{ARC Centre of Excellence for All Sky Astrophysics in 3 Dimensions (ASTRO 3D), Australia}

\author{G.\,S.\,Da Costa}
\affiliation{Research School of Astronomy and Astrophysics, Australian National University, Canberra, ACT 2611, Australia}
\alsoaffiliation{ARC Centre of Excellence for All Sky Astrophysics in 3 Dimensions (ASTRO 3D), Australia}

\author{M.\,Bessell}
\affiliation{Research School of Astronomy and Astrophysics, Australian National University, Canberra, ACT 2611, Australia}
\alsoaffiliation{ARC Centre of Excellence for All Sky Astrophysics in 3 Dimensions (ASTRO 3D), Australia}

\author{M.\,McKenzie}
\affiliation{Research School of Astronomy and Astrophysics, Australian National University, Canberra, ACT 2611, Australia}
\alsoaffiliation{ARC Centre of Excellence for All Sky Astrophysics in 3 Dimensions (ASTRO 3D), Australia}
\alsoaffiliation{Carnegie Science Observatories, 813 Santa Barbara St., Pasadena, CA 91101, USA}

\author{G.\,Cordoni}
\affiliation{Research School of Astronomy and Astrophysics, Australian National University, Canberra, ACT 2611, Australia}
\alsoaffiliation{ARC Centre of Excellence for All Sky Astrophysics in 3 Dimensions (ASTRO 3D), Australia}

\author{N.\,Christlieb}
\affiliation{Zentrum für Astronomie der Universität Heidelberg, Landessternwarte, Königstuhl 12, 69117 Heidelberg, Germany}

\author{S.\,Buder}
\affiliation{Research School of Astronomy and Astrophysics, Australian National University, Canberra, ACT 2611, Australia}
\alsoaffiliation{ARC Centre of Excellence for All Sky Astrophysics in 3 Dimensions (ASTRO 3D), Australia}


\begin{document}

\begin{abstract}
We present a chemo-dynamical study conducted with 2dF$+$AAOmega of $\sim 6000$ \textit{Gaia} DR3 non-variable candidate metal-poor stars that lie in the direction of the Galactic plane. Our spectral analysis reveals 15 new extremely metal-poor (EMP) stars, with the lowest metallicity at $\FeH = -4.0 \pm 0.2$\,dex. Two of the EMP stars are also carbon enhanced, with the largest enhancement of $\XFe{C} = 1.3 \pm 0.1$ occurring in a dwarf. Using our $\XFe{C}$ results, we demonstrate that the number of carbon-depleted stars decreases with lower metallicities, and the fraction of carbon-enhanced stars increases, in agreement with previous studies. 

Our dynamical analysis reveals that the fraction of prograde and retrograde disk stars, defined as $z_{\rm max} < 3$\,kpc, with $J_{\phi}/J_{\rm tot} > 0.75$ and $J_{\phi}/J_{\rm tot} < -0.75$ respectively, changes as metallicities decrease. Disk stars on retrograde orbits make up $\sim 10$\% of all the stars in our sample with metallicities below $-2.1$\,dex. Interestingly, the portion of retrograde disk stars compared with the number of kinematically classified halo stars is approximately constant at $4.6$\,\% for all metallicities below $-1.5$\,dex. We also see that $J_{\phi}$ increases from $380 \pm 50$ to $1320 \pm 90$\,km s$^{-1}$ kpc across metallicity range $-1.5$ to $-1.1$, consistent with the spin-up of the Galactic disk. Over the metallicity range $-3.0 < \FeH < -2.0$, the slopes of the metallicity distribution functions for the prograde and retrograde disk stars are similar and comparable to that for the halo population. However, detailed chemical analyses based on high resolution spectra are needed to distinguish the accreted versus in-situ contributions. Finally, we show that our spectroscopic parameters reveal serious systematics in the metallicities published in recent studies that apply various machine learning techniques to \textit{Gaia} XP spectra.
\end{abstract}

\section{Introduction}
According to the Standard Big Bang Nucleosynthesis model, only  hydrogen, helium and traces of light elements (e.g. lithium) were produced during the Big Bang. These baryons clustered with the structures defined by dark matter, eventually collapsing to form the first stars at $z >10$ \citep[e.g.][]{beers_discovery_2005, bromm_formation_2009, frebel_near-field_2015, fraser_mass_2017}. The first stars, often referred to as Population III (or Pop. III), dominated the star formation history until their supernovae enriched the interstellar medium, allowing the formation of the next generations of stars: the Pop. II and Pop. I stars \citep[e.g.][]{klessen_first_2023}. Direct observations of Pop. III stars in the high-redshift Universe \citep[e.g.][]{oh_he_2001, scannapieco_detectability_2003, greif_observational_2009, zackrisson_spectral_2011, zackrisson_detecting_2012, rydberg_detection_2013, mas-ribas_boosting_2016, riaz_unveiling_2022} has so far been unsuccessful, with instruments like the James Web Space Telescope requiring a Pop. III star to be gravitationally-lensed for it to be detectable \citep{zackrisson_detection_2024}. The likelihood of low-mass Pop. III stars ($\rm{M} < 0.8\,\rm{M}_\odot$) forming, which could live until the present-day, is unknown \citep[e.g.][]{ishiyama_where_2016, chandra_searching_2021}, but if they exist, forthcoming massive spectroscopic surveys such as 4MOST and WEAVE \citep[e.g.][]{de_jong_4most_2019, jin_wide-field_2024} will be able to put constraints on their existence. 

One approach to understanding the properties of Pop. III stars is through the study of extremely metal-poor stars (EMP), defined as $[\textrm{Fe}/\textrm{H}]$\footnote{$[\textrm{X}/\textrm{H}] = \textrm{log}(N_{\textrm{X}}/N_{\textrm{H}})_{\star} - \textrm{log}(N_{\textrm{X}}/N_{\textrm{H}})_{\odot}$, where $N_{\textrm{X}}$ is the number density for element X.} $\leq -3.0$ dex \citep{beers_discovery_2005}. These stars may be direct descendants of Pop. III stars, with their chemical composition offering insights into the first stars and the evolution of the early Galaxy. Numerous stellar archaeological studies have provided insights into EMP stars by studying them in  Galactic regions dominated by old stars such as the halo and/or bulge \citep[e.g.][]{christlieb_stellar_2008, schorck_stellar_2009, frebel_stellar_2010, salvadori_mining_2010, caffau_topos_2013, howes_extremely_2015, howes_embla_2016, dacosta_skymapper_2019, arentsen_pristine_2020, yong_high-resolution_2021, ishigaki_origin_2021, li_s_2022}.

Recently, it was discovered that a population of EMP stars confined to disk-like orbits exists \citep[e.g.][]{sestito_tracing_2019, sestito_pristine_2020, kielty_pristine_2021, fernandez-alvar_pristine_2021, cordoni_exploring_2021, chiti_metal-poor_2021}, with the majority on prograde rather than retrograde orbits. For example, studies such as \citet{sestito_pristine_2020}, which used the Pristine survey \citep{starkenburg_pristine_2017}, showed that at the $5.0\sigma$ confidence level, the majority of their 1027 metal-poor stars (with $\FeH \leq -2.5$\,dex) are classified as prograde disk stars. The follow-up study \citet{sestito_exploring_2021} used the NIHAO-UHD simulations to show that there is a distinct population of prograde metal-poor stars in the models. Similarly, the observational study \citet{hong_candidate_2023} showed that from their sample of 11.5 million stars from the SkyMapper Southern Survey Data Release 2 \citep{onken_skymapper_2019} and Stellar Abundance and Galactic Evolution Survey \citep{fan_stellar_2023}, the majority of their disk EMPs are confined to prograde orbits with low eccentricities.

To understand the true nature of these stars, high-resolution spectroscopy of EMP prograde disk stars is needed to gain a detailed perspective from their chemistry. For example, the star SDSS J102915+172927, also known as the ``Caffau'' star \citep{caffau_extremely_2011, caffau_primordial_2012} at $\FeH{} = -4.89 \pm 0.10$ has prograde disk orbit with $z_{\rm max} < 3$\,kpc and $e = 0.12 \pm 0.01$ \citep{sestito_tracing_2019}. This star has no significant [C/Fe] enhancement, with an upper limit of $\XFe{C} < 0.6$ derived with 3D hydrodynamical model atmospheres, assuming local thermodynamical equilibrium (LTE) \citep{lagae_raising_2023}, a surprise as most stars at this metallicity range are carbon-enhanced \citep[e.g.][]{placco_carbon-enhanced_2014, arentsen_pristine_2021}. The nature of this star challenges models for early star formation that require [C/Fe] enhancement to allow low-mass star formation \citep{schneider_formation_2012}. 

Another EMP star confined on a prograde orbit is P1836849 \citep{dovgal_probing_2024}, which has $\FeH{} = -3.3 \pm 0.1$\,dex. Compared with other metal-poor stars, P1836849 has very low abundances (with respect to iron) of $\alpha$-elements (Na, Mg, Si), together with large abundances for Cr and Mn. Comparisons with other EMPs in prograde disk orbits \citep[e.g.][]{cordoni_exploring_2021, yong_high-resolution_2021} shows that there is no obvious common formation origin for these stars when considering the differences in chemistry and kinematics.

These recent revelations provide a new perspective on the diverse process that contributed to the formation of the early Galaxy. Among the various possibilities, the disk may have formed through in-plane accretions \citep[e.g.][]{dinescu_absolute_2002, majewski_exploring_2012, myeong_sausage_2018}, in-situ events \citep[e.g.][]{alfaro_topography_2022, conroy_birth_2024}, or through stellar migration \citep[e.g.][]{grenon_kinematics_1999, schonrich_chemical_2009, minchev_evolution_2012}. EMP stars on both prograde and retrograde disk orbits are thus important, as they can be used as tracers for the early build-up of the Galaxy \citep{sestito_exploring_2021}, but the significant lack of EMP disk stars makes this difficult. In this paper, we expand the pool of EMP disk stars by presenting a new survey using 2dF coupled with the AAOmega spectrograph \citep{saunders_aaomega_2004, sharp_performance_2006}, focused on the Galactic disk using stars selected from \textit{Gaia} DR3 \citep{gaia_collaboration_gaia_2023}. In what follows we present our sample selection (Section \ref{sec:observations}), the determination and validation of stellar parameters (Section \ref{sec:analysis}), a chemo-dynamical analysis (Section \ref{sec:results (chemodynamics)}) and a discussion of the results (Section \ref{sec:discussion}). A detailed comparison of our spectroscopic parameters against those inferred from other analyses of \textit{Gaia} XP spectra is provided in \ref{sec:gaia xp compare}.

\section{Observations}
\label{sec:observations}

\subsection{Sample selection}
\label{subsec:gaia xp spectroscopy}
\textit{Gaia} Data Release 3 includes 220 million low-resolution spectra from the Blue (BP) and Red (RP) spectrophotometers \citep{gaia_collaboration_gaia_2023}, which cover the wavelength regions $330-680$\,nm and $640-1050$\,nm, respectively, with a spectral resolving power ranging from 30 to 100 \citep{de_angeli_gaia_2023}. This wide wavelength coverage allows analysis of \textit{Gaia} BP/RP (XP hereafter) spectra to provide reasonably good inferences on stellar parameters like metallicity and effective temperature from the spectral energy distributions (SED) \citep[e.g.][]{yao_188000_2023}. 

Numerous catalogues of stellar parameters have been published based on \textit{Gaia} XP spectra \citep[e.g.][]{andrae_gaia_2023, andrae_robust_2023, fouesneau_gaia_2023, martin_pristine_2023, li_aspgap_2023, viswanathan_pristine_2024, xylakis-dornbusch_metallicities_2024}. For our work,
we adopted the \citet{zhang_parameters_2023} catalogue.  These authors developed a data-driven model to provide estimates of stellar parameters ($\Teff$, $\logg$ and $\FeH$) and reddening for the entire \textit{Gaia} XP spectra dataset.

We used that catalogue to select metal-poor star candidates, initially regardless of location on the sky, stellar parameters, evolutionary state, and reddening E(B$-$V), but adopting the suggested basic reliability cut $\mathtt{quality\_flags} < 8$ in order to maximise completeness at the cost of potential contamination from misclassified more metal-rich stars. The adopted reliability cut indicates, broadly, that the quality of the fit was acceptable and that the parallax estimate is within $10\,\sigma$ agreement with the actual \textit{Gaia} measurement.  Nevertheless, uncertainties in $\Teff$, $\logg$ or $\FeH$ may be significant.

In order to mitigate the effects of metallicity uncertainties on our selection, we specifically chose stars with $\FeH + \sigma_{\FeH}< -1$. We included the error term in our selection to adopt the 1 $\sigma$ upper bound on the metallicity. This process resulted in a selection of 3\,979\,676 stars.

We intended to observe the sample at Siding Spring Observatory (SSO) using the 2-degree field fibre-fed multi-object system (2dF) instrument \citep{lewis_anglo-australian_2002}. For this, we further filtered the target list according to a blending criterion, based on experience with typical seeing at SSO\@. In particular we required: 

\begin{eqnarray}
    s > 5 - 3 \Delta G_{BP} + (\Delta G_{BP})^2,
\end{eqnarray}

where $s$ is the separation between the target star and its neighbours in units of arcsec, and $\Delta G_{BP}$ is the difference in brightness (in units of mag). The required separation for a contaminating star fainter by 1\,mag is 3\,arcsec, and for successively brighter contaminating stars the required separations are 5, 9, 15, 23, 33, 45 and 60\,arcsec. Applying this criterion reduced our catalogue to 3\,230\,740 stars.

We then broadly grouped stars according to metallicity (actually $\FeH + \sigma_{\FeH}$) in bins of 0.5 dex, and used this as a primary priority marker. At a finer level, targets were further prioritised by brightness, with those at fainter magnitudes having lower priority than those with brighter magnitudes. This priority impacts the allocation of fibres in 2dF, where targets with greater priority (e.g.\ brighter targets) are more likely to be given allocations than those with lower priority (e.g.\ fainter targets). Further, 
in order to avoid observing known globular cluster members, we deselected stars within the tidal radius of all objects listed in the \citet{harris_new_2010} catalogue of globular clusters (with the exception of NGC\,362, see next Section). 
We tiled the sky with overlapping fields of $2 \deg$ diameter, and prioritised the selection of fields according to the number of candidate metal-poor stars with $\FeH < -1$. Fields close to the Galactic disk were preferentially selected over others. To ensure maximum use of the 2dF fibres, targets with lower priority were used to fill out the fibres once the higher priority targets were allocated.
We did not use reddening either for selecting fields or for prioritising candidate stars. 

\subsection{Medium resolution spectroscopy}
\label{subsec:aaomega spectroscopy}
Stars selected from the \citet{zhang_parameters_2023} sample were followed up using 2dF+AAOmega on the AAT at SSO. The 2dF instrument allows us to target up to 392 targets simultaneously within a 2-degree field \citep{sharp_performance_2006}. The real number of fibres allocated to targets is typically lower than this ($\sim 300$) due to the need for fibres placed on the sky (typically 20--30) for sky subtraction, avoidance of collisions with other fibres (i.e.\ in crowded regions), and fibres that are broken.

\begin{figure}[tp]
    \centering
    \includegraphics[width=1\linewidth]{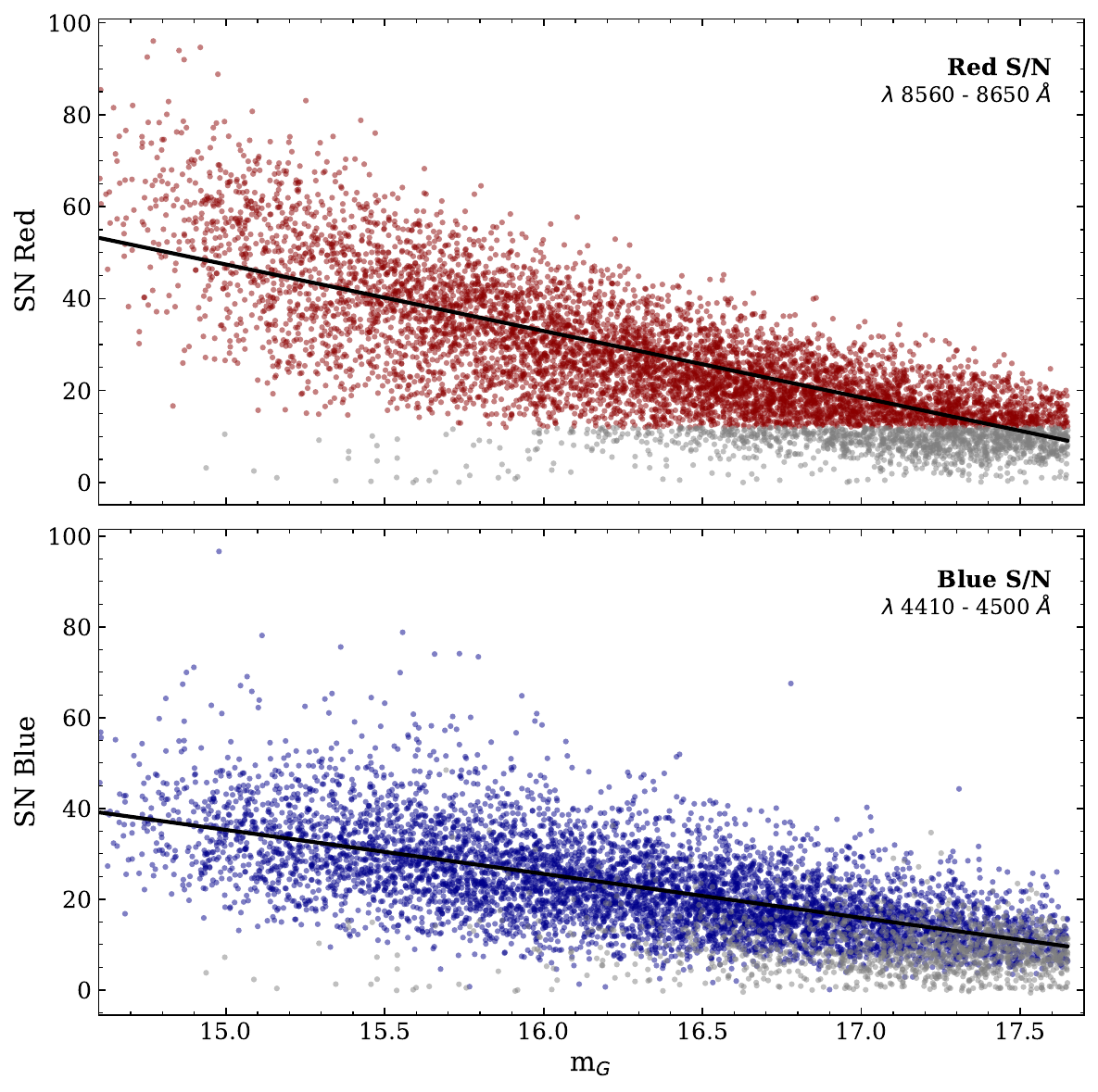}
    \caption{S/N ratios for our sample in both the blue and red detectors. Stars with S/N$_{\textrm{red}} \leq 12$ were removed due to insufficient data quality, shown by the grey points. The fit to the S/N data is shown by the black line. \textit{Upper panel}: S/N ratios of the red arm, measured from the continuum at 8560 -- 8650\,Å. \textit{Lower panel}: S/N ratios of the blue arm, measured from the continuum at 4410 -- 4500\,Å.}
    \label{fig:example quality plot}
\end{figure}

The AAOmega spectrograph \citep[e.g.][]{sharp_performance_2006} was used with the 580V grating in the blue arm covering the wavelength region $3700 \leq \lambda \leq 5800$\,\AA{} at $R \approx 1300$, and the 1700D grating for the red arm covering the wavelength region $8450 \leq \lambda \leq 9000$\,\AA{} at $R \approx 10,000$. These were chosen as the 580V grating provides coverage of the Ca H \& K lines (3933 and 3969\,\AA{}) for metallicity estimates \citep[e.g.][]{beers_search_1992, frebel_nucleosynthetic_2005, norris_he_2007, roederer_search_2014}, the CH G-band ($\sim 4300$\,Å) for carbon abundances, and the Mg\,I\,b lines (5167, 5172, 5183\,Å) with relatively good throughput. The 1700D grating provides coverage of the Ca\,II triplet (8498, 8542 and 8662\,Å), another excellent indicator of metallicity \citep[e.g.][]{armandroff_metallicities_1991, starkenburg_nir_2010, carrera_near-infrared_2013, howes_extremely_2015, da_costa_ca_2016}, as well as accurate radial velocities. These constraints allowed us to probe down to an apparent $G$ magnitude of about 17.5\,mag, well matched by how faint XP spectra go for reliable parameter determination. This also returns a number of metal-poor candidates which broadly matches the available number of fibres. To avoid the impact of scattered light, we selected targets within a range of 2.5\,mag. At $m_G \sim 16$ across 4410--4500\,\AA{}, we achieved a signal-to-noise (S/N) ratio of $\sim 25$, and for 8560 -- 8650\,\AA{} S/N\,$\sim 30$ (see Fig.\,\ref{fig:example quality plot}). Typical exposure times for our fields are $4 \times 1800$\,s, with $\sim 2$'' seeing.

\begin{figure*}[tp]
    \centering
    \includegraphics[width=1\linewidth]{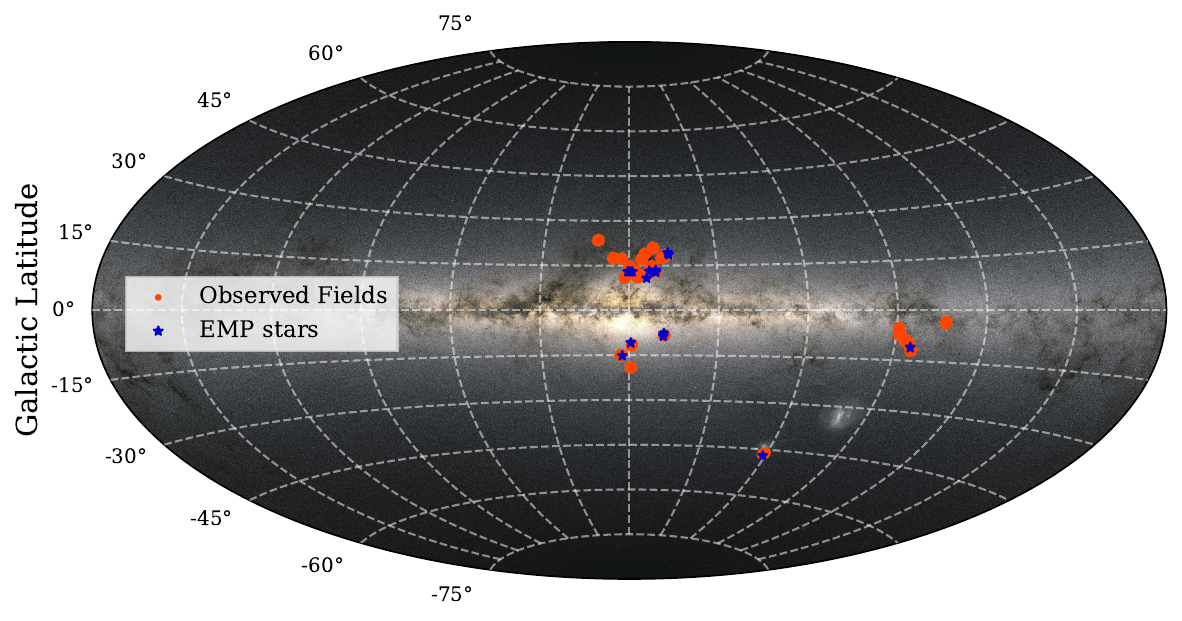}
    \caption{Location of fields observed on the AAT, shown in Galactic coordinates with Aitoff projection. The \textit{Gaia} DR3 image is used as the background image \citep{gaia_collaboration_gaia_2023}. Red circles mark observed fields, and blue star symbols indicate the vetted EMP stars. Fields near the centre of the figure are at $l \sim 0^{\circ}$, fields on the far-right are at $l\sim 265^{\circ}$, and the field near the SMC is at $l \sim 300^{\circ}$. The field near the SMC is a calibration field incorporating the globular cluster NGC\,362.}
    \label{fig:aat fields}
\end{figure*}

With this setup, 8911 metal-poor star candidates from our sample were observed over 74 hours across 10 nights in 2023 covering 28 fields. The list of fields is in Table\,\ref{tab:aat fields}, and their locations are shown by the red circles in Fig.\,\ref{fig:aat fields}. We have 21 fields above and below the Galactic centre ($b \pm 20^{\circ}$ at $l \sim 0$), with six fields at $l \sim 265^{\circ}$ and $b \sim -8^{\circ}$. The field near the Small Magellanic Cloud (SMC; $l \sim 300^{\circ}$ and $b \sim -46^{\circ}$) contains the globular cluster NGC\,362, targeted as part of our validation (see Section \ref{subsec:gc validation}), but also including field metal-poor star targets. As discussed below, 15 EMPs were found in the data, shown by the blue star symbols in Fig.\ \ref{fig:aat fields}.

\begin{table}[tp]
    \centering
    \caption{First three fields observed at the AAT, identified by the field centre coordinates in equatorial and Galactic coordinates, alongside the number of stars observed in each field. Field ra\_0103-7059 is the field located by the SMC, containing globular cluster NGC\,362 for validation. Full table available on the electronic version of the paper.}
    \begin{tabular}{lccrrc}
        \hline
        Field ID & RA & Dec & $l$ [deg] & $b$ [deg] & \# Stars \\
        \hline
        ra\_0103-7050 & 01:03:14.3 & $-$70:50:56 &    301.5 &  $-$46.2 &     263 \\
        ra\_0752-5047 & 07:52:37.1 & $-$50:47:13 &    264.4 &  $-$11.8 &     315 \\
        ra\_0803-3737 & 08:03:09.5 & $-$37:37:29 &    253.9 &   $-$3.5 &      74 \\
        \hline
    \end{tabular}
    \label{tab:aat fields}
\end{table}

\begin{figure*}[tp]
    \centering
    \includegraphics[width=1\linewidth]{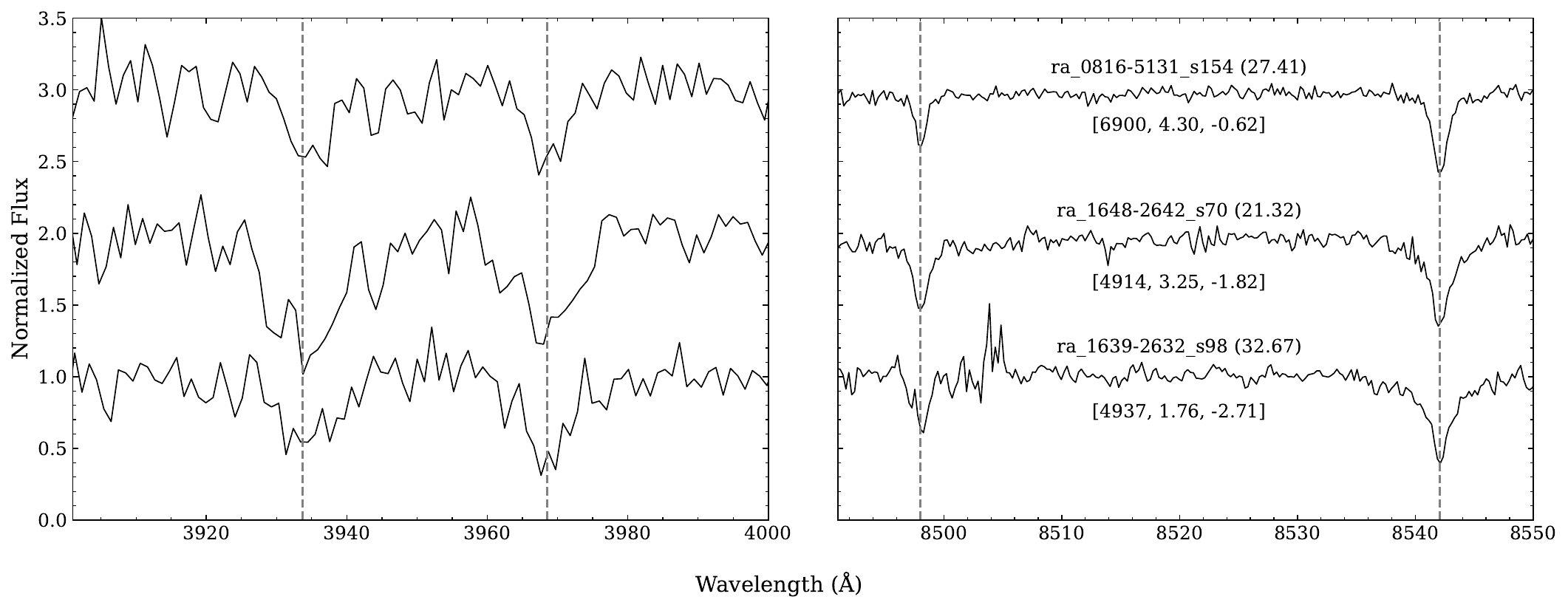}
    \caption{The Ca\,II H and K region alongside the Ca\,II triplet region for stars \textit{ra\_0816-5131\_s154}, \textit{ra\_1648-2642\_s70} and \textit{ra\_1639-2632\_s98}. The names with their red S/N are written above the spectra in the right panel. $\Teff$, $\logg$ and $\FeH$ values for each star are beneath the spectra on the same panel. All three stars have $\textrm{m$_\textrm{G}$} \approx 16.25$\,mag. Star \textit{ra\_0816-5131\_s154} (top) shows an example of a relatively hot spectrum at $T_{\textrm{eff}} = 6900$\,K, star \textit{ra\_1648-2642\_s70} (middle) is an example cool spectrum at $T_{\textrm{eff}} = 4914$\,K, and star \textit{ra\_1639-2632\_s98} (bottom) is an example very metal-poor star at $\FeH=-2.71$. \textit{Left panel}: Ca\,II H and K region from the 580V arm for the three stars. Vertical dashed lines show the Ca\,II H line at 3933.66 \AA{}, and the Ca\,II K line at 3968.47 \AA{}. \textit{Right panel}: Ca\,II triplet region from the 1700D arm for the three stars. Vertical dashed lines show the first two of the Ca\,II triplet lines, at 8498.02 \AA{} and 8542.09 \AA{} respectively.}
    \label{fig:example spec}
\end{figure*}

The data was reduced using version 8.03b of the 2dF reduction pipeline \texttt{2dfdr}\footnote{\url{https://aat.anu.edu.au/science/software/2dfdr}}. Here, fibre flat-field exposures were used to define fibre positions on the detector, whilst arc lamp exposures were used to wavelength calibrate the extracted spectra \citep[e.g.][]{howes_embla_2016, da_costa_ca_2016}. Accurate sky subtraction is crucial for the Ca\,II triplet region, so we performed this using \textsc{skyflux(cor)}, which takes the strengths of night sky emission lines present in the raw spectra and uses them to normalise the fibre throughputs. Cosmic rays were then rejected using the \textsc{bclean} method, a simple ``zap-ray'' method that is more general and efficient than the Laplacian edge detection methods like \textsc{lacosmic} and \textsc{pycosmic}, but is less effective \citep{dokkum_cosmicray_2001}. The wavelength-calibrated sky-subtracted spectra from the individual exposures were then combined using flux weighting with rejection to remove any remaining cosmic rays and to improve the overall signal-to-noise ratio. Fig.\,\ref{fig:example quality plot} shows the distribution of S/N values, calculated using the \textsc{snr\_derived} method in \texttt{specutils}\footnote{\url{https://specutils.readthedocs.io/en/stable/}} \citep{astropy-specutils_development_team_specutils_2019}. For the blue detector, we measured the S/N from the continuum at 4410--4500\,Å, while for the red detector, we measured it from the continuum at 8560--8650\,Å. Stars with S/N$_{\textrm{red}} \leq 12$\footnote{S/N of 12 was chosen from visual inspection of the spectra.} were removed due to poor data quality in both the blue and red arms. Stars with the $\mathtt{phot\_variable\_flag} = \mathtt{VARIABLE}$ flag in the \textit{Gaia} DR3 dataset were also removed, reducing the final sample to 5795 stars. Examples of reduced AAOmega spectra are shown in Fig.\,\ref{fig:example spec}. 

\section{Analysis Techniques}
\label{sec:analysis}

\subsection{Stellar Parameters}
\label{subsec:stellar params}
To accurately extract stellar parameters for the sample, we developed a methodology that self-consistently derives $\Teff$ and $\logg$, then uses the template fitting code \texttt{RVSpecFit}\footnote{\url{https://github.com/segasai/rvspecfit}} \citep{koposov_accurate_2011, koposov_rvspecfit_2019} to get $\FeH$ and $\XFe{C}$. The code initialises with parameters from the \textit{Gaia} XP analysis of \citet{zhang_parameters_2023}, iterating through the process by determining $\FeH$ for a given $\Teff$ and $\logg$, then redetermining $\Teff$ and $\logg$ with the new metallicity. This repeats until reaching convergence, typically in 2--3 iterations. $\XFe{C}$ is then determined independently using the converged stellar parameters. Uncertainties for stellar parameters are described in Section \ref{subsec:param uncertainty}. Below we describe our procedure in more detail.

\subsubsection{Effective Temperature}
\label{subsec:teff methods}
To compute stellar effective temperatures we use \textit{Gaia} DR3 photometry with the \texttt{colte}\footnote{\url{https://github.com/casaluca/colte}} code \citep{casagrande_galah_2021}. This uses colour-$\Teff$ relations derived implementing the InfraRed Flux Method in the \textit{Gaia} and 2MASS system which have very mild dependency on metallicity and surface gravity. Reddening corrections used the the \citet{schlegel_maps_1998} extinction map, rescaled as described in \citet{casagrande_skymapper_2019}. To ensure consistency with our bolometric corrections (see Section \ref{subsec:logg methods}), in \texttt{colte} we adopted the `COD' extinction law (based on \citealt{cardelli_relationship_1989} and \citealt{odonnell_r_1994}). 

\subsubsection{Surface Gravity}
\label{subsec:logg methods}
A precise parallax measurement together with an estimate of the luminosity and stellar mass can yield a highly precise estimate of surface gravity. Low-precision parallaxes, however, can be biased due to sampling effects, and require a more involved treatment that relies on knowing the true distribution of stellar distances to derive $\logg$ values \citep[e.g.][]{bailer-jones_estimating_2021}. 

We therefore opted for a two-pronged approach based on parallax quality, which we describe below. Distances and therefore absolute \textit{Gaia} DR3 magnitudes were derived using the inverse of these processes once $\logg$ has been found (see Section \ref{subsec:distances}).

For stars with good parallax ($\pi \geq 3 \sigma$), we used the absolute bolometric magnitude, $\Teff$ and stellar mass (assuming 0.8\,$M_{\odot}$) directly to get the surface gravity. For the bolometric magnitude, we used  \texttt{bcutil}\footnote{\url{https://github.com/casaluca/bolometric-corrections}} \citep{casagrande_use_2018} with the \textit{Gaia} DR3 $G$ filter, which computes synthetic bolometric corrections for given values of $\Teff$, $\FeH$, $\logg$ (using the initial value) and $E(B-V)$, while adopting the \citet{cardelli_relationship_1989} and \citet{odonnell_r_1994} extinction laws.

For stars lacking good parallaxes ($\pi < 3 \sigma$), we instead opted to employ isochrones calculated by the Dartmouth evolution code \citep{dotter_dartmouth_2008}\footnote{\url{http://stellar.dartmouth.edu/models/index.html}}. The base isochrones cover a range of metallicities from $\FeH=0.5$ down to $-4.0$\,dex with a metallicity stepsize of 0.5\,dex. We interpolated these isochrones to a finer stepsize of 0.1\,dex using \texttt{iso\_interp\_feh}\footnote{\url{http://stellar.dartmouth.edu/models/programs.html}}. For all models, helium enrichment was set as $Y = 0.245+1.5Z$. We then selected the following isochrones: for $\FeH$ from 0.0 to $-0.2$\,dex, we assumed solar ages with [$\alpha$/Fe] = 0.0. For $\FeH$ from $-0.2$ to $-0.5$\,dex, [$\alpha$/Fe] = 0.2 was adopted, again assuming solar ages. Isochrones with $\FeH$ from $-0.6$ to $-4.0$\,dex had an assumed age of 12 Gyr with $[\alpha/\mathrm{Fe}] = 0.4$\,dex. Although the goal of this work is to identify and analyse metal-poor stars, which presumably are old, we note that for metal-rich stars selecting an older isochrone rather than a younger one results in an increase of $\sim 0.3$\,dex in $\logg$. In order to allow for a more precise $\Teff$-$\logg$ relationship, we resampled the isochrones in the $\Teff$-$\logg$ plane to double the sampling. We further assumed stars with $\pi < 3 \sigma$ to be giants. In doing so, equivalent evolutionary points were used to restrict the isochrones to luminosities brighter than the turn-off.

\subsubsection{Metallicity and Radial Velocity}
\label{subsec:feh methods}
We derived metallicities, radial velocities and carbon abundances with our code, which performs a least-squares fit to a grid of synthetic template spectra with a Nelder-Mead search\footnote{A numerical method used to find the maximum or minimum of an objective function within multidimensional space \citep{Nelder_Mead_1965}.}. The templates are fitted to observations through multiplying by a polynomial continuum over the wavelengths $3800$--$5500$\,\AA{} for the blue arm, and $8400$--$8800$\,\AA{} for the red arm. This is done for a set of stellar parameters, whereby $\Teff$ and $\logg$ are determined as per the previous sections, while $\FeH$, [$\alpha$/Fe], and radial velocity RV are determined using \texttt{RVSpecFit} (see \citet{li_southern_2019} for more detail). We found through trial and error that we achieved good outcomes for our spectra when continuum fitting using fifth order Chebyshev polynomials for both arms, noting that the code by default uses radial basis functions.

The \texttt{RVSpecFit} code comes prepacked with the PHOENIX synthetic grid library \citep{husser_new_2013}, but this lacked the ability to vary and fit [C/Fe]. To overcome this, we simultaneously fit both the blue and red arms as follows. For the 580V arm, we computed spectra that spanned a range of $\XY{C}{Fe}$ using the LTE Turbospectrum code \citep{plez_turbospectrum_2012}, while for the 1700D arm, we calculated spectra using the pySME code \citep{wehrhahn_pysme_2023}. This utilities NLTE corrections that are necessary to accurately calculate the cores of the Ca-II triplet lines to determine $\FeH$ abundances. Fits to the blue and red spectra were iterated five times using updated stellar parameters at each step. Grids of synthetic template spectra were processed using a two-step interpolation procedure. First, spectra were broadened to the observed resolving power and sampled at appropriate steps in wavelength. Then a Delaunay triangulation \citep[see e.g.][]{amidror_scattered_2002} was performed to interpolate between templates, providing smoothly changing spectral templates as a function of stellar atmospheric parameters. As old, metal-poor stars have generally long since spun down to rotational velocities well below the instrumental resolution of AAOmega, we opted to neglect rotational broadening. 

\subsubsection{Carbon abundances}
\label{subsec:carb abundance}
\texttt{RVSpecFit} has the facility to fit $\FeH$ alongside another dimension in abundance. Typically this is [$\alpha$/Fe], but as our 580V spectra contained the CH G-band ($\sim 4300$\,\AA), we fitted for [C/Fe] instead after determining the other stellar parameters. For this, we calculated $\chi^2$ for the CH G-band region while simultaneously fitting the continuum slope. The range $4150 \leq \lambda \leq 4450$\,\AA{} was considered for each star over a grid of synthetic spectra, while adopting the same $\Teff$, $\logg$ and $\FeH$ values determined in the previous sections. Abundances were measured from $-5.0 \leq \rm{[C/Fe]} \leq -\FeH + 0.5$ at 0.01\,dex stepsizes, then we took the lowest $\chi^2$ as the accepted $\XFe{C}$ value.

\subsection{Stellar Parameter Error Analysis and Distances}
\label{subsec:param uncertainty}
With our data, we have two main sources of error: those that come directly from measurements on the spectrum (treated as an uncorrelated Gaussian error), and those that are related to the uncertainties in the stellar parameters that are propagated through the analysis.

For the stellar parameters, we perturbed input parameters (magnitudes, parallax and reddening) for each star 100 times randomly, sampling their uncertainties as we describe more in detail below. All errors excluding radial velocity were derived from the standard deviation of the randomisations. Median errors for each stellar parameter based on S/N quality is shown in Table\,\ref{tab:mean errors}. Within this procedure we also derived distances and distance errors for the stars, which we describe at the end of this section. 

\begin{table}[tp]
    \centering
    \caption{Median error on stellar parameters, representing Monte Carlo propagated systematic uncertainties for $\Teff$, $\FeH$, $\logg$ and $\XFe{C}$. Errors on radial velocity were derived directly from \texttt{RVSpecFit}. Errors are split into stars with high S/N ($\geq 25$) and those with low S/N ($< 25$). For $\XFe{C}$, the mean errors exclude stars with no detections.}
    \begin{tabular}{l|cc}
        \hline
        Parameter & Median error (S/N\,$< 25$) & Median error (S/N\,$\geq 25$) \\
        \hline
        $\Teff$ [K] & 150 & 110 \\
        $\FeH$ & 0.16 & 0.14 \\
        $\logg$ & 0.23 & 0.22 \\
        RV [\kms] & 40 & 10 \\
        $\XFe{C}$ & 0.12 & 0.093 \\
        \hline
    \end{tabular}
    \label{tab:mean errors}
\end{table}

\subsubsection{Stellar Parameter Errors}
Uncertainty in effective temperature was determined by running \texttt{colte} with randomised \textit{Gaia} magnitudes (using errors provided by \textit{Gaia} DR3) and reddening values (taking errors to be 10\%) to produce new $\Teff$ estimates from the BP$-$RP colour. Nominal values of $\FeH$ and $\logg$ derived from Section\,\ref{subsec:stellar params} were used for this.

After retrieving the randomised $\Teff$'s, we derive new bolometric corrections and $\logg$ values using the methods outlined in Section \ref{subsec:logg methods}. Given that the parallax was also randomised, several stars had their $\logg$ derived either using the randomised parallax directly ($\pi \geq 3\sigma$) or using the randomised $\Teff$ with initial $\FeH$ to interpolate over isochrones ($\pi < 3\sigma$).

In order to rapidly estimate how an uncertainty in $\Teff$ will affect the derived $\FeH$, we assumed a linear dependence between these two quantities. We estimated the slope using a test spectrum fit for each spectrum when perturbing $\Teff$ by $+100$\,K. From tests on both the Wide-Field Spectrograph (WiFeS) \citep{dopita_wide_2010} data and on the globular clusters (see Section \ref{section:validation}), we imposed a floor of the precision of 0.2\,dex in metallicity. We found that \texttt{RVSpecFit} metallicity errors were unreliable due to residual sky emission, sometimes resulting in extremely large uncertainties that were not representative of the actual quality of the fit. 

Uncertainties in the radial velocities were determined directly within \texttt{RVSpecFit} via the Nelder-Mead fit to the spectra. At the resolution of our spectra, the relatively small stellar parameter uncertainties on the radial velocity determination is smaller than the measurement error itself, and we therefore neglect it.

Uncertainties in $\XFe{C}$ were found by inputting randomised $\Teff$, $\logg$ and $\FeH$ into our explicit CH G-band fitting procedure described in Section\,\ref{subsec:carb abundance} to get randomised $\XFe{C}$ values. 

As for determining $2\sigma$ detection limits for $\XFe{C}$, we explicitly integrated the likelihood as a function of the $\XFe{C}$ abundance. We used synthetic spectra covering $-5.0 \le \XFe{C} \le -\FeH + 0.5$ in steps of 0.01\,dex, calculating the likelihood for each. We also calculated the likelihoods when comparing to the synthetic spectrum with $\XFe{C} = -5.0$, under the assumption that this had negligible CH absorption, and used this case as our null hypothesis. We adopted a $2 \sigma$ detection limit at the abundance where our synthetic spectra differ at the $2 \sigma$ level from the spectrum with $\XFe C = -5.0$. In cases where the best-fitting spectrum had an abundance below this $2 \sigma$ detection limit, we report this abundance value as the upper limit to the fit. The fits of the CH G-band for our 15 EMPs are shown in \ref{append:gband fits}.

\subsubsection{Distance}
\label{subsec:distances}
Distances and their errors were determined by using the randomised parallax directly in the distance modulus if the parallax quality was good ($\pi \geq 3 \sigma$), or through isochrones with randomised $\FeH$, BP$-$RP and $E(B-V)$ values if the parallax quality was poor ($\pi < 3 \sigma$). The median was then taken as the final value, with the uncertainties being the standard deviation of the distribution. 

\begin{figure*}[tp]
    \centering
    \includegraphics[width=0.7\linewidth]{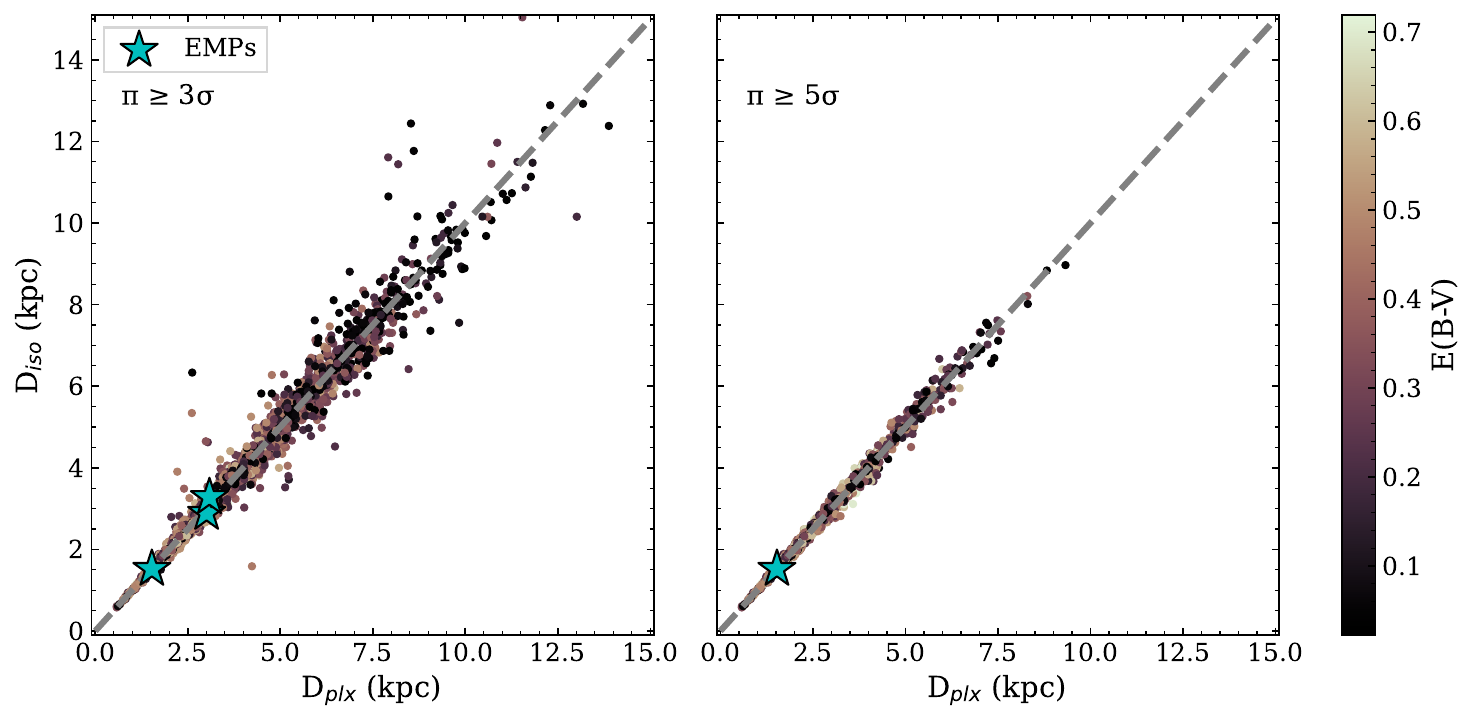}
    \caption{Distance comparison between those derived from inverting parallax ($\rm{D_{plx}}$) and from isochrones ($\rm{D_{iso}}$), colour-coded by $E(B-V)$. Stars with parallax quality $\pi \geq 3 \sigma$ is shown on the left panel. The right panel shows the comparison for stars with $\pi \geq 5 \sigma$.}
    \label{fig:dist_compare}
\end{figure*}

For stars with $\pi \geq 3 \sigma$, we show in Fig.\,\ref{fig:dist_compare} that the two techniques correspond well with each other. The scatter decreases for stars with $\pi \geq 5 \sigma$, showing that our isochrone distances are reliable.

\subsection{Methodology Validation} 
\label{section:validation}
The tools we developed in Section \ref{subsec:stellar params} and \ref{subsec:param uncertainty} were tested with internal consistency checks: first involving the HR diagram with isochrones, and then on several independent datasets. These datasets involve comparing stellar parameters with data from WiFeS (Section\,\ref{subsec:wifes validation}), with stars in globular clusters observed with the same instrumental setup (Section\,\ref{subsec:gc validation}), and with those from other \textit{Gaia} XP surveys (Section\,\ref{subsec:gaia xp spectroscopy}). No mention of our stars is made in any high-resolution literature.

\subsection{HR Diagram}
\begin{figure*}
    \centering
    \includegraphics[width=1\linewidth]{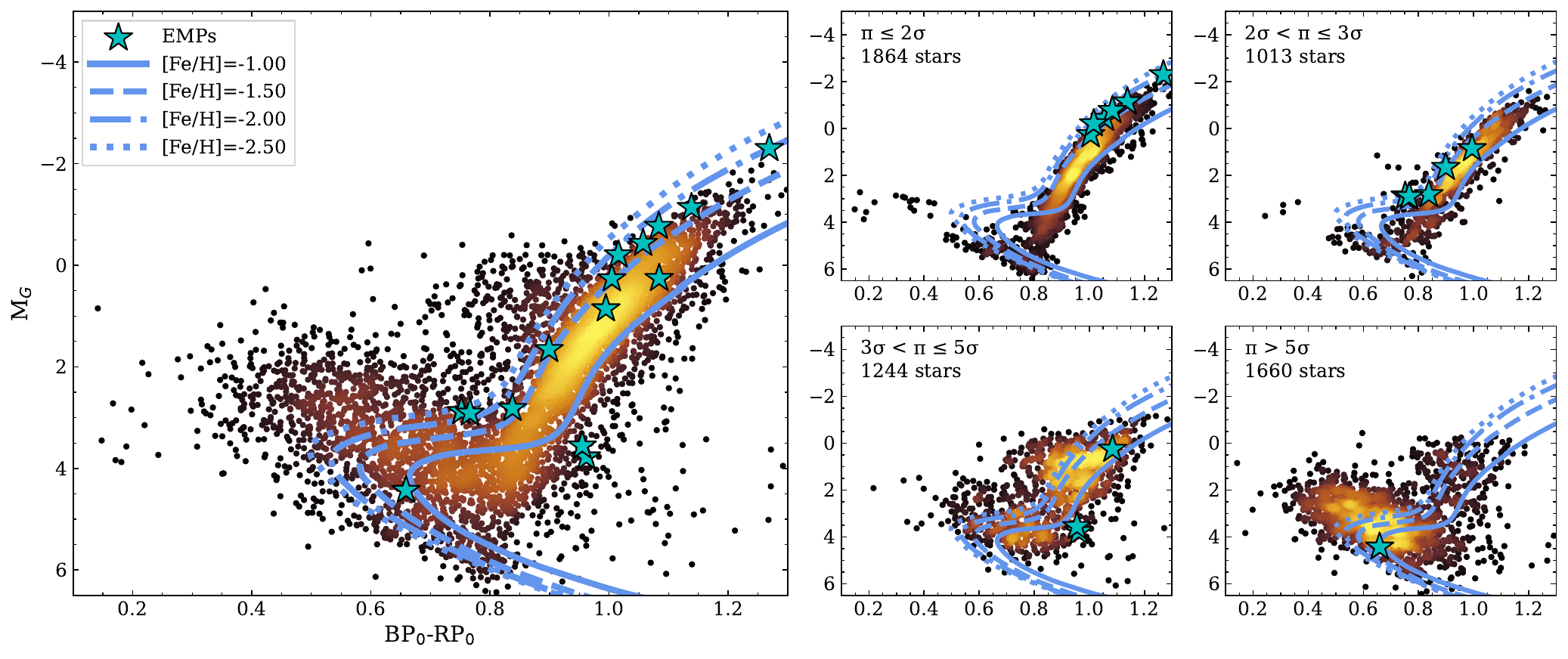}
    \caption{HR diagram of the full sample using $\rm{BP_0 - RP_0}$ (corrected for reddening, detailed in Section \ref{subsec:teff methods}) and $\rm{M_G}$ (using distances described in Section \ref{subsec:distances}), sub-sectioned according to parallax quality. Stars are coloured by their Kernel Density Estimator density, where lighter colours indicate higher density. Our 15 vetted EMP stars (see Section \ref{sec:results (chemodynamics)} for description on the vetting process) are shown by the blue star symbols. Dartmouth isochrones transformed to the \textit{Gaia} DR2 system are overplotted in blue from $\FeH =-1.00$\,dex to $\FeH =-2.50$\,dex in 0.50\,dex step-size, assuming an age of 12 Gyr. \textit{Left panel}: HR diagram for the full sample. \textit{Upper-middle panel}: HR diagram for stars with parallax quality $\leq 2\sigma$. \textit{Lower-middle panel}: HR diagram for stars with parallax quality $3\sigma < \pi \leq 5\sigma$. \textit{Upper-right panel}: HR diagram for stars with parallax quality $2\sigma < \pi \leq 3\sigma$. \textit{Lower-right panel}: HR diagram for stars with parallax quality $\geq 5\sigma$.}
    \label{fig:aat hr}
\end{figure*}

The sample in this work covers a broad range of evolutionary stages. We show this through the Hertzsprung-Russell (HR) diagram in Fig.\,\ref{fig:aat hr} segmented by parallax ($\pi$) quality, where it is evident that both turn-off and red giant branch (RGB) stars are included in the survey. The majority of the stars on the RGB have poor parallaxes ($\pi \leq 2\sigma$ and $2\sigma < \pi \leq 3\sigma$), while those at the turn-off generally have good parallaxes ($\pi > 5\sigma$). BP$-$RP colours and \textit{Gaia} DR3 magnitudes were corrected for reddening using the Schlegel dustmap as described in Section \ref{subsec:teff methods}, and $\rm{M_G}$ was calculated using either parallax measurements or isochrone fits as described in Section \ref{subsec:distances}.

\begin{table*}[tp]
    \centering
    \caption{The initial and shifted stellar parameters for three EMPs in Fig.\,\ref{fig:aat hr}. The first two stars (ra\_1639-2632-s419 and ra\_1659-2154-s261) were shifted from $\rm{(BP_0 - RP_0, M_G)} \approx (0.95, 3.60)$ to $\rm{(BP_0 - RP_0, M_G)} = (0.86, 2.43)$, whilst the third star (ra\_1604-2712-s188) was shifted from $\rm{(BP_0 - RP_0, M_G)} \approx (0.66, 4.50)$ to $\rm{(BP_0 - RP_0, M_G)} = (0.62, 4.41)$.}
    \begin{tabular}{l|cccc|cccc}
        \hline
          & \multicolumn{4}{c|}{Initial} & \multicolumn{4}{c}{Shifted} \\
        Star & $E(B-V)_{i}$ & $T_{\textrm{eff},i}$ & $\FeH_i$ & $\chi^{2}_{i}$ & $E(B-V)_{f}$ & $T_{\textrm{eff}, f}$ & $\FeH_f$ & $\chi^{2}_{f}$ \\
        \hline
        ra$\_$1639-2632-s419 & 0.37 & 5016 & $-$3.02 & 4.47 & 0.45 & 5354 & $-$2.62 & 4.31 \\
        ra$\_$1659-2154-s261 & 0.35 & 5044 & $-$3.19 & 5.72 & 0.43 & 5354 & $-$2.72 & 5.52 \\
        ra$\_$1604-2712-s188 & 0.069 & 6151 & $-$3.26 & 3.62 & 0.11 & 6266 & $-$3.24 & 3.43 \\
        \hline
    \end{tabular}
    \label{tab:silly emps}
\end{table*}

The location of three EMP stars on the HR diagram is inconsistent with their derived metallicities (see Table\,\ref{tab:emp params}). The likely reason for this mismatch is the adoption of erroneous reddening values for the stars. We found that if we allow the reddening to be a free parameter, we are then able to locate the stars in the HR diagram at locations consistent with their metallicities using higher reddening values. The higher reddening implies hotter effective temperatures and slightly higher metallicities.  With the larger reddenings, the first two stars increase their metallicities by $\sim 0.4$, with the third one increasing by $\sim 0.02$ dex. The initial and shifted parameters for the three stars, together with the initial and final reddening values are shown in Table\,\ref{tab:silly emps}. 

We note, however, that we do not use reddening as a free parameter for any other stars in the sample because of the difficulties in constraining it properly from spectral fits alone.  Indeed the adjustments for these three stars highlights the difficulties in constraining reddening, and this uncertainty is kept into account in our error budget.

\subsubsection{WiFeS}
\label{subsec:wifes validation}
\begin{figure*}[tp]
    \centering
    \includegraphics[width=1\linewidth]{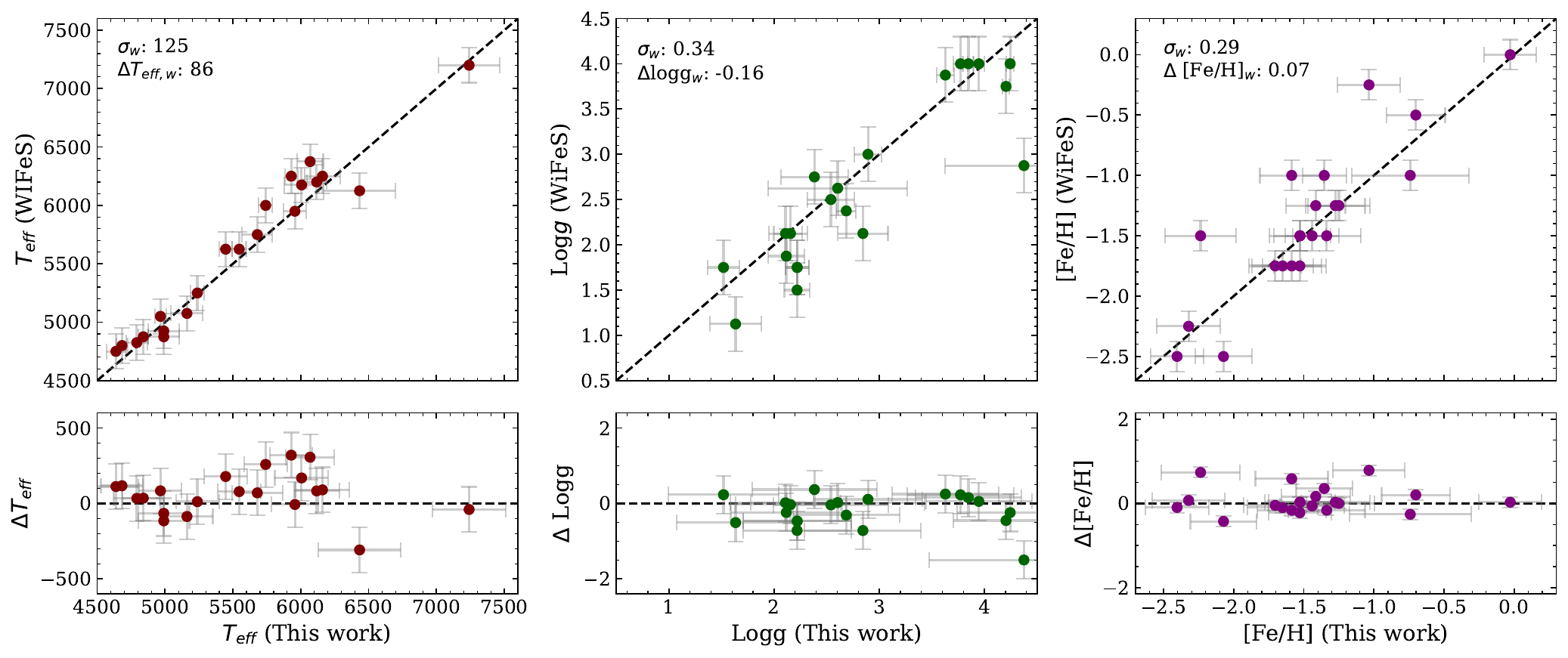}
    \caption{\textit{Upper panels}: Direct comparison of $\Teff$, $\logg$ and $\FeH$ values for 21 sample stars derived independently from the WiFeS and AAOmega spectra using different spectral analysis codes. The standard deviation and mean difference values are written into each subplot. The dashed line is the 1:1 relation. \textit{Lower panels}: The difference between the parameters derived from the AAOmega and WiFeS spectra; the dashed line is for zero difference.}
    \label{fig:wifes vs aat}
\end{figure*}

WiFeS is an Integral Field Unit (IFU) instrument on the 2.3m telescope at SSO. We observed 20 relatively bright stars from our sample with WiFeS using the B3000 grating, which covers 3200--5900\,\AA\ at $R \approx 3000$. Stellar parameters were derived for these stars from the WiFeS spectra using the \texttt{fitter} code (see \citealt{norris_most_2013}, with updates as described by \citealt{da_costa_spectroscopic_2023}). The accurate spectrophotometric calibration of the WiFeS spectra make them an ideal case to validate our stellar parameters. Fig.\,\ref{fig:wifes vs aat} shows the comparison for $\Teff$, $\logg$ and $\FeH$ together with their differences. The WiFeS stars are listed in Table\,\ref{tab:wife stars} with both sets of stellar parameters.

\begin{table*}[tp]
    \centering
    \caption{First three stars observed with the WiFeS spectrograph ordered by RA. Source ID column refers to the \textit{Gaia} DR3 source ID. If the star has a `good' parallax of $\pi \geq 3 \sigma$, then it is given True. Otherwise it is given False. Stellar parameters from the WiFeS spectra are given as well as those from our analysis. Full table available on the electronic version of the paper.}
    \setlength{\tabcolsep}{5.3pt}
    \begin{tabular}{l|ccccccccccc}
        \hline
        Source ID & RA & Dec & m$_{\textrm{G}}$ & $\pi \geq 3 \sigma$ & $T_{\rm eff, WiFeS}$ & $T_{\rm eff}$ & $\logg_{\rm WiFeS}$ & $\logg$ & $\FeH_{\rm WiFeS}$ & $\FeH$ \\
        \hline
        6042093888185179904 & 16:05:04.12 & $-$27:59:11.08 & 15.16 & True & 5950 & 5960 $\pm$ 60 & 2.125 & 2.8 $\pm$ 0.2 & $-$1.50 & $-$1.5 $\pm$ 0.2 \\
        6046160015317721216 & 16:40:01.16 & $-$26:43:55.94 & 15.23 & True & 7200 & 7200 $\pm$ 200 & 4.000 & 3.95 $\pm$ 0.05 & $-$0.25 & $-$1.0 $\pm$ 0.2 \\
        6034101641251466112 & 16:44:35.48 & $-$26:49:20.58 & 15.01 & True & 6250 & 6200 $\pm$ 100 & 3.000 & 2.9 $\pm$ 0.1 & $-$1.00 & $-$1.6 $\pm$ 0.2 \\
        \hline
    \end{tabular}
    \label{tab:wife stars}
\end{table*}

For $\Teff$ the agreement between the WiFeS values and our spectral analysis is good ($\sigma_w = 125$\,K, $\Delta T_{\rm{eff}, w} = 86$\,K)\footnote{$\sigma_w$ is the weighted standard deviation, whilst $\Delta T_{\rm{eff}, w}$ is the weighted mean difference, in this case for $\Teff$. Weights come from the uncertainty in the stellar parameters.}, particularly for $\Teff < 5500$\,K, where the scatter is low. At higher temperatures, the scatter increases with increasing temperatures with the WiFeS data overestimating $\Teff$ compared to our values. We note that \citet{da_costa_spectroscopic_2023} also saw this when they compared WiFeS $\Teff$'s with GALAH DR3 $\Teff$'s (see their fig.\,2).

For $\logg$, the comparison is again good with $\sigma_w = 0.34$ and $\Delta \logg_w = -0.16$\,dex. The WiFeS $\logg$ values are quantised at the 0.125 dex level (so precision is at best half of this), but \citet{dacosta_skymapper_2019} showed that the true estimate for the WiFeS $\logg$ errors is likely around 0.3--0.35\,dex. The $\logg$ random errors for our sample are $\sim$0.3\,dex also, so we would expect a scatter of $\sim0.42$\,dex in the differences. The $\sigma_w$ value of 0.34 is close to this value from which we conclude that it is likely that our analysis yields similar precisions for $\logg$ as the WiFeS analysis. It is worth noting that $\logg$ values from WiFeS are independent from parallax (derived directly from spectrophotometric fits), so the lack of any systematic differences ($\Delta \logg_w = -0.16$) makes our approach reliable and consistent.

The $\FeH$ comparison between WiFeS and us ($\sigma = 0.29$\,dex, $\Delta \FeH_w = 0.07$\,dex) is also good. In particular, the agreement is satisfactory across the entire metallicity range with no obvious systematic trend at the lowest metallicities.

\subsubsection{Globular clusters}
\label{subsec:gc validation}
In \citet{da_costa_ca_2016}, a study of RGB stars in Galactic globular and open clusters was performed using observational data collected using the same instrumental setup as this work (see Section \ref{subsec:aaomega spectroscopy}). For consistency, we reduced the raw 2dF data from the \citet{da_costa_ca_2016} observations ourselves using the same version of \texttt{2dfdr} (version 8.03b) as for our main sample. These globular cluster spectra therefore provide an excellent data set on which to test our analysis code. 

\begin{figure*}[tp]
    \centering
    \includegraphics[width=1\linewidth]{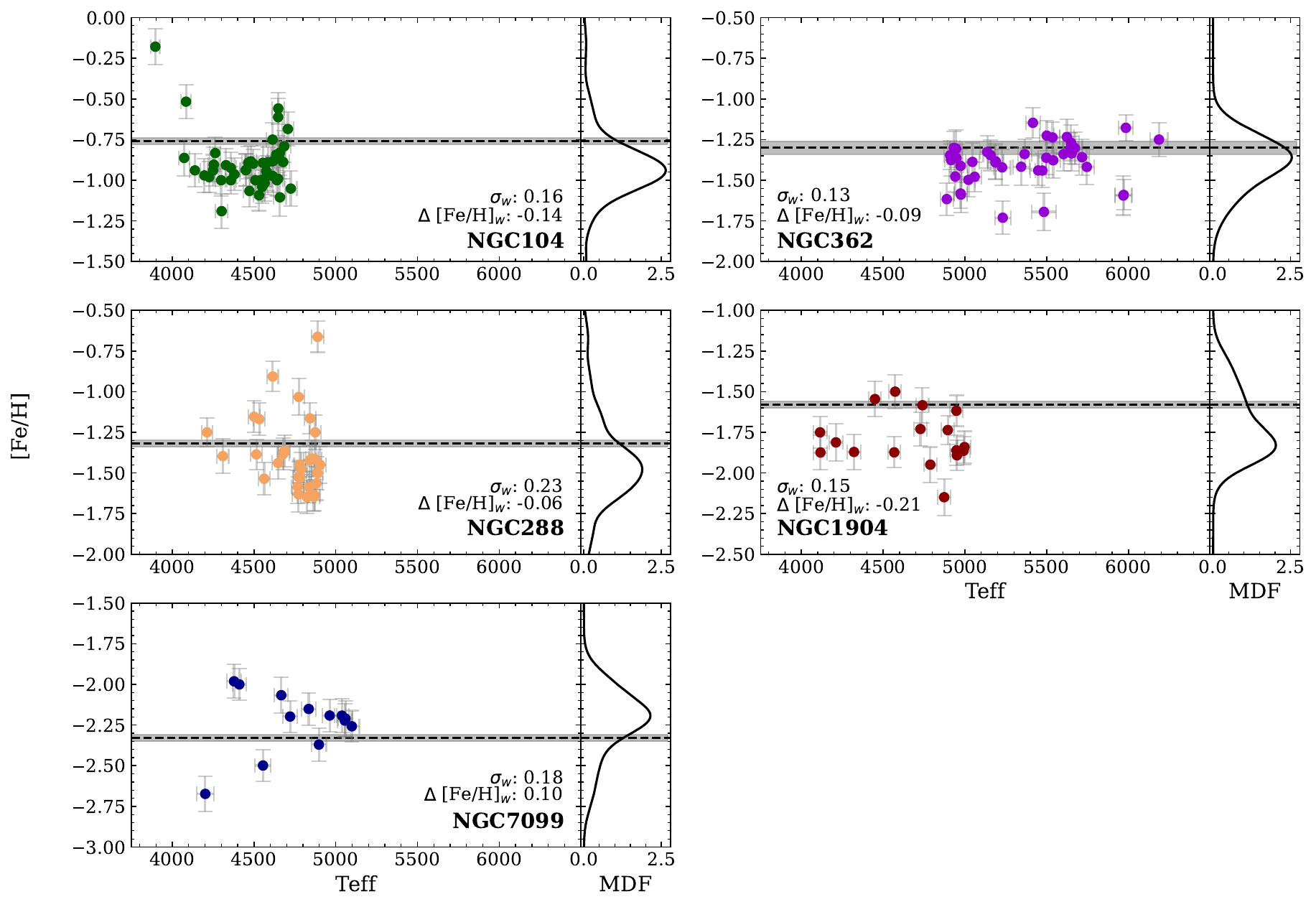}
    \caption{Stellar parameter results for globular clusters NGC\,104 (47), NGC\,362 (48), NGC\,288 (30), NGC\,1904 (17) and NGC\,7099 (14). Each subplot has an accompanying metallicity distribution function (MDF) using the kernel density estimator of the cluster's metallicity. The horizontal dashed lines represent the reference values provided by \citet{carretta_intrinsic_2009}. The errors in this are shown by the shaded region. The weighted standard deviation ($\sigma_W$) and mean difference ($\Delta \FeH_w$) for each cluster against the reference value is printed on the subplots.}
    \label{fig:gc results}
\end{figure*}

The globular cluster results are shown in Fig\,\ref{fig:gc results}, with the addition of NGC\,362, which was observed as part of our program for validation purposes. The \citet{carretta_intrinsic_2009} metallicities for the clusters are shown by the black dashed line.

Overall, we find that the metallicity bias and scatter for each cluster is less than 0.25\,dex, which implies that our analysis is acceptably accurate. Crucially, we do not find significant biases for the warmest or coolest stars observed.

The results for all five of our sample clusters (NGC\,104, NGC\,362, NGC\,288, NGC\,1904 and NGC\,7099) are in good agreement with literature. For any individual cluster star, the uncertainty we get from our analysis is a good representation of the true uncertainty in our main sample.

\subsubsection{\textit{Gaia} XP Surveys}
\label{subsec:xp validation}
Surveys targeting metal-poor stars like Pristine or SkyMapper have proven to provide reliable estimates of metallicities thanks to their use of metallicity-sensitive photometric colours \citep[e.g.][]{starkenburg_pristine_2017, dacosta_skymapper_2019, casagrande_skymapper_2019, chiti_stellar_2021}. Newer surveys utilising \textit{Gaia} XP data take advantage of more complex tools like forward modelling and machine learning to derive stellar parameters, which we explore in more detail in \ref{sec:gaia xp compare}. Here we illustrate a comparison of our data to that of \citet{martin_pristine_2023}, who used \textit{Gaia} XP spectra to calculate synthetic Pristine photometry to infer metallicity estimates.

\begin{figure}[tp]
    \centering
    \includegraphics[width=1\linewidth]{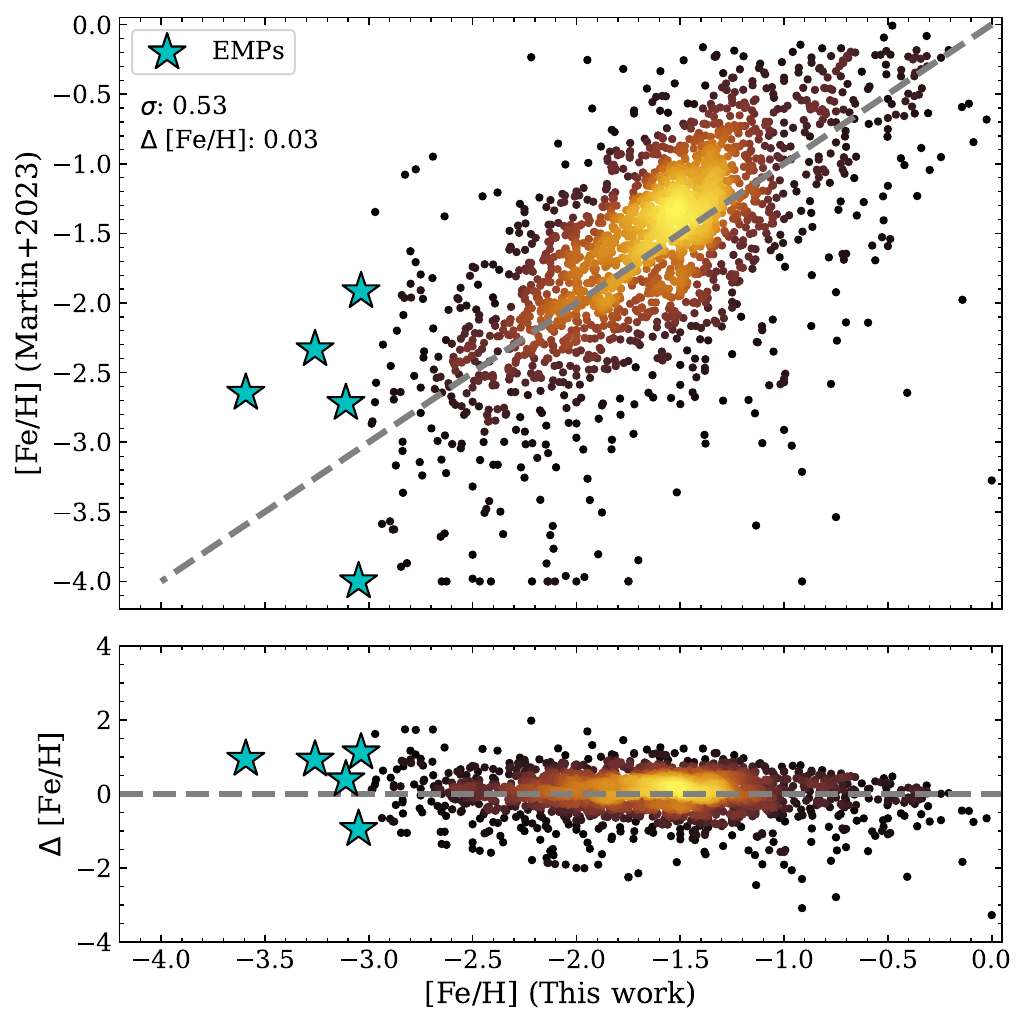}
    \caption{Metallicity comparison between our data and that of \citet{martin_pristine_2023} for 2125 stars in common, 5 of which are identified in our sample as EMPs. Standard deviation and mean difference are shown on the plot. Colour represents density, as in Fig. \ref{fig:aat hr}.}
    \label{fig:aat martin}
\end{figure}

A cross-match of their catalogue with our data revealed 2125 stars in common, including 5 of our EMP stars.  One of these is also an EMP star in \citet{martin_pristine_2023}.  Fig.\,\ref{fig:aat martin} shows the metallicity estimates comparison.  Overall, there is a good correlation between the two estimates with a low mean difference ($\mu = 0.03$) but a substantial standard deviation ($\sigma = 0.53$). Given the uncertainties in both metallicity determinations, the agreement is satisfactory. To understand how the \citet{martin_pristine_2023} metallicities inferred from the photometric indices derived \textit{Gaia} XP spectra change with magnitude, we selected first stars with $19 \leq S/N \leq 21$ from our spectra. We then separated these stars into `bright' ($m_G \leq 16$; 24 stars) and `faint' ($m_G > 16$; 59 stars) groups.  We then saw that the scatter in the metallicity differences for the bright group  ($\sigma = 0.50$) is lower than for the faint group ($\sigma = 0.64$). This implies that the metallicities inferred from \textit{Gaia} XP colours downgrade with fainter magnitudes, whilst our metallicities remains consistent at fixed S/N. Our technique is apparently more robust for fainter stars than the XP spectra, driven by its degrading S/N quality at lower magnitudes.

\section{Chemodynamics of the metal poor Galaxy}
\label{sec:results (chemodynamics)}
\subsection{Chemistry}
\begin{figure*}
    \centering
    \includegraphics[width=1\linewidth]{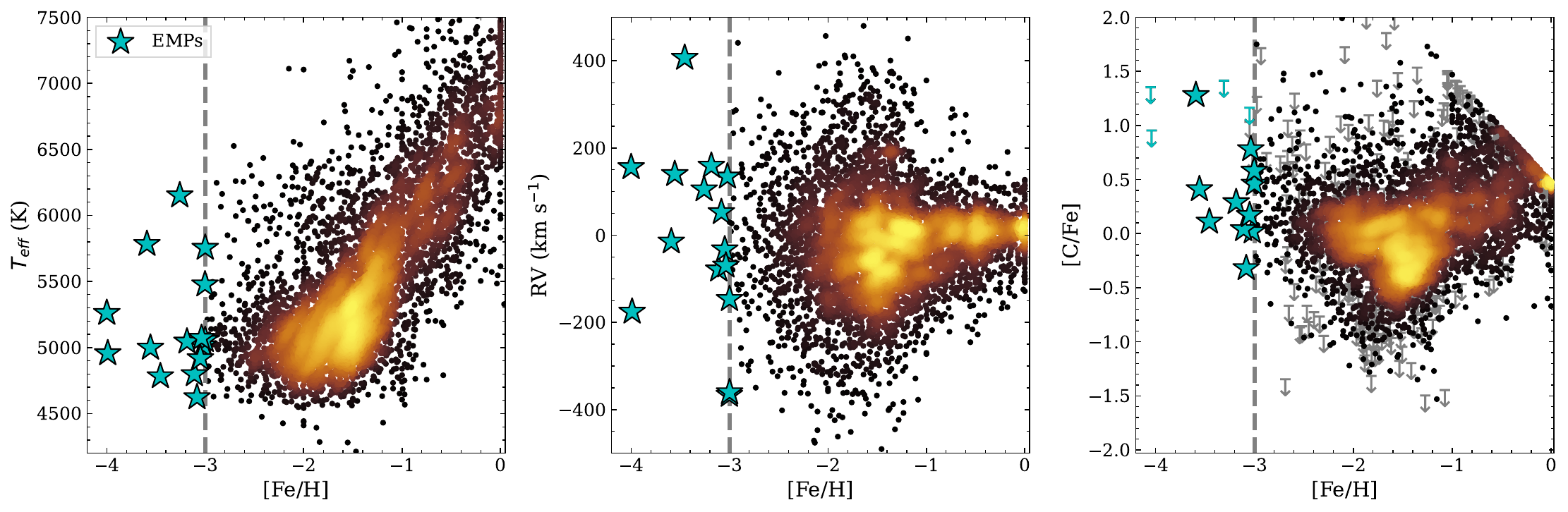}
    \caption{Stellar parameter results for our sample, showing effective temperature (\textit{left}), heliocentric radial velocity in \kms\ (\textit{centre}) and [C/Fe] (\textit{right}) results against $\FeH$.  In the right panel downward arrows indicate upper limits in cases where the CH G-band could not be detected. EMPs with $\FeH{} < -3$ have been individually vetted; the vertical dashed line indicates this limit for clarity. Note that the stars on the diagonal line seen in upper right of the $\XFe{C}$ panel represents the synthetic grid edges.}
    \label{fig:aat params}
\end{figure*}

The stellar parameters for the stars in our sample cover a wide parameter space for $\Teff$, $\logg$, $\FeH$ and $\XFe{C}$. The derived values for the full sample are shown in Fig.\,\ref{fig:aat params} for effective temperature, radial velocity and [C/Fe], as a function of $\FeH$. From the sample, we have identified 15 new candidate EMPs, vetted manually by visual inspection of the spectra and of their synthetic fits. Stars with $\FeH < -3.0$\,dex that did not pass the vetting procedure were removed from the dataset due to poor fits. The lowest metallicity in the sample belongs to star ra\_1633-2814\_s284, with $\FeH = -4.0 \pm 0.20$. Stellar parameters for the vetted EMP sample are given in Table\,\ref{tab:emp params}. These stars are denoted with blue star symbols across the Figures in this paper; a vertical grey-dashed line drawn at $\FeH =-3.0$\,dex in the panels of Fig.\,\ref{fig:aat params} shows the separation of the vetted EMPs from the rest of the dataset.

\begin{table*}[tp]
    \centering
    \caption{Stellar parameters for our 15 EMP stars, ordered by RA. The \textit{Gaia} DR3 source ID follows our star IDs. The $\XFe{C}$ values presented are our non-corrected values, with the evolutionary corrections, $\Delta \XFe{C}$, shown in the final column. The correction only applies to RGB stars ($\logg < 3.0$). Dwarfs automatically have a value of zero. Kinematic assignments are given (see Section \ref{subsec:aat kinematics}), with HAL being halo, RET being retrograde disk, PRO being prograde disk, and GSE being Gaia Sausage-Enceladus.}
    \setlength{\tabcolsep}{2.3pt}
    \begin{tabular}{ll|cccccccccc}
        \hline
        Star ID & Gaia ID & RA & Dec & m$_{\textrm{G}}$ & $\pi \geq 3 \sigma$ & $T_{\rm eff}$ & $\logg$ & $\FeH$ & $\XFe{C}$ & $\Delta \XFe{C}$ & Pop \\
        \hline
        ra\_0103-7050\_s236 & 4691686763639138688 & 01:03:18.24 & $-$70:07:30.7 & 16.44 & False & 4620 $\pm$ 10 & 1.0 $\pm$ 0.4 & $-$3.1 $\pm$ 0.1 & $-$0.32 $\pm$ 0.02 & 0.74 & HAL \\
        ra\_0752-5047\_s47  & 5513982132481251456 & 07:58:33.25 & $-$50:44:34.3 & 16.38 & True  & 4780 $\pm$ 90 & 2.4 $\pm$ 0.7 & $-$3.5 $\pm$ 0.2 & 0.09 $\pm$ 0.07 & 0.01 & HAL \\
        ra\_1604-2712\_s188 & 6043056235732990848 & 16:00:33.27 & $-$26:59:31.2 & 15.54 & True  & 6150 $\pm$ 40 & 4.31 $\pm$ 0.05 & $-$3.3 $\pm$ 0.1 & $<$ 1.4 & 0.00 & HAL \\
        ra\_1604-2712\_s292 & 6042817710432143104 & 16:02:07.43 & $-$27:42:03.7 & 16.77 & False & 5780 $\pm$ 60 & 3.4 $\pm$ 0.2 & $-$3.6 $\pm$ 0.1 & 1.3 $\pm$ 0.1 & 0.00 & HAL \\
        ra\_1633-2814\_s284 & 6044482989511923584 & 16:33:14.64 & $-$28:03:05.1 & 17.13 & False & 5300 $\pm$ 200 & 2.4 $\pm$ 0.7 & $-$4.0 $\pm$ 0.2 & $<$ 1.3 & 0.00 & GSE \\
        ra\_1633-2814\_s130 & 6044035900595959424 & 16:34:49.20 & $-$28:28:09.8 & 16.88 & False & 5000 $\pm$ 200 & 1.8 $\pm$ 0.7 & $-$3.6 $\pm$ 0.1 & 0.4 $\pm$ 0.2 & 0.02 & GSE \\
        ra\_1639-2632\_s419 & 6046167303889429888 & 16:40:04.93 & $-$26:37:02.1 & 17.18 & True  & 5000 $\pm$ 100 & 3.8 $\pm$ 0.9 & $-$3.0 $\pm$ 0.2 & 0.0 $\pm$ 0.1 & 0.00 & PRO \\
        ra\_1648-2642\_s91  & 6033877134724691328 & 16:51:13.49 & $-$27:27:24.6 & 16.56 & False & 5060 $\pm$ 70 & 2.0 $\pm$ 0.2 & $-$3.0 $\pm$ 0.1 & 0.47 $\pm$ 0.08 & 0.02 & HAL \\
        ra\_1659-2154\_s347 & 4126178792119445888 & 16:56:39.11 & $-$22:14:04.6 & 17.28 & False & 5500 $\pm$ 100 & 3.1 $\pm$ 0.4 & $-$3.0 $\pm$ 0.1 & 0.6 $\pm$ 0.2 & 0.00 & HAL \\
        ra\_1659-2154\_s261 & 4126309672676244096 & 16:58:15.46 & $-$21:37:03.5 & 16.97 & True  & 5000 $\pm$ 100 & 3.6 $\pm$ 0.4 & $-$3.2 $\pm$ 0.3 & 0.30 $\pm$ 0.05 & 0.00 & PRO \\
        ra\_1659-2154\_s114 & 4127849985388551808 & 17:00:18.31 & $-$21:05:57.2 & 17.15 & False & 4960 $\pm$ 70 & 1.7 $\pm$ 0.2 & $-$4.0 $\pm$ 0.1 & $<$ 0.9 & 0.10 & HAL \\
        ra\_1752-4300\_s155 & 5956522780887593344 & 17:49:24.09 & $-$42:38:14.4 & 15.63 & False & 4800 $\pm$ 50 & 1.4 $\pm$ 0.3 & $-$3.1 $\pm$ 0.1 & 0.05 $\pm$ 0.06 & 0.50 & PRO \\
        ra\_1752-4300\_s269 & 5956280128069527808 & 17:54:29.31 & $-$43:00:13.7 & 16.53 & False & 4920 $\pm$ 40 & 1.7 $\pm$ 0.5 & $-$3.0 $\pm$ 0.1 & 0.17 $\pm$ 0.05 & 0.20 & HAL \\
        ra\_1832-3457\_s438 & 6734926160922018048 & 18:30:01.04 & $-$34:21:53.9 & 17.29 & False & 5750 $\pm$ 40 & 3.4 $\pm$ 0.4 & $-$3.0 $\pm$ 0.1 & $<$ 1.1 & 0.00 & HAL \\
        ra\_1853-3255\_s51  & 6732834065149293312 & 18:53:59.72 & $-$33:29:07.2 & 15.58 & False & 5070 $\pm$ 20 & 2.1 $\pm$ 0.2 & $-$3.0 $\pm$ 0.1 & 0.79 $\pm$ 0.03 & 0.01 & HAL \\ 
        \hline
    \end{tabular}
    \label{tab:emp params}
\end{table*}

The left panel of Fig.\,\ref{fig:aat params} shows that in general, hotter stars are more metal-rich than their cooler counterparts. This trend is due to our survey selecting primarily RGB stars, with evolutionary effects prohibiting hot metal-poor stars being present. In general, hotter stars tend to be more massive and younger, therefore having higher metallicities \citep{nordstrom_geneva-copenhagen_2004, holmberg_geneva-copenhagen_2007, holmberg_geneva-copenhagen_2009, casagrande_new_2011}. We see the largest concentration of stars at $\Teff \sim 5100$, before the number of stars decreases with lower metallicities when keeping this $\Teff$ fixed. A `typical' star in our sample has $\Teff \sim 5200$ and $\FeH \sim -1.5$ (according to the highest concentrations in the plot). Our EMP candidates cover a sizeable temperature range from $T_{\textrm{eff}} \approx 4600$\,K to $T_{\textrm{eff}} \approx 6200$\,K with metallicities as low as $\FeH$\,$\approx -4.0$\,dex, emphasising the diversity of the population that our survey is sampling. 

The heliocentric radial velocity results in the middle panel of Fig \ref{fig:aat params} are broadly consistent with the expectations for a sample that moves from thick-disk dominance at higher metallicities with $\sigma(rv) \sim 45\,$\kms\ \citep[e.g.][]{zwitter_radial_2008, kordopatis_chemodynamics_2020}, to an increasing halo dominance at lower metallicities with $\sigma(rv) \sim 150$\,\kms\ \citep[e.g.][]{carollo_two_2007, carollo_structure_2010}. At the lowest metallicities ($\FeH < -2.5$) there are fewer stars but the dispersion apparently remains large.

Our [C/Fe] results, shown in the right panel of Fig.\,\ref{fig:aat params}, highlights the significant variance we have in our carbon abundances. The distribution of $\XFe{C}$ abundances reflects the range of evolutionary stages present in our sample, together with the uncertainties in the determinations, as well as any population dependent variations in intrinsic $\XFe{C}$ values. 

\begin{figure}[tp]
    \centering
    \includegraphics[width=1\linewidth]{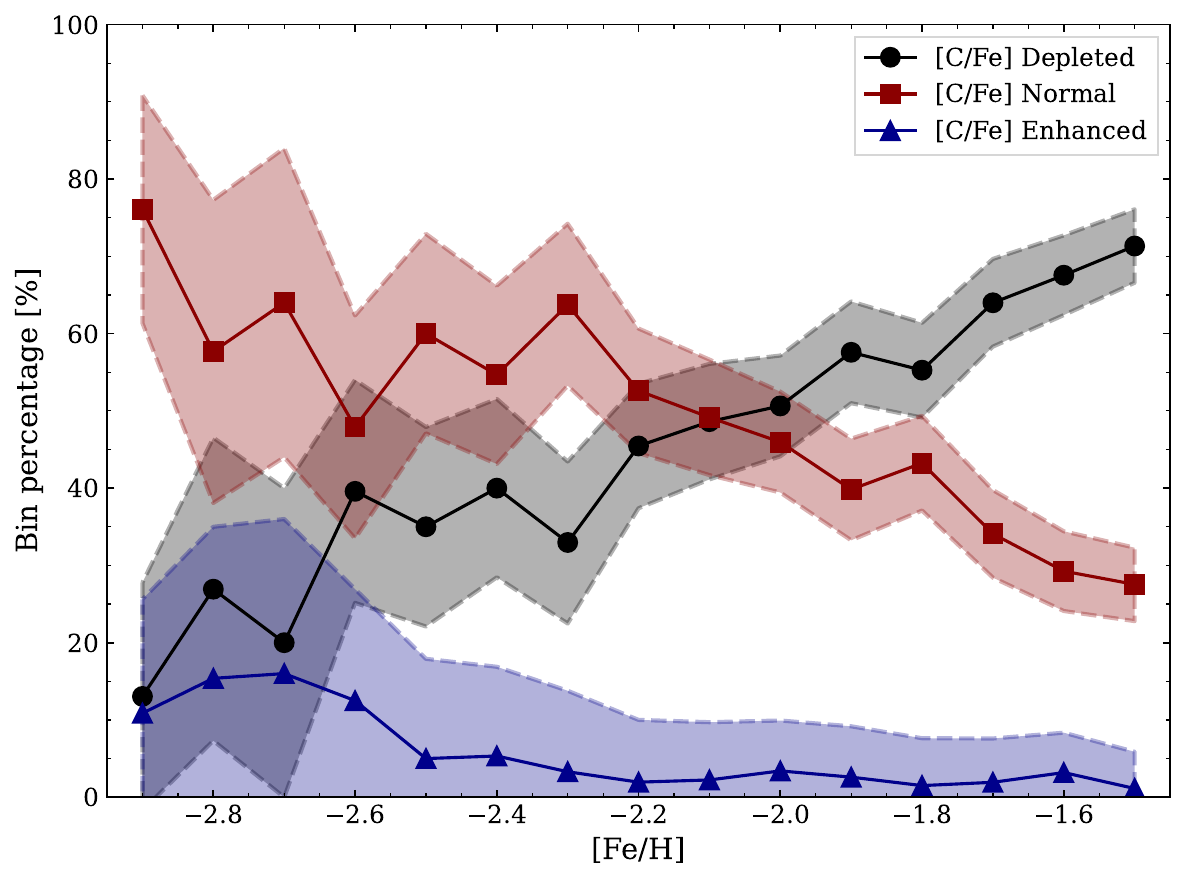}
    \caption{The [C/Fe] distribution for stars with $\FeH{} \leq -1.5$ based on their $\XFe{C}$ values: carbon-depleted ($\XFe{C} < 0$) in black circles, carbon-normal ($0 \leq \XFe{C} < 0.7$) in red squares, and carbon-enhanced ($\XFe{C} \geq 0.7$) in blue triangles. Only stars with detectable\,CH G-band are shown; those with upper limits are ignored. Bins are 0.1\,dex in size. The last bin ($< -2.9$) includes all stars below this metallicity. Shaded regions are errors on each frequency assuming Poisson statistics.}
    \label{fig:carb distribution}
\end{figure}

To further investigate the $\XFe{C}$ results, the sample was split into regions based on their $\XFe{C}$ values: carbon-depleted ($\XFe{C} < 0$), carbon-normal ($0 \leq \XFe{C} < 0.7$) and carbon-enhanced ($\XFe{C} \geq 0.7$). For the stars with $\FeH \leq -1.5$, we show in Fig.\,\ref{fig:carb distribution} the distributions of measured $\XFe{C}$ values as a function of [Fe/H] for each region. Stars with upper limits on $\XFe{C}$ were excluded. In total, for those stars with $\FeH \leq -1.5$, 2672 stars are carbon-depleted, 2735 are carbon-normal, and 388 are carbon-enhanced. We see that the portion of carbon-enhanced stars increases from 1.1\,\% to 16.0\,\% with decreasing metallicity. On the other-hand, the portion of carbon-depleted stars decreases from 71.3\% to 13.0\% when moving from $\FeH = -1.5$ to $-3.0$. This agrees with previous work that show similar trends \citep[e.g.][]{placco_carbon-enhanced_2014, singh_empirical_2020, arentsen_inconsistency_2022}. There are no obvious differences when separating the $\XFe{C}$ results into our different kinematic groupings (defined in Section \ref{subsec:orbit select}).

\begin{figure}[tp]
    \centering
    \includegraphics[width=1\linewidth]{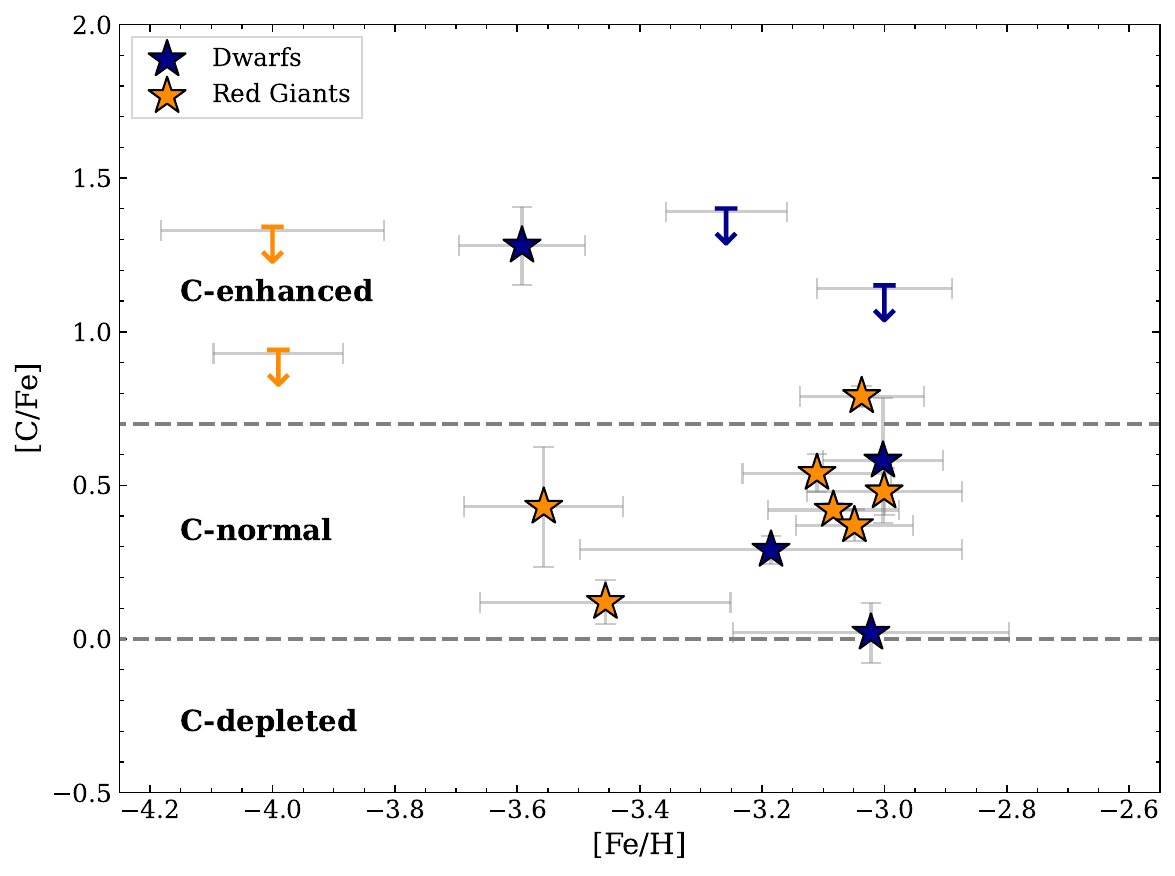}
    \caption{The [C/Fe] values for the 15 vetted EMP stars. Evolutionary mixing corrections have been applied to the carbon abundances for the RGB stars. Each star is marked either as a dwarf (log\,$g \geq 3.0$\,dex) or as a red giant (log\,$g < 3.0$\,dex). The plot is split into three regions based on [C/Fe] values. Uncertainties for both [C/Fe] and $\FeH$ are present. Downward arrows indicate $2 \sigma$ upper limits on [C/Fe] abundances.}
    \label{fig:emp carb vs z}
\end{figure}

The [C/Fe] results for our EMP stars are explored further in Fig.\ \ref{fig:emp carb vs z}. Defining dwarfs as $\logg \geq 3.0$ and giants as $\logg < 3.0$, we have two dwarfs and three giants with only upper limits for the $\XFe{C}$ values; these are represented by the downward arrows in the figure.  Excluding these five non-detections, and after applying evolutionary corrections to the abundance ratios for the RGB stars\footnote{\url{https://vplacco.pythonanywhere.com/}}, we see that eight EMPs have normal carbon values ($0 \leq \textrm{[C/Fe]} < 0.7$), two are carbon-enhanced ($\textrm{[C/Fe]} \geq 0.7$), and none are carbon-depleted ($\textrm{[C/Fe]} \leq 0$). For the carbon-enhanced stars, one star is a giant with $\XFe{C} = 0.79 \pm 0.03$ (the evolutionary correction is negligible for this star), and the other is a carbon-enhanced dwarf with $\XFe{C} = 1.3 \pm 0.1$. For the dwarf, given its low $\FeH$ abundance of $-3.6 \pm 0.1$, it may have obtained its abundances from a Pop. III progenitor that significantly enriched the surrounding gas with carbon. This is the explanation offered by, for example, \citet{christlieb_stellar_2002} for star HE0107--25240, and by \citet{nordlander_lowest_2019} for star SMSS J160540.18--144323.1, both of which are extremely carbon-rich ultra-metal-poor stars. In both cases, the enrichment in carbon may have arisen from a Pop. III `fallback' supernova \citep[e.g.][]{umeda_nucleosynthesis_2002, nomoto_nucleosynthesis_2013}. Both carbon-enhanced stars could also have received their carbon enhancement from mass transfer from a (now white dwarf) companion during its AGB star phase \citep[e.g.][]{whitehouse_dwarf_2018}. Tests to investigate which of the two carbon-enhancement mechanisms is likely include measuring abundances of the slow neutron capture process elements like Ba, enhanced in the mass transfer scenario, but not in the Pop. III fallback scenario. The CH G-band fits to all vetted EMP stars are shown in Fig.\,\ref{fig:emp carb fits 1} and Fig.\,\ref{fig:emp carb fits 2}

\subsection{Orbital Dynamics}
\label{subsec:aat kinematics}
To determine the dynamical properties of our sample, we used the Python code \texttt{galpy}\footnote{\url{https://github.com/jobovy/galpy}} that adopts the potentials in \citet{bovy_galpy_2015}, a simple and accurate model for the gravitational potential of the Milky Way. We transform celestial coordinates to galactocentric coordinates 
by assuming that the Sun is 20.8\,pc above the plane of the MW \citep{bennett_vertical_2019}, and has a peculiar velocity of $(U, V, W) = (11.1, 12.24, 7.25)$\,\kms\ \citep{schonrich_local_2010}. The Solar circular velocity was taken as 238\,\kms\ at the Solar radius of 8.2\,kpc \citep{bland-hawthorn_galaxy_2016}. Orbits were then integrated using a 4th order symplectic integrator in \texttt{galpy}, given distances derived from Section \ref{subsec:distances}, stellar coordinates and proper motions from \textit{Gaia} DR3, and heliocentric radial velocities from Section \ref{subsec:feh methods}. Orbital integration is computed over $\pm 1$\,Gyr using randomised distances, radial velocities and proper motions. The resulting distributions are then used to get uncertainties in the kinematic parameters.

\subsubsection{Orbital selection}
\label{subsec:orbit select}
\begin{figure*}[tp]
    \centering
    \includegraphics[width=1\linewidth]{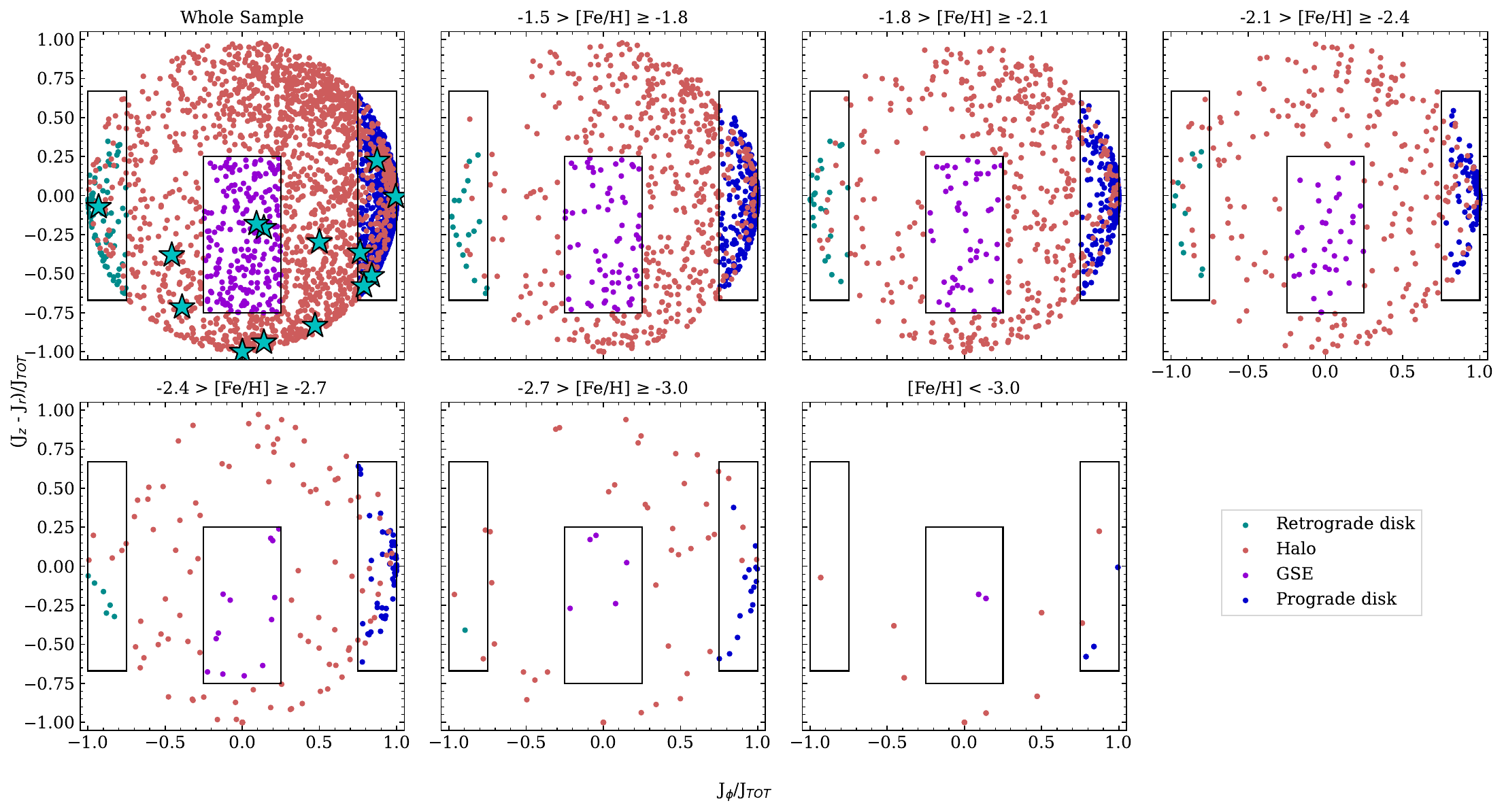}
    \caption{Action momentum space for the entire sample (\textit{upper left}), then separated into metallicity bins below $\FeH = -1.5$. The x-axis, $J_\phi/J_{\textrm{TOT}}$, is the azimuthal component of the action vector. The y-axis, $(J_z - J_r)/J_{\textrm{TOT}}$, is the difference between the star's vertical action and its radial action. Both axes are normalised by the total angular momentum $J_{\textrm{TOT}}$. Stars with $J_\phi/J_{\textrm{TOT}} \leq -0.75$ and $z_{\rm max} < 3$\,kpc are assigned as retrograde disk (cyan), whilst those with $J_\phi/J_{\textrm{TOT}} \geq 0.75$ and $z_{\rm max} < 3$\,kpc are assigned as prograde disk (blue). Stars with $-0.25 \leq J_\phi/J_{\textrm{TOT}} \leq 0.75$ and $-0.75 \leq (J_{z} - J_{r})/J_{\textrm{TOT}} \leq 0.25$ are assigned to GSE orbits (purple). Everything else is assigned as halo (orange). ``Disk'' stars that meet the $J_\phi/J_{\textrm{TOT}}$ requirement but not the $z_{\rm max} < 3$ requirement are assigned as halo. Note that in the final panel, two EMPs overlap with nearly identical values of $J_\phi/J_{\textrm{TOT}} \sim 0.0$ and $(J_{z} - J_{r})/J_{\textrm{TOT}} \sim -1.0$.}
    \label{fig:aat action}
\end{figure*}

\begin{figure*}
    \centering
    \includegraphics[width=1\linewidth]{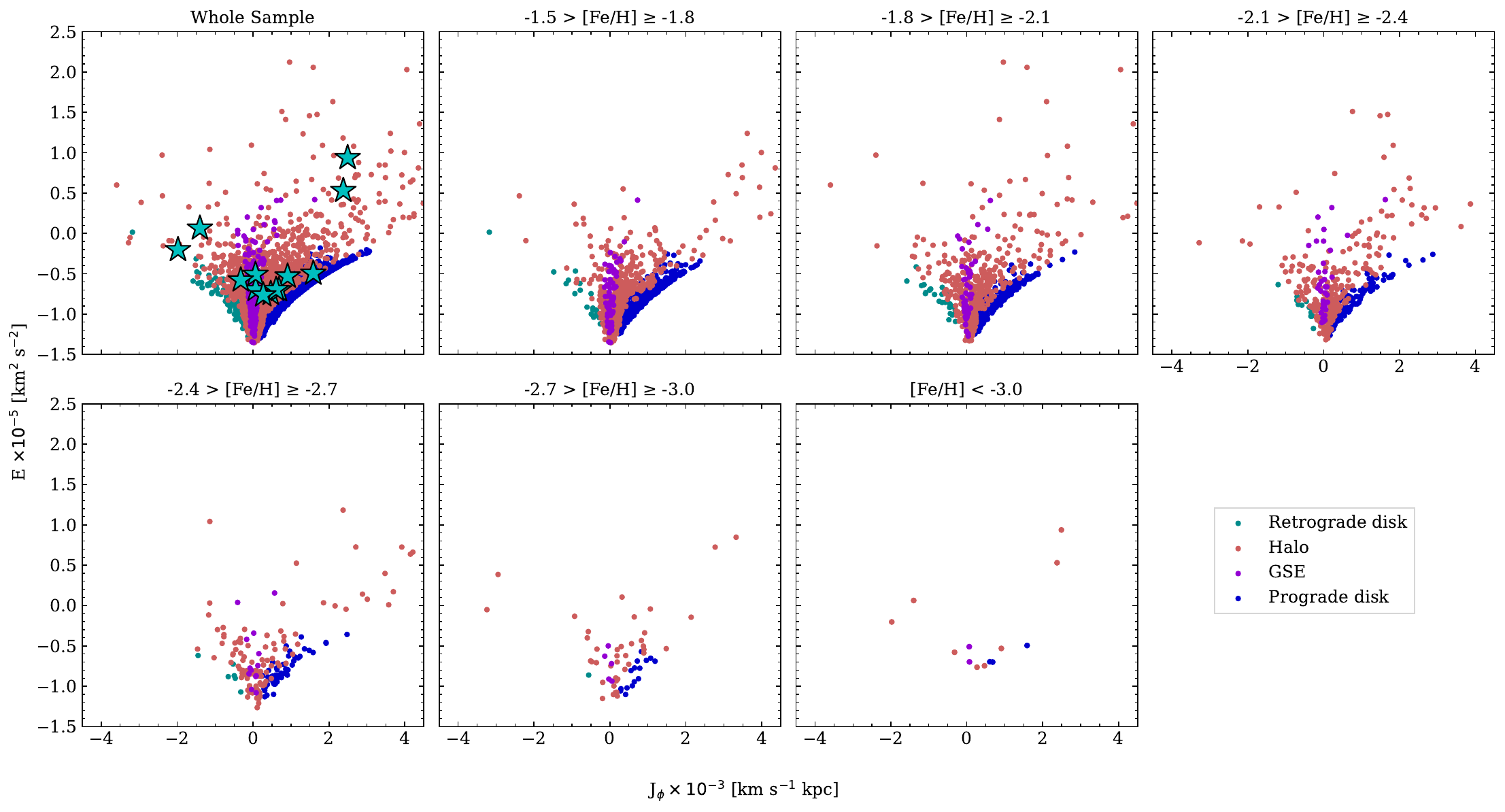}
    \caption{The angular momentum against energy plot for our sample. Same layout as Fig. \ref{fig:aat action}.}
    \label{fig:lz vs E}
\end{figure*}

Based on the resulting orbital parameters, we classified each star in the full sample into prograde, retrograde, halo and Gaia-Sausage-Enceladus (GSE) \citep{helmi_merger_2018, belokurov_co-formation_2018} orbits using the action momentum plot shown in Fig.\,\ref{fig:aat action}. This uses a combination of the azimuthal ($J_\phi$), vertical ($J_z$) and radial ($J_r$) components of the action vector (normalised by the total angular momentum, $J_{\textrm{TOT}}$). The equivalent angular moment vs energy plot is shown in Fig. \ref{fig:lz vs E}, where we show that our current selection holds in Lz vs E space. 

As was done in \citet{sestito_exploring_2021}, the plot is segmented into four different kinematic classifications based on values of $J_\phi/J_{\textrm{TOT}}$. Stars assigned with prograde disk orbits ($\rm{N_{pro}/N_{TOT}} = 62.9\%$) have $J_\phi/J_{\textrm{TOT}} \geq 0.75$ and $z_{\rm max} < 3$\,kpc. 

The majority of these stars belong to the thick disk, with only 8.0\% of the total prograde population have $z_{\rm max} < 0.3$\,kpc representative of the thin disk. Those assigned retrograde disk orbits ($\rm{N_{ret}/N_{TOT}} = 1.4\%$) have $J_\phi/J_{\textrm{TOT}} \leq -0.75$ and $z_{\rm max} < 3$\,kpc, with one star having $z_{\rm max} < 0.3$\,kpc.

The relative weight of retrograde against prograde disk populations is $\rm{N_{ret}/(N_{ret}+N_{pro})} = 2.2\%$ across all metallicities in our sample. Stars with $-0.25 \leq J_\phi/J_{\textrm{TOT}} \leq 0.75$ and $-0.75 \leq (J_{z} - J_{r})/J_{\textrm{TOT}} \leq 0.25$, regardless of their $z_{\rm max}$, are assigned to the GSE ($\rm{N_{gse}/N_{TOT}} = 3.9\%$), with the remainder classified as halo stars ($\rm{N_{hal}/N_{TOT}} = 32.0\%$). 

The relative weight of GSE against halo populations is $\rm{N_{gse}/(N_{gse}+N_{hal})} = 11.4\%$. A large portion of the sample is located within the prograde disk, followed by halo stars passing through the disk. Kinematic classifications for our 15 EMPs are shown at the end of Table\,\ref{tab:emp params}.

\begin{figure}[tp]
    \centering
    \includegraphics[width=1\linewidth]{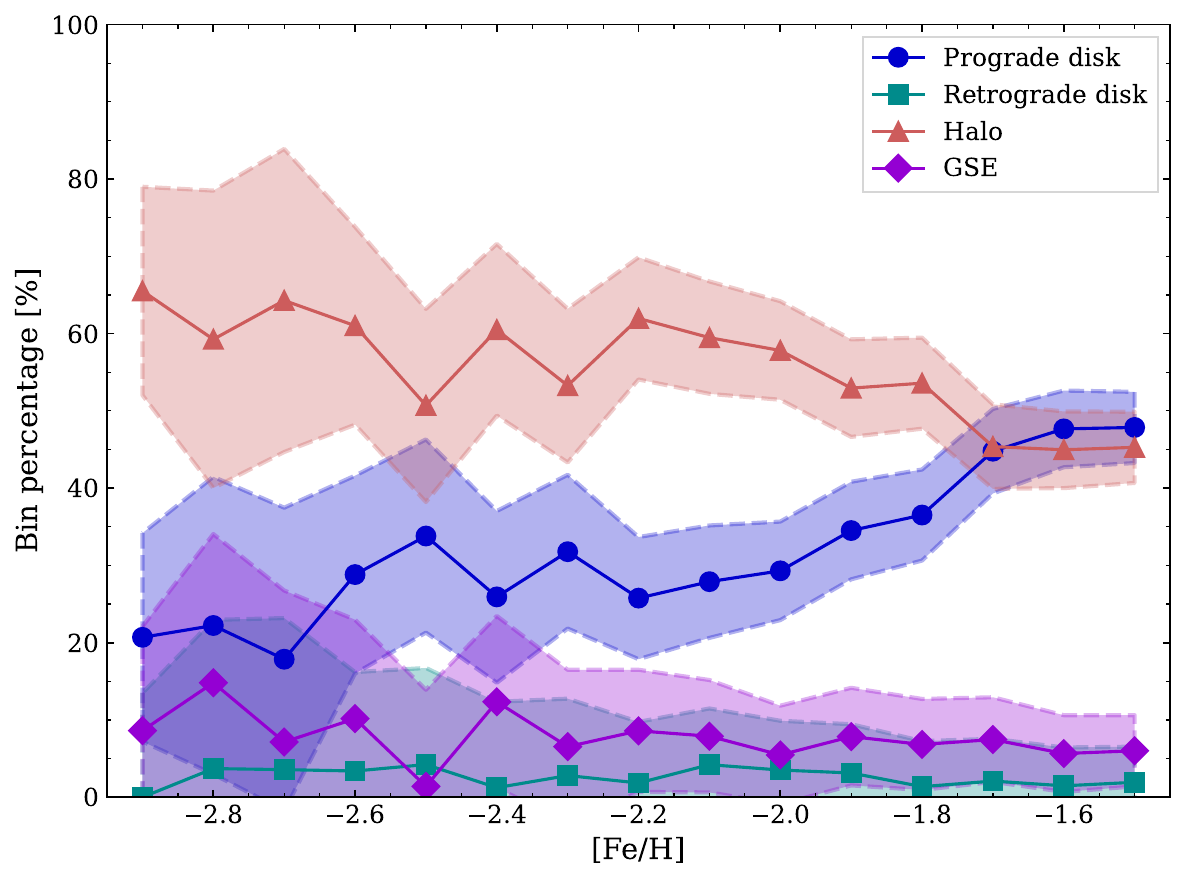}
    \caption{Orbital population distributions for stars with $\FeH{} \leq -1.5$. Bins are 0.1 dex in size. Each bin is normalised by the total number of stars in that bin. Prograde disk is shown in blue circles, retrograde disk in cyan squares, halo in orange triangles, and GSE in purple diamonds. Layout and structure of plot is identical to Fig.\ref{fig:carb distribution}.}
    \label{fig:pop distribution}
\end{figure}

The relative portions of these population groups change over decreasing metallicities, with clear trends present as shown in Fig.\ref{fig:pop distribution}. At the $\FeH{}=-1.5$ bin, the prograde disk is dominant with 48.6\,\% of the total population at this metallicity, before steadily decreasing to 20.7\,\% in the lowest metallicity bin. Over the same metallicity range, the retrograde disk is initially at 1.9\,\% in the higher metallicity bins, then peaks at 4.2\,\% in the $\FeH{} = -2.1$ bin, but given the errors, the fraction remains relatively constant throughout. Taking the fractions of retrograde against prograde stars from our criteria above, in the highest metallicity bin we have $\rm{N_{ret}/(N_{ret} + N_{pro})} = 3.8\%$, increasing up to 13.1\,\% in the $\FeH = -2.1$ bin, then remaining relatively constant at 10.3\,\% for metallicity bins below this. The portion of retrograde disk stars against halo stars is relatively constant for all metallicities at $\rm{N_{ret}/N_{hal}} = 4.6\,\%$, indicating similar formation origins.

The halo population is at its lowest in the $\FeH = -1.5$ bin at 45.3\,\% of the total in this metallicity bin, then becoming increasingly prominent in lower metallicity bins, peaking at 65.5\,\% in the $\FeH < -2.9$ bin. The GSE stars remain relatively consistent from $\FeH = -1.5$ to -2.3 at 6.9\,\%, then increasing to reach peak percentage at 14.8\,\% in the $\FeH = -2.8$ bin. Again, the lack of stars in these low metallicity bins makes it hard to draw any definitive conclusions. The portion of GSE against halo stars is constant at $\rm{N_{gse}/(N_{gse} + N_{hal})} = 11.7\,\%$ from $\FeH = -1.5$ to -2.3, then become more variable below this, peaking to 20\,\% in the $\FeH = -2.8$ bin. Note that these ratios are likely dependent upon what selection cuts are used, so results may vary.

\begin{figure*}[tp]
    \centering
    \includegraphics[width=1\linewidth]{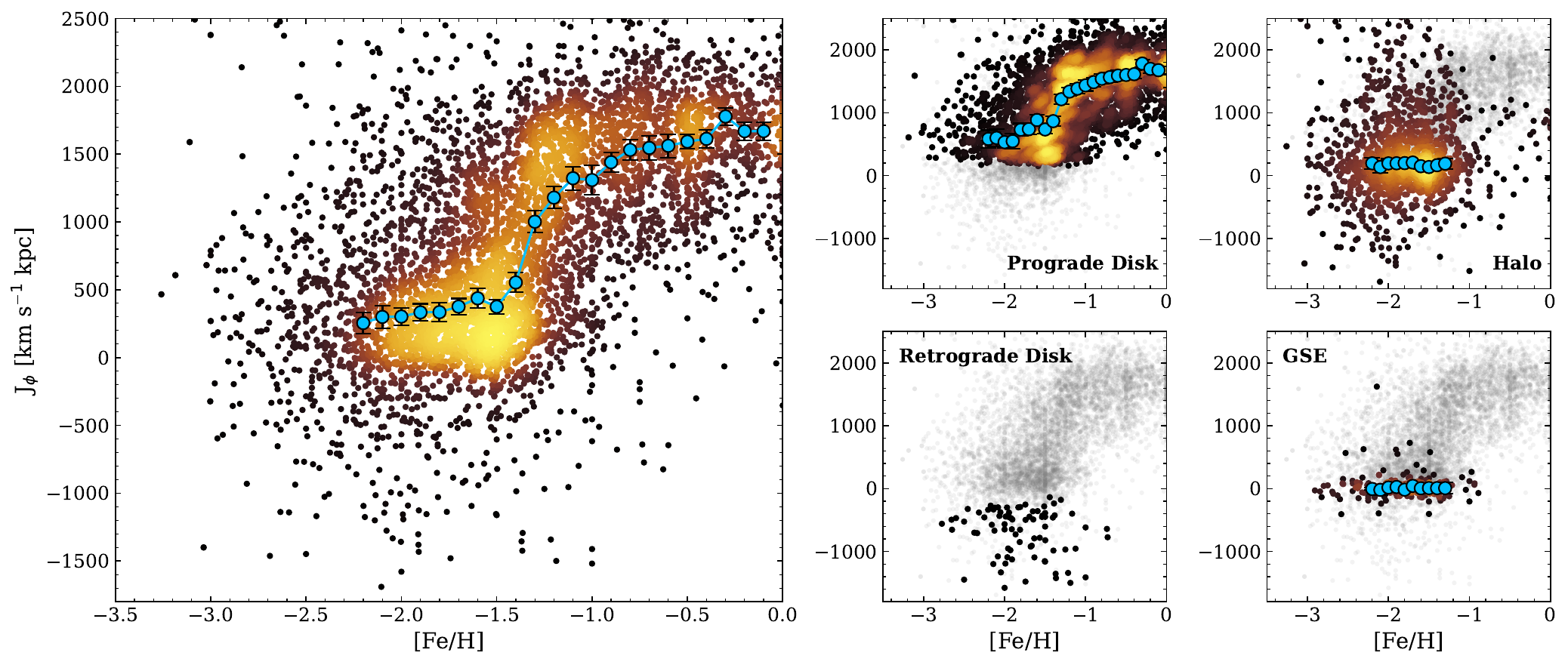}
    \caption{Variation of the vertical angular momentum with respect to metallicity for our whole sample (\textit{left panel}), the prograde disk (\textit{upper-middle panel}), the retrograde disk (\textit{lower-middle panel}), halo (\textit{upper-right panel}) and GSE (\textit{lower-right panel}) populations. The blue points over-plotted in each panel is the median and standard median error of each $\FeH$ bin ($0.05$ in size) from $-2.2$ to $-0.1$. For the halo and GSE panels, this was only done up to $-1.2$ due to the lack of metal-rich stars. Given the lack of any stars in the retrograde sample, no medians were calculated. The grey points in the smaller panels is the whole data sample under-plotted.}
    \label{fig:z vs jphi}
\end{figure*}

We investigate the evolution of the vertical angular momentum, $J_{\phi}$, in Fig. \ref{fig:z vs jphi}. By binning the data into $\FeH$ bins from $-2.2$ to $-0.1$ with $0.05$ bin size, then taking the median and standard median error of each (blue points over-plotted in the figure), we see three distinct features in the sample: one at $\FeH < -1.5$, with $J_\phi$ consistent across metallicities from $J_\phi = 300 \pm 80$\,km s$^{-1}$ kpc at $\FeH = -2.2$, to $440 \pm 70$\,km s$^{-1}$ kpc at $\FeH = -1.6$. From $\FeH = -1.5$ to $-1.1$, $J_\phi$ increases from $380 \pm 50$\,km s$^{-1}$ kpc to $1320 \pm 90$\,km s$^{-1}$, before slowly increasing at the higher metallicities from $J_\phi = 1300 \pm 100$\,km s$^{-1}$ at $\FeH = -1.0$, to $1670 \pm 70$\,km s$^{-1}$ at $\FeH = -0.1$.

Looking at the specific kinematic groupings, the halo group has small $J_\phi$ variation, from $200 \pm 90$\,km s$^{-1}$ at $\FeH = -2.2$, to $300 \pm 200$\,km s$^{-1}$ at $\FeH = -1.2$. Over the same range, the GSE also has small variation, from $J_\phi = 0 \pm 20$\,km s$^{-1}$ to $55 \pm 80$\,km s$^{-1}$.

The prograde disk shows the most variation, particularly from $\FeH = -1.5$ to $-1.1$, where $J_\phi$ increases from $730 \pm 70$\,km s$^{-1}$ to $1380 \pm 80$\,km s$^{-1}$. At the higher metallicities, the $J_\phi$ values steadily increase to $1680 \pm 60$\,km s$^{-1}$ at $\FeH = -0.1$. Given the lack of stars in the retrograde disk, we can not say anything about its behaviour.

The first feature is predominately seen in the halo group, with the second and third features shown in the prograde disk sample. These features suggests we are seeing the `spin-up' of the disk, where it transitions from being chaotic to the ordered disk we see today. This transitional phase, shown by the significant increase in $J_\phi$ with metallicity between $-1.5$ and $-1.1$, is inline with the \textit{Aurora} hypothesis suggested by \citet{belokurov_dawn_2022}. The details of this will be explored more in our discussion section (Section \ref{sec:discussion}).

\subsection{Metallicity Distribution Function}
\label{subsec:mdf}
\begin{figure}[tp]
    \centering
    \includegraphics[width=1\linewidth]{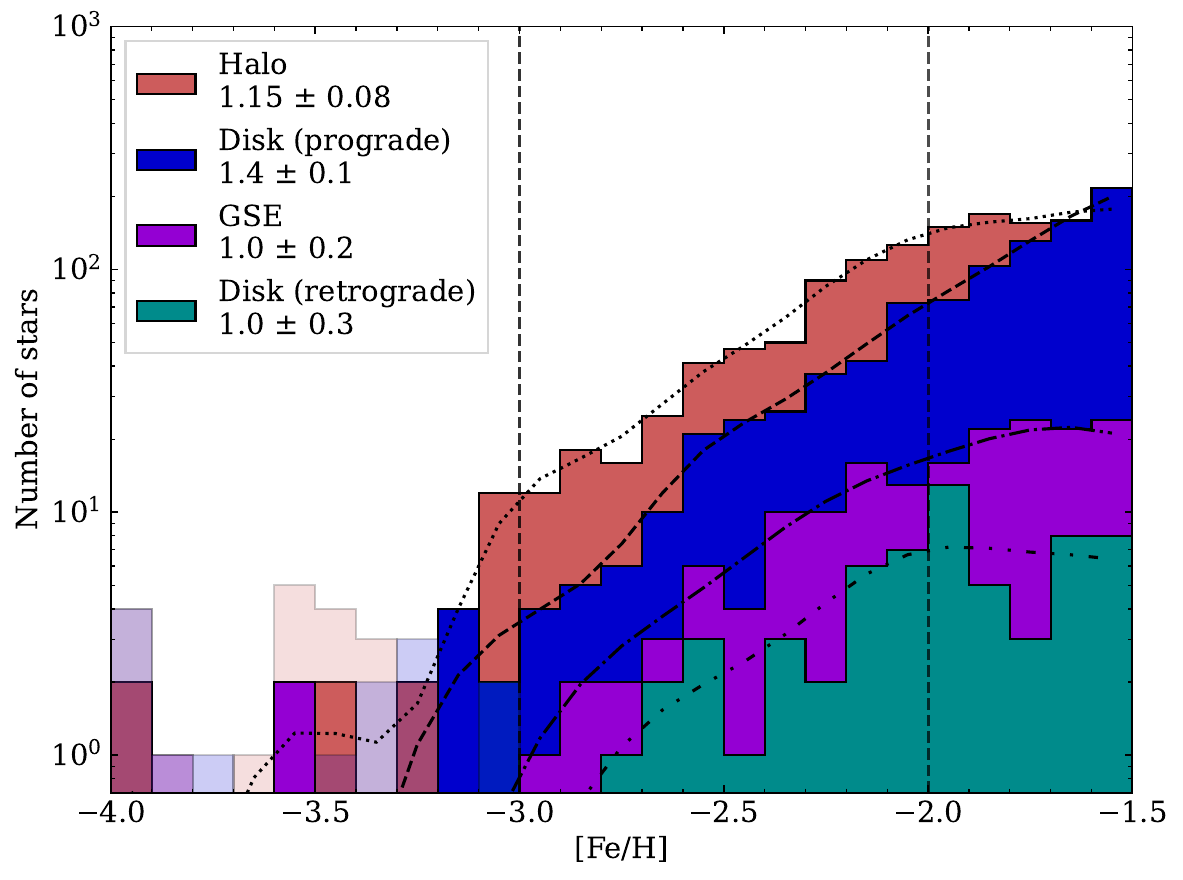}
    \caption{The MDFs of our sample stars in the halo, prograde disk, retrograde disk and GSE classifications. The bin size is 0.1\,dex. The slopes of the distributions from $\FeH = -2.0$ to $-3.0$ are written within the legend. KDE fits are shown by the dot-dot, dash-dash, dot-dash and dot-dot-dash black lines for halo, prograde disk, GSE and retrograde disk populations, respectively. The faded distributions below $\FeH \leq -3.0$ are the stars that did not passing the vetting process, with solid histograms the 15 EMPs.}
    \label{fig:aat mdf}
\end{figure}

The metallicity distribution function (MDF) for our kinematically selected sub-samples is shown in Fig.\,\ref{fig:aat mdf} using stars with $\FeH = -1.5$ to $-4.0$. Stars with $\FeH > -1.5$ were excluded due to our original selection on \citet{zhang_parameters_2023} to only include stars with $\FeH_{\rm Zhang} \leq -1.5$ (see Section \ref{subsec:gaia xp spectroscopy} for more detail). The population designations used assignments performed in Fig.\,\ref{fig:aat action}, and the bin size is set to be 0.1\,dex. This section will discuss both the apparent peak and the apparent slope in the MDFs for each kinematic group in detail, with comparisons made to literature if possible. All mention of `disk' stars (both prograde and retrograde) are those meeting both the $J_{\phi}/J_{\rm{TOT}}$ and $z_{\rm max}$ requirements. Those just meeting the $J_{\phi}/J_{\rm{TOT}}$ requirement are considered as halo.

The first inference from Fig.\,\ref{fig:aat mdf} is that the halo MDF levels off and starts decreasing above $\FeH \sim -2.0$, consistent with our current understanding of its MDF peaking at $\FeH \approx -1.7$ \citep[e.g.][]{youakim_pristine_2020}.

Although not shown on Fig.\,\ref{fig:aat mdf}, the prograde disk `peaks' at $\FeH = -1.07 \pm 0.05$, but given we deliberately selected metal-poor stars from Gaia XP with $\FeH \leq -1.5$, we are systematically excluding the metal-rich stars, so this value is likely to be significantly biased. Regarding the retrograde disk population, we noticed that the MDF does not keep increasing above $-2.0$, an interesting result which suggests that the retrograde disk population is predominantly metal-poor. Thus, that the retrograde population may primarily be comprised of stars with $\FeH \sim -2.0$, potentially also being members of the \textit{Aurora} population \citep{belokurov_dawn_2022}.

For the GSE, we see that the MDF peaks at $\FeH = -1.75 \pm 0.02$, inconsistent with studies like \citet{feuillet_skymapper-gaia_2020} finding $\FeH = -1.17$; \citet{feuillet_selecting_2021} getting $\FeH = -1.15$, or \citet{limberg_reconstructing_2022} finding the peak at $\FeH = -1.22$. If we instead used the selection criteria defined in \citet{feuillet_selecting_2021} to define a different GSE sample, then again we find that the MDF peaks at $-1.72 \pm 0.02$. Therefore, the discrepancies are likely due to contamination from halo stars that we did not consider. 

The slopes of the MDFs from $\FeH = -2.0$ to $-3.0$ (binning the data into 7 bins) were calculated for all four kinematic groups. For the halo, we derived a slope of $\Delta(\textrm{LogN})/\Delta\FeH = 1.15 \pm 0.08$\footnote{Errors for all MDF slopes were taken from least squares fits.}, and to be consistent with the literature, we also derived slopes for other metallicity ranges. For \citet{youakim_pristine_2020}, who used a metallicity range from $\FeH = -2.5$ to $-3.4$ to derive a slope of $1.0 \pm 0.1$, in agreement with the value we find, $1.3 \pm 0.2$ for the same metallicity range. Similarly, \citet{dacosta_skymapper_2019},  determined a slope of $1.5 \pm 0.1$ over $\FeH = -2.75$ to $-4.00$, a value consistent within uncertainties from our slope of $1.8 \pm 0.8$ for the same metallicity interval. However, the relative lack of EMP stars in our sample leads to larger uncertainty in the slopes making definitive comparisons. 

For the prograde disk, the slope is $1.4 \pm 0.1$ from $\FeH = -2.0$ to $-3.0$, which is within 2$\sigma$ of the combined uncertainties of our halo value.  It is also within the uncertainty of the retrograde disk slope at $\Delta(\textrm{LogN})/\Delta\FeH = 1.0 \pm 0.3$. Recent studies have begun to reveal the unique nature of the disk-like kinematics of metal-poor stars in the prograde disk against those in the retrograde disk orbits \citep[e.g.][]{sestito_tracing_2019, sestito_pristine_2020, cordoni_exploring_2021, carter_ancient_2021, carollo_understanding_2023}. It is suggested that the retrograde disk was formed via accretion, but the metal-poor component of the prograde disk may have formed this way too. Our data can not answer this from kinematics alone, but we can attempt to answer where these stars came from, particularly with our comparable MDF slopes, which we discuss more in Section \ref{sec:discussion}.

The slope for the GSE is $\Delta(\textrm{LogN})/\Delta\FeH = 1.0 \pm 0.2$, similar to both the halo and the retrograde disk. This suggests that these three groups underwent similar formation mechanisms, likely of an accreted origin with similar star formation efficiencies. To validate our slope of the GSE MDF, we used data provided by \citet{bonifacio_topos_2021} in their Table\,2 to derived our own estimate of their MDF slope. At $\Delta(\textrm{LogN})/\Delta\FeH = 1.18 \pm 0.09$ over the same metallicity range, we agree within uncertainty with \citet{bonifacio_topos_2021}.

\section{Discussion}
\label{sec:discussion}
We have determined the chemodynamics for 5795 stars observed with 2dF-AAOmega in regions close to the Galactic plane. Our chemical analysis reveals 2571 stars with $\FeH < -1.5$, whereof roughly half belong to the Galactic halo, half to the prograde disk, and smaller portions to the retrograde disk and to the GSE. On the tail end of the metallicity distribution are 15 previously unrecognised EMP stars with $\FeH \leq 3$.

For stars with detections, the range of $\XFe{C}$ values among the EMPs possibly indicates a range of processes for the origin of carbon relative to iron, independent of the difference in kinematic groups.

For our EMPs on metal-poor prograde disk orbits, there are two possible formation scenarios for their populations: either through in-situ (formed directly within the Galaxy), or via an in-plane accretion event (formed using material external to the Galaxy such as from a merger event). Previous studies like \citet{sestito_exploring_2021} have shown that the retrograde disk was formed through accretion, but FIRE-2 simulations \citep{hopkins_fire-2_2018} suggest something similar occurs for the prograde disk. By simulating Milky Way-mass galaxies, \citet{santistevan_origin_2021} found that their stars with $\FeH < -2.0$ were predominately on prograde disk orbits, with their formation consistent with a merger with a LMC/SMC-mass gas rich merger 7--12.5\,Gyr ago. Observationally, the study by \citet{feuillet_old_2022} found a population of RGB and RR Lyare variable disk stars that match the chemical signatures of an accretion event. This has been done successfully for the GSE, with \citet{buder_galah_2022} using chemistry to isolate the accreted structure (see their Fig.\,9). Obtaining detailed chemistry of our EMP prograde disk stars will therefore allow us to investigate whether their chemical signatures align with possible merger events in the metal-poor prograde disk.

Another interesting revelation is that the retrograde disk population is a relatively constant fraction of the halo population at 4.6\% for all metallicities below $\FeH = -1.5$, as shown in Fig.\,\ref{fig:pop distribution}. This could suggest that the retrograde disk was part of the halo formation, which itself is dominated by accretion. Having detailed chemistry of the retrograde disk population for all metallicities below $\FeH = -1.5$ would allow us to compare with the halo at the same metallicities. The fraction of the retrograde disk population compared to the metal-poor prograde disk is also relatively constant at 10.3\,\% for metallicities below $\FeH = -2.1$. This may suggest that the metal-poor disk, both prograde and retrograde, likely shared similar formation processes as the halo. 

Fig. \ref{fig:z vs jphi} shows three features: one with low $J_\phi$ at $\FeH < -1.5$ (ranging from $300 \pm 80$ to $440 \pm 70$\,km s$^{-1}$ kpc), seen in the halo and GSE sample. Another with $J_\phi$ increasing from $380 \pm 50$ to $1320 \pm 90$\,km s$^{-1}$ kpc when $-1.5 \leq \FeH < -1.1$, seen in the prograde disk sample. Then $J_\phi$ flattening out when $\FeH \geq -1.1$, reaching to $1680 \pm 60$\,km s$^{-1}$ kpc at the highest metallicities, again seen in the prograde disk sample. We are seeing the `spin-up' of the prograde disk as metallicity increases, first proposed by \citet{belokurov_dawn_2022}, whereby the Galaxy transitions from chaotic, majority-accreted systems (low $J_\phi$), to more structured and defined in-situ systems (high $J_\phi$) seen in the galaxy today. The \textit{Aurora} population was defined as being the group of stars born before this period, occurring at consistent metallicity ranges seen in our sample.

A further notable discovery is that between $\FeH = -2.0$ and $-3.0$, there are strong similarities in the MDFs in Fig.\,\ref{fig:aat mdf} amongst all four kinematic populations. The prograde disk has the steepest slope, which is within errors of the retrograde disk, but the difference between the halo and the GSE is small at $2\sigma$ and $1.8\sigma$ respectively. The inability to distinctively distinguish the four populations suggests that the star formation processes over this metallicity range is similar despite the different environments. GSE, halo and retrograde disk were suggested to form through accretion, and so we cannot rule out that the metal-poor end of the prograde disk was also formed through an accretion event. Programs are currently in progress to follow up on these EMPs using higher-resolution instruments, allowing us to perform detailed chemical analyses to confirm this.

\section{Summary and Conclusions}
\label{sec:conclusion}
In this work, we have performed follow-up observations of 8911 metal-poor star candidates selected from the \textit{Gaia} DR3 \citet{zhang_parameters_2023} catalogue, using the medium resolution 2dF$+$AAOmega spectrograph, focusing towards the plane of the Galactic disk. Combining the capabilities of \texttt{colte}, isochrones and \texttt{RVSpecFit}, we have self-consistently derived the stellar parameters $\Teff$, $\logg$, $\FeH$, [C/Fe] and radial velocity for 5795 non-variable stars. Parameter uncertainties were found through Monte Carlo randomisations. Our parameters were tested and verified against parameters derived from WiFeS spectroscopy for 20 of our brightest stars, showing good agreement for all three stellar parameters (Fig.\,\ref{fig:wifes vs aat}). Tests with RGB in five globular clusters also showed good agreement with literature $\FeH$ values from \citet{carretta_intrinsic_2009} (Fig.\,\ref{fig:gc results}). Agreement with $\FeH$ values for 2125 stars in common with  \citet{martin_pristine_2023} was also excellent, with minimal scatter and small mean differences. We found that other, more complex approaches using \textit{Gaia} XP did not perform as well as \citet{martin_pristine_2023}.

The stellar parameters for the sample (Fig.\,\ref{fig:aat params}) revealed the presence of 15 previously unknown candidate EMP stars ($\FeH \leq -3.0$) that cover a variety of evolutionary stages from turn-off all the way to the top of the RGB (Fig.\,\ref{fig:aat hr}), with the lowest metallicity being $\FeH = -4.0 \pm 0.2$\,dex. A closer look at the [C/Fe] results for our 15 EMPs (Fig.\,\ref{fig:emp carb vs z}) revealed that most of our stars with detections are carbon-normal, with two carbon-enhanced stars. One is a giant with $\XFe{C} = 0.79 \pm 0.03$, and the other is a dwarf with $\XFe{C} = 1.3 \pm 0.1$. This dwarf, which has $\FeH = -3.6 \pm 0.1$ may have been enriched via a Pop. III `fallback' supernovae or through mass transfer from a companion star (now white dwarf) during its AGB phase. High resolution follow-up to detect (or not) slow neutron capture process elements, such as Ba, is required to first confirm, then differentiate these scenarios. For the whole sample, we saw that when we separated the stars into different regions of carbon enhancement (Fig.\,\ref{fig:carb distribution}), the number of carbon-depleted stars decreases with lower metallicities, while the number of carbon-enhanced stars increases over the same range. This agrees with results shown by e.g. \citet{placco_carbon-enhanced_2014}. 

We then analysed the orbital dynamics of the sample and used the selection cuts defined in \citet{sestito_exploring_2021} to assign kinematic populations (Fig.\,\ref{fig:aat action}). We select stars belonging to the prograde disk, retrograde disk (with both disks satisfying $z_{\rm max} < 3$\,kpc and their specific $J_{\phi}/J{\rm TOT}$ requirements), halo and GSE. From this, the growth and decline of each group across various metallicity bins is revealed (Fig.\,\ref{fig:pop distribution}). Of note is the decline of prograde star percentages with decreasing metallicity from $\rm{N_{pro}/N_{bin}} = 48.6\,\%$ in the $\FeH = -1.5$ bin, to 20.7\,\% in the $\FeH < -2.9$ bin. The peak percentage of the total number of retrograde disk stars in the $\FeH = -2.1$ bin is 4.2\,\%, then below this the portion of retrograde disk against prograde disk stars remains relatively constant at $\rm{N_{ret}/(N_{ret} + N_{pro})} = 10.3\,\%$. The portion of retrograde disk against halo stars is relatively consistent across all metallicities at $\rm{N_{ret}/(N_{ret} + N_{halo})} = 4.6\,\%$, indicating similar formations origins.

Using our $\FeH$ and $J_\phi$ measurements (Fig. \ref{fig:z vs jphi}), we observe the `spin-up' of the Milky Way from $\FeH = -1.5$ with $J_\phi = 380 \pm 50$\,km s$^{-1}$ kpc, to $\FeH = -1.1$ with $J_\phi = 1320 \pm 90$\,km s$^{-1}$ kpc. This is consistent with the \textit{Aurora} in-situ population first hypothesised by \citet{belokurov_dawn_2022}.

The metallicity distribution function for the halo, prograde disk, retrograde disk and GSE stars (Fig.\,\ref{fig:aat mdf}) were then evaluated. Between $\FeH = -2.0$\,dex and $\FeH = -3.0$\,dex, we derived the slopes (defined as d(logN)/d($\FeH$)) for the four regions, with the MDFs consistent amongst each other. 

The retrograde slope agrees with the halo and GSE, confirming its likely accreted origins, though we see that within uncertainty, it also agrees with the metal-poor prograde disk. This suggests that the metal-poor prograde disk may also be formed through accretion. Future studies involving measuring $\alpha$-element abundances of disk EMP stars may unearth the true nature of the disk, particularly the metal-weak prograde disk.

\begin{acknowledgement}
We acknowledge the traditional owners of the land on which the AAT and ANU stand, the Gamilaraay, the Ngunnawal and the Ngambri people. We pay our respects to Elders, past and present, and are proud to continue their tradition of surveying the night sky in the Southern hemisphere. We are grateful for the team operating and maintaining the AAT and 2dF$+$AAOmega spectrograph. Their efforts do not go unnoticed, as this work would not be possible without their dedication and commitment to keep the instruments and telescope running.

Parts of this research were supported by the Australian Research Council Centre of Excellence for All Sky Astrophysics in 3 Dimensions (ASTRO 3D), through project number CE170100013.

This work was supported by computational resources provided by the Australian Government through the National Computational Infrastructure (NCI) under the National Computational Merit Allocation Scheme (project y89).
T.N. acknowledges support from the Knut and Alice Wallenberg Foundation.

We thank the anonymous referee for their insightful comments and suggestions. Their feedback has helped improve the quality and strength of this manuscript.
\end{acknowledgement}

\section*{Data Availability}
The derived stellar parameters for both our survey sample (also including orbital parameters) and the data for the globular clusters is available here \url{https://zenodo.org/records/14292178}. 

\printendnotes

\bibliography{paper.bib}

\begin{thebibliography}{}
\expandafter\ifx\csname natexlab\endcsname\relax\def\natexlab#1{#1}\fi

\bibitem[{{Alfaro} {et~al.}(2022){Alfaro}, {Jim{\'e}nez}, {S{\'a}nchez-Gil}, {S{\'a}nchez}, {Gonz{\'a}lez}, \& {Ma{\'\i}z Apell{\'a}niz}}]{alfaro_topography_2022}
{Alfaro}, E.~J., {Jim{\'e}nez}, M., {S{\'a}nchez-Gil}, M.~C., {et~al.} 2022, \apj, 937, 114

\bibitem[{Amidror(2002)}]{amidror_scattered_2002}
Amidror, I. 2002, Journal of Electronic Imaging, 11, 157

\bibitem[{Andrae {et~al.}(2023{\natexlab{a}})Andrae, Rix, \& Chandra}]{andrae_robust_2023}
Andrae, R., Rix, H.-W., \& Chandra, V. 2023{\natexlab{a}}, \apjs, 267, 8

\bibitem[{Andrae {et~al.}(2023{\natexlab{b}})Andrae, Fouesneau, Sordo, Bailer-Jones, Dharmawardena, Rybizki, De~Angeli, Lindstrøm, Marshall, Drimmel, Korn, Soubiran, Brouillet, Casamiquela, Rix, Abreu~Aramburu, Álvarez, Bakker, Bellas-Velidis, Bijaoui, Brugaletta, Burlacu, Carballo, Chaoul, Chiavassa, Contursi, Cooper, Creevey, Dafonte, Dapergolas, de~Laverny, Delchambre, Demouchy, Edvardsson, Frémat, Garabato, García-Lario, García-Torres, Gavel, Gomez, González-Santamaría, Hatzidimitriou, Heiter, Jean-Antoine~Piccolo, Kontizas, Kordopatis, Lanzafame, Lebreton, Licata, Livanou, Lobel, Lorca, Magdaleno~Romeo, Manteiga, Marocco, Mary, Nicolas, Ordenovic, Pailler, Palicio, Pallas-Quintela, Panem, Pichon, Poggio, Recio-Blanco, Riclet, Robin, Santoveña, Sarro, Schultheis, Segol, Silvelo, Slezak, Smart, Süveges, Thévenin, Torralba~Elipe, Ulla, Utrilla, Vallenari, van Dillen, Zhao, \& Zorec}]{andrae_gaia_2023}
Andrae, R., Fouesneau, M., Sordo, R., {et~al.} 2023{\natexlab{b}}, \aap, 674, A27

\bibitem[{Ardern-Arentsen {et~al.}(2024)Ardern-Arentsen, Monari, Queiroz, Starkenburg, Martin, Chiappini, Aguado, Belokurov, Carlberg, Monty, Myeong, Schultheis, Sestito, Venn, Vitali, Yuan, Zhang, Buder, Lewis, Oliver, Wan, \& Zucker}]{ardern-arentsen_pristine_2024}
Ardern-Arentsen, A., Monari, G., Queiroz, A. B.~A., {et~al.} 2024, \mnras, 530, 3391

\bibitem[{Arentsen {et~al.}(2022)Arentsen, Placco, Lee, Aguado, Martin, Starkenburg, \& Yoon}]{arentsen_inconsistency_2022}
Arentsen, A., Placco, V.~M., Lee, Y.~S., {et~al.} 2022, \mnras, 515, 4082

\bibitem[{Arentsen {et~al.}(2020)Arentsen, Starkenburg, Martin, Hill, Ibata, Kunder, Schultheis, Venn, Zucker, Aguado, Carlberg, González~Hernández, Lardo, Longeard, Malhan, Navarro, Sánchez-Janssen, Sestito, Thomas, Youakim, Lewis, Simpson, \& Wan}]{arentsen_pristine_2020}
Arentsen, A., Starkenburg, E., Martin, N.~F., {et~al.} 2020, \mnras, 491, L11

\bibitem[{Arentsen {et~al.}(2021)Arentsen, Starkenburg, Aguado, Martin, Placco, Carlberg, González~Hernández, Hill, Jablonka, Kordopatis, Lardo, Mashonkina, Navarro, Venn, Buder, Lewis, Wan, \& Zucker}]{arentsen_pristine_2021}
Arentsen, A., Starkenburg, E., Aguado, D.~S., {et~al.} 2021, \mnras, 505, 1239

\bibitem[{Armandroff \& Da~Costa(1991)}]{armandroff_metallicities_1991}
Armandroff, T.~E., \& Da~Costa, G.~S. 1991, \aj, 101, 1329

\bibitem[{{Astropy-Specutils Development Team}(2019)}]{astropy-specutils_development_team_specutils_2019}
{Astropy-Specutils Development Team}. 2019, Astrophysics Source Code Library, ascl:1902.012

\bibitem[{Bailer-Jones {et~al.}(2021)Bailer-Jones, Rybizki, Fouesneau, Demleitner, \& Andrae}]{bailer-jones_estimating_2021}
Bailer-Jones, C. A.~L., Rybizki, J., Fouesneau, M., Demleitner, M., \& Andrae, R. 2021, \aj, 161, 147

\bibitem[{Beers \& Christlieb(2005)}]{beers_discovery_2005}
Beers, T.~C., \& Christlieb, N. 2005, \araa, 43, 531

\bibitem[{Beers {et~al.}(1992)Beers, Preston, \& Shectman}]{beers_search_1992}
Beers, T.~C., Preston, G.~W., \& Shectman, S.~A. 1992, \aj, 103, 1987

\bibitem[{Belokurov {et~al.}(2018)Belokurov, Erkal, Evans, Koposov, \& Deason}]{belokurov_co-formation_2018}
Belokurov, V., Erkal, D., Evans, N.~W., Koposov, S.~E., \& Deason, A.~J. 2018, \mnras, 478, 611

\bibitem[{Belokurov \& Kravtsov(2022)}]{belokurov_dawn_2022}
Belokurov, V., \& Kravtsov, A. 2022, \mnras, 514, 689

\bibitem[{Bennett \& Bovy(2019)}]{bennett_vertical_2019}
Bennett, M., \& Bovy, J. 2019, \mnras, 482, 1417

\bibitem[{Bland-Hawthorn \& Gerhard(2016)}]{bland-hawthorn_galaxy_2016}
Bland-Hawthorn, J., \& Gerhard, O. 2016, \araa, 54, 529

\bibitem[{Bonifacio {et~al.}(2021)Bonifacio, Monaco, Salvadori, Caffau, Spite, Sbordone, Spite, Ludwig, Di~Matteo, Haywood, François, Koch-Hansen, Christlieb, \& Zaggia}]{bonifacio_topos_2021}
Bonifacio, P., Monaco, L., Salvadori, S., {et~al.} 2021, \aap, 651, A79

\bibitem[{Bovy(2015)}]{bovy_galpy_2015}
Bovy, J. 2015, \apjs, 216, 29

\bibitem[{Bromm {et~al.}(2009)Bromm, Yoshida, Hernquist, \& McKee}]{bromm_formation_2009}
Bromm, V., Yoshida, N., Hernquist, L., \& McKee, C.~F. 2009, \nat, 459, 49

\bibitem[{Buder {et~al.}(2022)Buder, Lind, Ness, Feuillet, Horta, Monty, Buck, Nordlander, Bland-Hawthorn, Casey, de~Silva, D'Orazi, Freeman, Hayden, Kos, Martell, Lewis, Lin, Schlesinger, Sharma, Simpson, Stello, Zucker, Zwitter, Ciucă, Horner, Kobayashi, Ting, Wyse, \& Wyse}]{buder_galah_2022}
Buder, S., Lind, K., Ness, M.~K., {et~al.} 2022, \mnras, 510, 2407

\bibitem[{Caffau {et~al.}(2011)Caffau, Bonifacio, François, Sbordone, Monaco, Spite, Spite, Ludwig, Cayrel, Zaggia, Hammer, Randich, Molaro, \& Hill}]{caffau_extremely_2011}
Caffau, E., Bonifacio, P., François, P., {et~al.} 2011, \nat, 477, 67

\bibitem[{Caffau {et~al.}(2012)Caffau, Bonifacio, François, Spite, Spite, Zaggia, Ludwig, Steffen, Mashonkina, Monaco, Sbordone, Molaro, Cayrel, Plez, Hill, Hammer, \& Randich}]{caffau_primordial_2012}
---. 2012, \aap, 542, A51

\bibitem[{Caffau {et~al.}(2013)Caffau, Bonifacio, Sbordone, François, Monaco, Spite, Plez, Cayrel, Christlieb, Clark, Glover, Klessen, Koch, Ludwig, Spite, Steffen, \& Zaggia}]{caffau_topos_2013}
Caffau, E., Bonifacio, P., Sbordone, L., {et~al.} 2013, \aap, 560, A71

\bibitem[{Cardelli {et~al.}(1989)Cardelli, Clayton, \& Mathis}]{cardelli_relationship_1989}
Cardelli, J.~A., Clayton, G.~C., \& Mathis, J.~S. 1989, \apj, 345, 245

\bibitem[{Carollo {et~al.}(2023)Carollo, Christlieb, Tissera, \& Sillero}]{carollo_understanding_2023}
Carollo, D., Christlieb, N., Tissera, P.~B., \& Sillero, E. 2023, \apj, 946, 99

\bibitem[{Carollo {et~al.}(2007)Carollo, Beers, Lee, Chiba, Norris, Wilhelm, Sivarani, Marsteller, Munn, Bailer-Jones, Fiorentin, \& York}]{carollo_two_2007}
Carollo, D., Beers, T.~C., Lee, Y.~S., {et~al.} 2007, \nat, 450, 1020

\bibitem[{Carollo {et~al.}(2010)Carollo, Beers, Chiba, Norris, Freeman, Lee, Ivezić, Rockosi, \& Yanny}]{carollo_structure_2010}
Carollo, D., Beers, T.~C., Chiba, M., {et~al.} 2010, \apj, 712, 692

\bibitem[{Carrera {et~al.}(2013)Carrera, Pancino, Gallart, \& del Pino}]{carrera_near-infrared_2013}
Carrera, R., Pancino, E., Gallart, C., \& del Pino, A. 2013, \mnras, 434, 1681

\bibitem[{Carretta {et~al.}(2009)Carretta, Bragaglia, Gratton, D'Orazi, \& Lucatello}]{carretta_intrinsic_2009}
Carretta, E., Bragaglia, A., Gratton, R., D'Orazi, V., \& Lucatello, S. 2009, \aap, 508, 695

\bibitem[{Carter {et~al.}(2021)Carter, Conroy, Zaritsky, Ting, Bonaca, Naidu, Johnson, Cargile, Caldwell, Speagle, \& Han}]{carter_ancient_2021}
Carter, C., Conroy, C., Zaritsky, D., {et~al.} 2021, \apj, 908, 208

\bibitem[{Casagrande {et~al.}(2011)Casagrande, Schönrich, Asplund, Cassisi, Ramírez, Meléndez, Bensby, \& Feltzing}]{casagrande_new_2011}
Casagrande, L., Schönrich, R., Asplund, M., {et~al.} 2011, \aap, 530, A138

\bibitem[{Casagrande \& VandenBerg(2018)}]{casagrande_use_2018}
Casagrande, L., \& VandenBerg, D.~A. 2018, \mnras, 479, L102

\bibitem[{Casagrande {et~al.}(2019)Casagrande, Wolf, Mackey, Nordlander, Yong, \& Bessell}]{casagrande_skymapper_2019}
Casagrande, L., Wolf, C., Mackey, A.~D., {et~al.} 2019, \mnras, 482, 2770

\bibitem[{Casagrande {et~al.}(2021)Casagrande, Lin, Rains, Liu, Buder, Horner, Asplund, Lewis, Martell, Nordlander, Stello, Ting, Wittenmyer, Bland-Hawthorn, Casey, De~Silva, D'Orazi, Freeman, Hayden, Kos, Lind, Schlesinger, Sharma, Simpson, Zucker, \& Zwitter}]{casagrande_galah_2021}
Casagrande, L., Lin, J., Rains, A.~D., {et~al.} 2021, \mnras, 507, 2684

\bibitem[{Chandra \& Schlaufman(2021)}]{chandra_searching_2021}
Chandra, V., \& Schlaufman, K.~C. 2021, \aj, 161, 197

\bibitem[{Chiti {et~al.}(2021{\natexlab{a}})Chiti, Frebel, Mardini, Daniel, Ou, \& Uvarova}]{chiti_stellar_2021}
Chiti, A., Frebel, A., Mardini, M.~K., {et~al.} 2021{\natexlab{a}}, \apjs, 254, 31

\bibitem[{Chiti {et~al.}(2021{\natexlab{b}})Chiti, Mardini, Frebel, \& Daniel}]{chiti_metal-poor_2021}
Chiti, A., Mardini, M.~K., Frebel, A., \& Daniel, T. 2021{\natexlab{b}}, \apj, 911, L23

\bibitem[{Christlieb {et~al.}(2008)Christlieb, Schörck, Frebel, Beers, Wisotzki, \& Reimers}]{christlieb_stellar_2008}
Christlieb, N., Schörck, T., Frebel, A., {et~al.} 2008, \aap, 484, 721

\bibitem[{Christlieb {et~al.}(2002)Christlieb, Bessell, Beers, Gustafsson, Korn, Barklem, Karlsson, Mizuno–Wiedner, \& Rossi}]{christlieb_stellar_2002}
Christlieb, N., Bessell, M.~S., Beers, T.~C., {et~al.} 2002, \nat, 419, 904

\bibitem[{{Conroy} {et~al.}(2022){Conroy}, {Weinberg}, {Naidu}, {Buck}, {Johnson}, {Cargile}, {Bonaca}, {Caldwell}, {Chandra}, {Han}, {Johnson}, {Speagle}, {Ting}, {Woody}, \& {Zaritsky}}]{conroy_birth_2024}
{Conroy}, C., {Weinberg}, D.~H., {Naidu}, R.~P., {et~al.} 2022, arXiv e-prints, arXiv:2204.02989

\bibitem[{Cordoni {et~al.}(2021)Cordoni, Da Costa, Yong, Mackey, Marino, Monty, Nordlander, Norris, Asplund, Bessell, Casey, Frebel, Lind, Murphy, Schmidt, Gao, Xylakis-Dornbusch, Amarsi, \& Milone}]{cordoni_exploring_2021}
Cordoni, G., Da Costa, G.~S., Yong, D., {et~al.} 2021, \mnras, 503, 2539

\bibitem[{Da~Costa(2016)}]{da_costa_ca_2016}
Da~Costa, G.~S. 2016, \mnras, 455, 199

\bibitem[{Da~Costa {et~al.}(2023)Da~Costa, Bessell, Nordlander, Hughes, Buder, Mackey, Spitler, \& Zucker}]{da_costa_spectroscopic_2023}
Da~Costa, G.~S., Bessell, M.~S., Nordlander, T., {et~al.} 2023, \mnras, 520, 917

\bibitem[{Da Costa {et~al.}(2019)Da Costa, Bessell, Mackey, Nordlander, Asplund, Casey, Frebel, Lind, Marino, Murphy, Norris, Schmidt, \& Yong}]{dacosta_skymapper_2019}
Da Costa, G.~S., Bessell, M.~S., Mackey, A.~D., {et~al.} 2019, \mnras, 489, 5900

\bibitem[{De~Angeli {et~al.}(2023)De~Angeli, Weiler, Montegriffo, Evans, Riello, Andrae, Carrasco, Busso, Burgess, Cacciari, Davidson, Harrison, Hodgkin, Jordi, Osborne, Pancino, Altavilla, Barstow, Bailer-Jones, Bellazzini, Brown, Castellani, Cowell, Delchambre, De~Luise, Diener, Fabricius, Fouesneau, Frémat, Gilmore, Giuffrida, Hambly, Hidalgo, Holland, Kostrzewa-Rutkowska, van Leeuwen, Lobel, Marinoni, Miller, Pagani, Palaversa, Piersimoni, Pulone, Ragaini, Rainer, Richards, Rixon, Ruz-Mieres, Sanna, Sarro, Rowell, Sordo, Walton, \& Yoldas}]{de_angeli_gaia_2023}
De~Angeli, F., Weiler, M., Montegriffo, P., {et~al.} 2023, \aap, 674, A2

\bibitem[{de~Jong {et~al.}(2019)de~Jong, Agertz, Berbel, Aird, Alexander, Amarsi, Anders, Andrae, Ansarinejad, Ansorge, Antilogus, Anwand-Heerwart, Arentsen, Arnadottir, Asplund, Auger, Azais, Baade, Baker, Baker, Balbinot, Baldry, Banerji, Barden, Barklem, Barthélémy-Mazot, Battistini, Bauer, Bell, Bellido-Tirado, Bellstedt, Belokurov, Bensby, Bergemann, Bestenlehner, Bielby, Bilicki, Blake, Bland-Hawthorn, Boeche, Boland, Boller, Bongard, Bongiorno, Bonifacio, Boudon, Brooks, Brown, Brown, Brüggen, Brynnel, Brzeski, Buchert, Buschkamp, Caffau, Caillier, Carrick, Casagrande, Case, Casey, Cesarini, Cescutti, Chapuis, Chiappini, Childress, Christlieb, Church, Cioni, Cluver, Colless, Collett, Comparat, Cooper, Couch, Courbin, Croom, Croton, Daguisé, Dalton, Davies, Davis, de~Laverny, Deason, Dionies, Disseau, Doel, Döscher, Driver, Dwelly, Eckert, Edge, Edvardsson, Youssoufi, Elhaddad, Enke, Erfanianfar, Farrell, Fechner, Feiz, Feltzing, Ferreras, Feuerstein, Feuillet, Finoguenov, Ford, Fotopoulou,
  Fouesneau, Frenk, Frey, Gaessler, Geier, Gentile~Fusillo, Gerhard, Giannantonio, Giannone, Gibson, Gillingham, González-Fernández, Gonzalez-Solares, Gottloeber, Gould, Grebel, Gueguen, Guiglion, Haehnelt, Hahn, Hansen, Hartman, Hauptner, Hawkins, Haynes, Haynes, Heiter, Helmi, Aguayo, Hewett, Hinton, Hobbs, Hoenig, Hofman, Hook, Hopgood, Hopkins, Hourihane, Howes, Howlett, Huet, Irwin, Iwert, Jablonka, Jahn, Jahnke, Jarno, Jin, Jofre, Johl, Jones, Jönsson, Jordan, Karovicova, Khalatyan, Kelz, Kennicutt, King, Kitaura, Klar, Klauser, Kneib, Koch, Koposov, Kordopatis, Korn, Kosmalski, Kotak, Kovalev, Kreckel, Kripak, Krumpe, Kuijken, Kunder, Kushniruk, Lam, Lamer, Laurent, Lawrence, Lehmitz, Lemasle, Lewis, Li, Lidman, Lind, Liske, Lizon, Loveday, Ludwig, McDermid, Maguire, Mainieri, Mali, Mandel, Mandel, Mannering, Martell, Martinez~Delgado, Matijevic, McGregor, McMahon, McMillan, Mena, Merloni, Meyer, Michel, Micheva, Migniau, Minchev, Monari, Muller, Murphy, Muthukrishna, Nandra, Navarro, Ness, Nichani,
  Nichol, Nicklas, Niederhofer, Norberg, Obreschkow, Oliver, Owers, Pai, Pankratow, Parkinson, Paschke, Paterson, Pecontal, Parry, Phillips, Pillepich, Pinard, Pirard, Piskunov, Plank, Plüschke, Pons, Popesso, Power, Pragt, Pramskiy, Pryer, Quattri, Queiroz, Quirrenbach, Rahurkar, Raichoor, Ramstedt, Rau, Recio-Blanco, Reiss, Renaud, Revaz, Rhode, Richard, Richter, Rix, Robotham, Roelfsema, Romaniello, Rosario, Rothmaier, Roukema, Ruchti, Rupprecht, Rybizki, Ryde, Saar, Sadler, Sahlén, Salvato, Sassolas, Saunders, Saviauk, Sbordone, Schmidt, Schnurr, Scholz, Schwope, Seifert, Shanks, Sheinis, Sivov, Skúladóttir, Smartt, Smedley, Smith, Smith, Sorce, Spitler, Starkenburg, Steinmetz, Stilz, Storm, Sullivan, Sutherland, Swann, Tamone, Taylor, Teillon, Tempel, ter Horst, Thi, Tolstoy, Trager, Traven, Tremblay, Tresse, Valentini, van~de Weygaert, van~den Ancker, Veljanoski, Venkatesan, Wagner, Wagner, Walcher, Waller, Walton, Wang, Winkler, Wisotzki, Worley, Worseck, Xiang, Xu, Yong, Zhao, Zheng, Zscheyge, \&
  Zucker}]{de_jong_4most_2019}
de~Jong, R.~S., Agertz, O., Berbel, A.~A., {et~al.} 2019, The Messenger, 175, 3

\bibitem[{Dinescu {et~al.}(2002)Dinescu, Majewski, Girard, Méndez, Sandage, Siegel, Kunkel, Subasavage, \& Ostheimer}]{dinescu_absolute_2002}
Dinescu, D.~I., Majewski, S.~R., Girard, T.~M., {et~al.} 2002, \apj, 575, L67

\bibitem[{Dokkum(2001)}]{dokkum_cosmicray_2001}
Dokkum, P. G.~v. 2001, \pasp, 113, 1420

\bibitem[{Dopita {et~al.}(2010)Dopita, Rhee, Farage, McGregor, Bloxham, Green, Roberts, Neilson, Wilson, Young, Firth, Busarello, \& Merluzzi}]{dopita_wide_2010}
Dopita, M., Rhee, J., Farage, C., {et~al.} 2010, \apss, 327, 245

\bibitem[{Dotter {et~al.}(2008)Dotter, Chaboyer, Jevremović, Kostov, Baron, \& Ferguson}]{dotter_dartmouth_2008}
Dotter, A., Chaboyer, B., Jevremović, D., {et~al.} 2008, \apjs, 178, 89

\bibitem[{Dovgal {et~al.}(2024)Dovgal, Venn, Sestito, Hayes, McConnachie, Navarro, Placco, Starkenburg, Martin, Pazder, Chiboucas, Deibert, Gamen, Heo, Kalari, Martioli, Xu, Diaz, Gomez-Jimenez, Henderson, Prado, Quiroz, Robertson, Ruiz-Carmona, Simpson, Urrutia, Waller, Berg, Burley, Hartman, Ireland, Margheim, Perez, \& Thomas-Osip}]{dovgal_probing_2024}
Dovgal, A., Venn, K.~A., Sestito, F., {et~al.} 2024, \mnras, 527, 7810

\bibitem[{Fan {et~al.}(2023)Fan, Zhao, Wang, Zheng, Zhao, Li, Chen, Yuan, Li, Tan, Song, Zuo, Huang, Luo, Esamdin, Ma, Li, Song, Grupp, Zhao, Ehgamberdiev, Burkhonov, Feng, Bai, Zhang, Niu, Khodjaev, Khafizov, Asfandiyarov, Shaymanov, Karimov, Yuldashev, Lu, Zhaori, Hong, Hu, Liu, \& Xu}]{fan_stellar_2023}
Fan, Z., Zhao, G., Wang, W., {et~al.} 2023, \apjs, 268, 9

\bibitem[{Fernández-Alvar {et~al.}(2021)Fernández-Alvar, Kordopatis, Hill, Starkenburg, Viswanathan, Martin, Thomas, Navarro, Malhan, Sestito, González~Hernández, \& Carlberg}]{fernandez-alvar_pristine_2021}
Fernández-Alvar, E., Kordopatis, G., Hill, V., {et~al.} 2021, \mnras, 508, 1509

\bibitem[{Feuillet {et~al.}(2022)Feuillet, Feltzing, Sahlholdt, \& Bensby}]{feuillet_old_2022}
Feuillet, D.~K., Feltzing, S., Sahlholdt, C., \& Bensby, T. 2022, \apj, 934, 21

\bibitem[{Feuillet {et~al.}(2020)Feuillet, Feltzing, Sahlholdt, \& Casagrande}]{feuillet_skymapper-gaia_2020}
Feuillet, D.~K., Feltzing, S., Sahlholdt, C.~L., \& Casagrande, L. 2020, \mnras, 497, 109

\bibitem[{{Feuillet} {et~al.}(2021){Feuillet}, {Sahlholdt}, {Feltzing}, \& {Casagrande}}]{feuillet_selecting_2021}
{Feuillet}, D.~K., {Sahlholdt}, C.~L., {Feltzing}, S., \& {Casagrande}, L. 2021, \mnras, 508, 1489

\bibitem[{Fouesneau {et~al.}(2023)Fouesneau, Frémat, Andrae, Korn, Soubiran, Kordopatis, Vallenari, Heiter, Creevey, Sarro, Laverny, Lanzafame, Lobel, Sordo, Rybizki, Slezak, Álvarez, Drimmel, Garabato, Delchambre, Bailer-Jones, Hatzidimitriou, Lorca, Fustec, Pailler, Mary, Robin, Utrilla, Aramburu, Bakker, Bellas-Velidis, Bijaoui, Blomme, Bouret, Brouillet, Brugaletta, Burlacu, Carballo, Casamiquela, Chaoul, Chiavassa, Contursi, Cooper, Dafonte, Demouchy, Dharmawardena, García-Lario, García-Torres, Gomez, González-Santamaría, Piccolo, Kontizas, Lebreton, Licata, Lindstrøm, Livanou, Romeo, Manteiga, Marocco, Martayan, Marshall, Nicolas, Ordenovic, Palicio, Pallas-Quintela, Pichon, Poggio, Recio-Blanco, Riclet, Santoveña, Schultheis, Segol, Silvelo, Smart, Süveges, Thévenin, Elipe, Ulla, Dillen, Zhao, \& Zorec}]{fouesneau_gaia_2023}
Fouesneau, M., Frémat, Y., Andrae, R., {et~al.} 2023, \aap, 674, A28

\bibitem[{Fraser {et~al.}(2017)Fraser, Casey, Gilmore, Heger, \& Chan}]{fraser_mass_2017}
Fraser, M., Casey, A.~R., Gilmore, G., Heger, A., \& Chan, C. 2017, \mnras, 468, 418

\bibitem[{Frebel(2010)}]{frebel_stellar_2010}
Frebel, A. 2010, Astronomische Nachrichten, 331, 474

\bibitem[{Frebel \& Norris(2015)}]{frebel_near-field_2015}
Frebel, A., \& Norris, J.~E. 2015, \araa, 53, 631

\bibitem[{Frebel {et~al.}(2005)Frebel, Aoki, Christlieb, Ando, Asplund, Barklem, Beers, Eriksson, Fechner, Fujimoto, Honda, Kajino, Minezaki, Nomoto, Norris, Ryan, Takada-Hidai, Tsangarides, \& Yoshii}]{frebel_nucleosynthetic_2005}
Frebel, A., Aoki, W., Christlieb, N., {et~al.} 2005, \nat, 434, 871

\bibitem[{{Gaia Collaboration} {et~al.}(2023){Gaia Collaboration}, Vallenari, Brown, Prusti, de~Bruijne, Arenou, Babusiaux, Biermann, Creevey, Ducourant, Evans, Eyer, Guerra, Hutton, Jordi, Klioner, Lammers, Lindegren, Luri, Mignard, Panem, Pourbaix, Randich, Sartoretti, Soubiran, Tanga, Walton, Bailer-Jones, Bastian, Drimmel, Jansen, Katz, Lattanzi, van Leeuwen, Bakker, Cacciari, Castañeda, De~Angeli, Fabricius, Fouesneau, Frémat, Galluccio, Guerrier, Heiter, Masana, Messineo, Mowlavi, Nicolas, Nienartowicz, Pailler, Panuzzo, Riclet, Roux, Seabroke, Sordo, Thévenin, Gracia-Abril, Portell, Teyssier, Altmann, Andrae, Audard, Bellas-Velidis, Benson, Berthier, Blomme, Burgess, Busonero, Busso, Cánovas, Carry, Cellino, Cheek, Clementini, Damerdji, Davidson, de~Teodoro, Nuñez~Campos, Delchambre, Dell'Oro, Esquej, Fernández-Hernández, Fraile, Garabato, García-Lario, Gosset, Haigron, Halbwachs, Hambly, Harrison, Hernández, Hestroffer, Hodgkin, Holl, Janßen, Jevardat~de Fombelle, Jordan, Krone-Martins,
  Lanzafame, Löffler, Marchal, Marrese, Moitinho, Muinonen, Osborne, Pancino, Pauwels, Recio-Blanco, Reylé, Riello, Rimoldini, Roegiers, Rybizki, Sarro, Siopis, Smith, Sozzetti, Utrilla, van Leeuwen, Abbas, Ábrahám, Abreu~Aramburu, Aerts, Aguado, Ajaj, Aldea-Montero, Altavilla, Álvarez, Alves, Anders, Anderson, Anglada~Varela, Antoja, Baines, Baker, Balaguer-Núñez, Balbinot, Balog, Barache, Barbato, Barros, Barstow, Bartolomé, Bassilana, Bauchet, Becciani, Bellazzini, Berihuete, Bernet, Bertone, Bianchi, Binnenfeld, Blanco-Cuaresma, Blazere, Boch, Bombrun, Bossini, Bouquillon, Bragaglia, Bramante, Breedt, Bressan, Brouillet, Brugaletta, Bucciarelli, Burlacu, Butkevich, Buzzi, Caffau, Cancelliere, Cantat-Gaudin, Carballo, Carlucci, Carnerero, Carrasco, Casamiquela, Castellani, Castro-Ginard, Chaoul, Charlot, Chemin, Chiaramida, Chiavassa, Chornay, Comoretto, Contursi, Cooper, Cornez, Cowell, Crifo, Cropper, Crosta, Crowley, Dafonte, Dapergolas, David, David, de~Laverny, De~Luise, De~March, De~Ridder,
  de~Souza, de~Torres, del Peloso, del Pozo, Delbo, Delgado, Delisle, Demouchy, Dharmawardena, Di~Matteo, Diakite, Diener, Distefano, Dolding, Edvardsson, Enke, Fabre, Fabrizio, Faigler, Fedorets, Fernique, Fienga, Figueras, Fournier, Fouron, Fragkoudi, Gai, Garcia-Gutierrez, Garcia-Reinaldos, García-Torres, Garofalo, Gavel, Gavras, Gerlach, Geyer, Giacobbe, Gilmore, Girona, Giuffrida, Gomel, Gomez, González-Núñez, González-Santamaría, González-Vidal, Granvik, Guillout, Guiraud, Gutiérrez-Sánchez, Guy, Hatzidimitriou, Hauser, Haywood, Helmer, Helmi, Sarmiento, Hidalgo, Hilger, Hładczuk, Hobbs, Holland, Huckle, Jardine, Jasniewicz, Jean-Antoine~Piccolo, Jiménez-Arranz, Jorissen, Juaristi~Campillo, Julbe, Karbevska, Kervella, Khanna, Kontizas, Kordopatis, Korn, Kóspál, Kostrzewa-Rutkowska, Kruszyńska, Kun, Laizeau, Lambert, Lanza, Lasne, Le~Campion, Lebreton, Lebzelter, Leccia, Leclerc, Lecoeur-Taibi, Liao, Licata, Lindstrøm, Lister, Livanou, Lobel, Lorca, Loup, Madrero~Pardo, Magdaleno~Romeo,
  Managau, Mann, Manteiga, Marchant, Marconi, Marcos, Marcos~Santos, Marín~Pina, Marinoni, Marocco, Marshall, Martin~Polo, Martín-Fleitas, Marton, Mary, Masip, Massari, Mastrobuono-Battisti, Mazeh, McMillan, Messina, Michalik, Millar, Mints, Molina, Molinaro, Molnár, Monari, Monguió, Montegriffo, Montero, Mor, Mora, Morbidelli, Morel, Morris, Muraveva, Murphy, Musella, Nagy, Noval, Ocaña, Ogden, Ordenovic, Osinde, Pagani, Pagano, Palaversa, Palicio, Pallas-Quintela, Panahi, Payne-Wardenaar, Peñalosa~Esteller, Penttilä, Pichon, Piersimoni, Pineau, Plachy, Plum, Poggio, Prša, Pulone, Racero, Ragaini, Rainer, Raiteri, Rambaux, Ramos, Ramos-Lerate, Re~Fiorentin, Regibo, Richards, Rios~Diaz, Ripepi, Riva, Rix, Rixon, Robichon, Robin, Robin, Roelens, Rogues, Rohrbasser, Romero-Gómez, Rowell, Royer, Ruz~Mieres, Rybicki, Sadowski, Sáez~Núñez, Sagristà~Sellés, Sahlmann, Salguero, Samaras, Sanchez~Gimenez, Sanna, Santoveña, Sarasso, Schultheis, Sciacca, Segol, Segovia, Ségransan, Semeux, Shahaf,
  Siddiqui, Siebert, Siltala, Silvelo, Slezak, Slezak, Smart, Snaith, Solano, Solitro, Souami, Souchay, Spagna, Spina, Spoto, Steele, Steidelmüller, Stephenson, Süveges, Surdej, Szabados, Szegedi-Elek, Taris, Taylor, Teixeira, Tolomei, Tonello, Torra, Torra, Torralba~Elipe, Trabucchi, Tsounis, Turon, Ulla, Unger, Vaillant, van Dillen, van Reeven, Vanel, Vecchiato, Viala, Vicente, Voutsinas, Weiler, Wevers, Wyrzykowski, Yoldas, Yvard, Zhao, Zorec, Zucker, \& Zwitter}]{gaia_collaboration_gaia_2023}
{Gaia Collaboration}, Vallenari, A., Brown, A. G.~A., {et~al.} 2023, \aap, 674, A1

\bibitem[{Greif {et~al.}(2009)Greif, Johnson, Klessen, \& Bromm}]{greif_observational_2009}
Greif, T.~H., Johnson, J.~L., Klessen, R.~S., \& Bromm, V. 2009, \mnras, 399, 639

\bibitem[{Grenon(1999)}]{grenon_kinematics_1999}
Grenon, M. 1999, Astrophysics and Space Science, 265, 331

\bibitem[{{Harris}(2010)}]{harris_new_2010}
{Harris}, W.~E. 2010, arXiv e-prints, arXiv:1012.3224

\bibitem[{Helmi {et~al.}(2018)Helmi, Babusiaux, Koppelman, Massari, Veljanoski, \& Brown}]{helmi_merger_2018}
Helmi, A., Babusiaux, C., Koppelman, H.~H., {et~al.} 2018, \nat, 563, 85

\bibitem[{Holmberg {et~al.}(2007)Holmberg, Nordström, \& Andersen}]{holmberg_geneva-copenhagen_2007}
Holmberg, J., Nordström, B., \& Andersen, J. 2007, Astronomy \& Astrophysics, 475, 519, number: 2 Publisher: EDP Sciences

\bibitem[{Holmberg {et~al.}(2009)Holmberg, Nordström, \& Andersen}]{holmberg_geneva-copenhagen_2009}
---. 2009, Astronomy \& Astrophysics, 501, 941, number: 3 Publisher: EDP Sciences

\bibitem[{{Hong} {et~al.}(2024){Hong}, {Beers}, {Lee}, {Huang}, {Hirai}, {Cabrera Garcia}, {Shank}, {Xu}, {Yuan}, {Mardini}, {Catapano}, {Zhao}, {Fan}, {Zheng}, {Wang}, {Tan}, {Zhao}, \& {Li}}]{hong_candidate_2023}
{Hong}, J., {Beers}, T.~C., {Lee}, Y.~S., {et~al.} 2024, \apjs, 273, 12

\bibitem[{Hopkins {et~al.}(2018)Hopkins, Wetzel, Kereš, Faucher-Giguère, Quataert, Boylan-Kolchin, Murray, Hayward, Garrison-Kimmel, Hummels, Feldmann, Torrey, Ma, Anglés-Alcázar, Su, Orr, Schmitz, Escala, Sanderson, Grudić, Hafen, Kim, Fitts, Bullock, Wheeler, Chan, Elbert, \& Narayanan}]{hopkins_fire-2_2018}
Hopkins, P.~F., Wetzel, A., Kereš, D., {et~al.} 2018, \mnras, 480, 800

\bibitem[{Howes {et~al.}(2015)Howes, Casey, Asplund, Keller, Yong, Nataf, Poleski, Lind, Kobayashi, Owen, Ness, Bessell, Da~Costa, Schmidt, Tisserand, Udalski, Szymański, Soszyński, Pietrzyński, Ulaczyk, Wyrzykowski, Pietrukowicz, Skowron, Kozłowski, \& Mróz}]{howes_extremely_2015}
Howes, L.~M., Casey, A.~R., Asplund, M., {et~al.} 2015, \nat, 527, 484

\bibitem[{Howes {et~al.}(2016)Howes, Asplund, Keller, Casey, Yong, Lind, Frebel, Hays, Alves-Brito, Bessell, Casagrande, Marino, Nataf, Owen, Da~Costa, Schmidt, \& Tisserand}]{howes_embla_2016}
Howes, L.~M., Asplund, M., Keller, S.~C., {et~al.} 2016, \mnras, 460, 884

\bibitem[{Husser {et~al.}(2013)Husser, Berg, Dreizler, Homeier, Reiners, Barman, \& Hauschildt}]{husser_new_2013}
Husser, T.-O., Berg, S. W.-v., Dreizler, S., {et~al.} 2013, \aap, 553, A6

\bibitem[{Ishigaki {et~al.}(2021)Ishigaki, Hartwig, Tarumi, Leung, Tominaga, Kobayashi, Magg, Simionescu, \& Nomoto}]{ishigaki_origin_2021}
Ishigaki, M.~N., Hartwig, T., Tarumi, Y., {et~al.} 2021, \mnras, 506, 5410

\bibitem[{Ishiyama {et~al.}(2016)Ishiyama, Sudo, Yokoi, Hasegawa, Tominaga, \& Susa}]{ishiyama_where_2016}
Ishiyama, T., Sudo, K., Yokoi, S., {et~al.} 2016, \apj, 826, 9

\bibitem[{Jin {et~al.}(2024)Jin, Trager, Dalton, Aguerri, Drew, Falcón-Barroso, Gänsicke, Hill, Iovino, Pieri, Poggianti, Smith, Vallenari, Abrams, Aguado, Antoja, Aragón-Salamanca, Ascasibar, Babusiaux, Balcells, Barrena, Battaglia, Belokurov, Bensby, Bonifacio, Bragaglia, Carrasco, Carrera, Cornwell, Domínguez-Palmero, Duncan, Famaey, Fariña, Gonzalez, Guest, Hatch, Hess, Hoskin, Irwin, Knapen, Koposov, Kuchner, Laigle, Lewis, Longhetti, Lucatello, Méndez-Abreu, Mercurio, Molaeinezhad, Monguió, Morrison, Murphy, Peralta de Arriba, Pérez, Pérez-Ràfols, Picó, Raddi, Romero-Gómez, Royer, Siebert, Seabroke, Som, Terrett, Thomas, Wesson, Worley, Alfaro, Allende Prieto, Alonso-Santiago, Amos, Ashley, Balaguer-Núñez, Balbinot, Bellazzini, Benn, Berlanas, Bernard, Best, Bettoni, Bianco, Bishop, Blomqvist, Boeche, Bolzonella, Bonoli, Bosma, Britavskiy, Busarello, Caffau, Cantat-Gaudin, Castro-Ginard, Couto, Carbajo-Hijarrubia, Carter, Casamiquela, Conrado, Corcho-Caballero, Costantin, Deason,
  de Burgos, De Grandi, Di Matteo, Domínguez-Gómez, Dorda, Drake, Dutta, Erkal, Feltzing, Ferré-Mateu, Feuillet, Figueras, Fossati, Franciosini, Frasca, Fumagalli, Gallazzi, García-Benito, Gentile Fusillo, Gebran, Gilbert, Gledhill, González Delgado, Greimel, Guarcello, Guerra, Gullieuszik, Haines, Hardcastle, Harris, Haywood, Helmi, Hernandez, Herrero, Hughes, Iršič, Jablonka, Jarvis, Jordi, Kondapally, Kordopatis, Krogager, La Barbera, Lam, Larsen, Lemasle, Lewis, Lhomé, Lind, Lodi, Longobardi, Lonoce, Magrini, Maíz Apellániz, Marchal, Marco, Martin, Matsuno, Maurogordato, Merluzzi, Miralda-Escudé, Molinari, Monari, Morelli, Mottram, Naylor, Negueruela, Oñorbe, Pancino, Peirani, Peletier, Pozzetti, Rainer, Ramos, Read, Rossi, Röttgering, Rubiño-Martín, Sabater, San Juan, Sanna, Schallig, Schiavon, Schultheis, Serra, Shimwell, Simón-Díaz, Smith, Sordo, Sorini, Soubiran, Starkenburg, Steele, Stott, Stuik, Tolstoy, Tortora, Tsantaki, Van der Swaelmen, van Weeren, Vergani, Verheijen,
  Verro, Vink, Vioque, Walcher, Walton, Wegg, Weijmans, Williams, Wilson, Wright, Xylakis-Dornbusch, Youakim, Zibetti, \& Zurita}]{jin_wide-field_2024}
Jin, S., Trager, S.~C., Dalton, G.~B., {et~al.} 2024, \mnras, 530, 2688

\bibitem[{Kielty {et~al.}(2021)Kielty, Venn, Sestito, Starkenburg, Martin, Aguado, Arentsen, Fabbro, González~Hernández, Hill, Jablonka, Lardo, Mashonkina, Navarro, Sneden, Thomas, Youakim, Bialek, \& Sánchez-Janssen}]{kielty_pristine_2021}
Kielty, C.~L., Venn, K.~A., Sestito, F., {et~al.} 2021, \mnras, 506, 1438

\bibitem[{{Klessen} \& {Glover}(2023)}]{klessen_first_2023}
{Klessen}, R.~S., \& {Glover}, S. C.~O. 2023, \araa, 61, 65

\bibitem[{Koposov(2019)}]{koposov_rvspecfit_2019}
Koposov, S.~E. 2019, Astrophysics Source Code Library, ascl:1907.013

\bibitem[{Koposov {et~al.}(2011)Koposov, Gilmore, Walker, Belokurov, Evans, Fellhauer, Gieren, Geisler, Monaco, Norris, Okamoto, Peñarrubia, Wilkinson, Wyse, \& Zucker}]{koposov_accurate_2011}
Koposov, S.~E., Gilmore, G., Walker, M.~G., {et~al.} 2011, \apj, 736, 146

\bibitem[{Kordopatis {et~al.}(2020)Kordopatis, Recio-Blanco, Schultheis, \& Hill}]{kordopatis_chemodynamics_2020}
Kordopatis, G., Recio-Blanco, A., Schultheis, M., \& Hill, V. 2020, \aap, 643, A69

\bibitem[{Lagae {et~al.}(2023)Lagae, Amarsi, Díaz, Lind, Nordlander, Hansen, \& Heger}]{lagae_raising_2023}
Lagae, C., Amarsi, A.~M., Díaz, L. F.~R., {et~al.} 2023, \aap, 672, A90

\bibitem[{Lewis {et~al.}(2002)Lewis, Cannon, Taylor, Glazebrook, Bailey, Baldry, Barton, Bridges, Dalton, Farrell, Gray, Lankshear, McCowage, Parry, Sharples, Shortridge, Smith, Stevenson, Straede, Waller, Whittard, Wilcox, \& Willis}]{lewis_anglo-australian_2002}
Lewis, I.~J., Cannon, R.~D., Taylor, K., {et~al.} 2002, \mnras, 333, 279

\bibitem[{{Li} {et~al.}(2024){Li}, {Wong}, {Hogg}, {Rix}, \& {Chandra}}]{li_aspgap_2023}
{Li}, J., {Wong}, K. W.~K., {Hogg}, D.~W., {Rix}, H.-W., \& {Chandra}, V. 2024, \apjs, 272, 2

\bibitem[{Li {et~al.}(2019)Li, Koposov, Zucker, Lewis, Kuehn, Simpson, Ji, Shipp, Mao, Geha, Pace, Mackey, Allam, Tucker, Da~Costa, Erkal, Simon, Mould, Martell, Wan, De~Silva, Bechtol, Balbinot, Belokurov, Bland-Hawthorn, Casey, Cullinane, Drlica-Wagner, Sharma, Vivas, Wechsler, Yanny, \& {S5 Collaboration}}]{li_southern_2019}
Li, T.~S., Koposov, S.~E., Zucker, D.~B., {et~al.} 2019, \mnras, 490, 3508

\bibitem[{Li {et~al.}(2022)Li, Ji, Pace, Erkal, Koposov, Shipp, Da~Costa, Cullinane, Kuehn, Lewis, Mackey, Simpson, Zucker, Ferguson, Martell, Bland-Hawthorn, Balbinot, Tavangar, Drlica-Wagner, De~Silva, \& Simon}]{li_s_2022}
Li, T.~S., Ji, A.~P., Pace, A.~B., {et~al.} 2022, \apj, 928, 30

\bibitem[{Limberg {et~al.}(2022)Limberg, Souza, Pérez-Villegas, Rossi, Perottoni, \& Santucci}]{limberg_reconstructing_2022}
Limberg, G., Souza, S.~O., Pérez-Villegas, A., {et~al.} 2022, \apj, 935, 109

\bibitem[{Majewski {et~al.}(2012)Majewski, Nidever, Smith, Damke, Kunkel, Patterson, Bizyaev, \& Pérez}]{majewski_exploring_2012}
Majewski, S.~R., Nidever, D.~L., Smith, V.~V., {et~al.} 2012, \apjl, 747, L37

\bibitem[{{Martin} {et~al.}(2023){Martin}, {Starkenburg}, {Yuan}, {Fouesneau}, {Ardern-Arentsen}, {De Angeli}, {Gran}, {Montelius}, {Rusterucci}, {Andrae}, {Bellazzini}, {Montegriffo}, {Esselink}, {Zhang}, {Venn}, {Viswanathan}, {Aguado}, {Battaglia}, {Bayer}, {Bonifacio}, {Caffau}, {C{\^o}t{\'e}}, {Carlberg}, {Fabbro}, {Fern{\'a}ndez Alvar}, {Gonz{\'a}lez Hern{\'a}ndez}, {Gonz{\'a}lez Rivera de La Vernhe}, {Hill}, {Ibata}, {Jablonka}, {Kordopatis}, {Lardo}, {McConnachie}, {Navarrete}, {Navarro}, {Recio-Blanco}, {S{\'a}nchez Janssen}, {Sestito}, {Thomas}, {Vitali}, \& {Youakim}}]{martin_pristine_2023}
{Martin}, N.~F., {Starkenburg}, E., {Yuan}, Z., {et~al.} 2023, arXiv e-prints, arXiv:2308.01344

\bibitem[{Mas-Ribas {et~al.}(2016)Mas-Ribas, Dijkstra, \& Forero-Romero}]{mas-ribas_boosting_2016}
Mas-Ribas, L., Dijkstra, M., \& Forero-Romero, J.~E. 2016, \apj, 833, 65

\bibitem[{Minchev {et~al.}(2012)Minchev, Famaey, Quillen, Matteo, Combes, Vlajić, Erwin, \& Bland-Hawthorn}]{minchev_evolution_2012}
Minchev, I., Famaey, B., Quillen, A.~C., {et~al.} 2012, Astronomy \& Astrophysics, 548, A126, publisher: EDP Sciences

\bibitem[{Myeong {et~al.}(2018)Myeong, Evans, Belokurov, Sanders, \& Koposov}]{myeong_sausage_2018}
Myeong, G.~C., Evans, N.~W., Belokurov, V., Sanders, J.~L., \& Koposov, S.~E. 2018, \apjl, 863, L28

\bibitem[{Nelder \& Mead(1965)}]{Nelder_Mead_1965}
Nelder, J.~A., \& Mead, R. 1965, The Computer Journal, 7, 308–313

\bibitem[{Nomoto {et~al.}(2013)Nomoto, Kobayashi, \& Tominaga}]{nomoto_nucleosynthesis_2013}
Nomoto, K., Kobayashi, C., \& Tominaga, N. 2013, \araa, 51, 457

\bibitem[{Nordlander {et~al.}(2019)Nordlander, Bessell, Da~Costa, Mackey, Asplund, Casey, Chiti, Ezzeddine, Frebel, Lind, Marino, Murphy, Norris, Schmidt, \& Yong}]{nordlander_lowest_2019}
Nordlander, T., Bessell, M.~S., Da~Costa, G.~S., {et~al.} 2019, \mnras, 488, L109

\bibitem[{Nordström {et~al.}(2004)Nordström, Mayor, Andersen, Holmberg, Pont, Jørgensen, Olsen, Udry, \& Mowlavi}]{nordstrom_geneva-copenhagen_2004}
Nordström, B., Mayor, M., Andersen, J., {et~al.} 2004, Astronomy \& Astrophysics, 418, 989, number: 3 Publisher: EDP Sciences

\bibitem[{Norris {et~al.}(2007)Norris, Christlieb, Korn, Eriksson, Bessell, Beers, Wisotzki, \& Reimers}]{norris_he_2007}
Norris, J.~E., Christlieb, N., Korn, A.~J., {et~al.} 2007, \apj, 670, 774

\bibitem[{Norris {et~al.}(2013)Norris, Yong, Bessell, Christlieb, Asplund, Gilmore, Wyse, Beers, Barklem, Frebel, \& Ryan}]{norris_most_2013}
Norris, J.~E., Yong, D., Bessell, M.~S., {et~al.} 2013, \apj, 762, 28

\bibitem[{O'Donnell(1994)}]{odonnell_r_1994}
O'Donnell, J.~E. 1994, \apj, 422, 158

\bibitem[{Oh {et~al.}(2001)Oh, Haiman, \& Rees}]{oh_he_2001}
Oh, S.~P., Haiman, Z., \& Rees, M.~J. 2001, \apj, 553, 73

\bibitem[{Onken {et~al.}(2019)Onken, Wolf, Bessell, Chang, Costa, Luvaul, Mackey, Schmidt, \& Shao}]{onken_skymapper_2019}
Onken, C.~A., Wolf, C., Bessell, M.~S., {et~al.} 2019, \pasa, 36, e033

\bibitem[{Placco {et~al.}(2014)Placco, Frebel, Beers, \& Stancliffe}]{placco_carbon-enhanced_2014}
Placco, V.~M., Frebel, A., Beers, T.~C., \& Stancliffe, R.~J. 2014, \apj, 797, 21

\bibitem[{Plez(2012)}]{plez_turbospectrum_2012}
Plez, B. 2012, Astrophysics Source Code Library, ascl:1205.004

\bibitem[{Riaz {et~al.}(2022)Riaz, Hartwig, \& Latif}]{riaz_unveiling_2022}
Riaz, S., Hartwig, T., \& Latif, M.~A. 2022, \apj, 937, L6

\bibitem[{Roederer {et~al.}(2014)Roederer, Preston, Thompson, Shectman, Sneden, Burley, \& Kelson}]{roederer_search_2014}
Roederer, I.~U., Preston, G.~W., Thompson, I.~B., {et~al.} 2014, \aj, 147, 136

\bibitem[{Rydberg {et~al.}(2013)Rydberg, Zackrisson, Lundqvist, \& Scott}]{rydberg_detection_2013}
Rydberg, C.-E., Zackrisson, E., Lundqvist, P., \& Scott, P. 2013, \mnras, 429, 3658

\bibitem[{Salvadori {et~al.}(2010)Salvadori, Ferrara, Schneider, Scannapieco, \& Kawata}]{salvadori_mining_2010}
Salvadori, S., Ferrara, A., Schneider, R., Scannapieco, E., \& Kawata, D. 2010, \mnras, 401, L5

\bibitem[{{Santistevan} {et~al.}(2021){Santistevan}, {Wetzel}, {Sanderson}, {El-Badry}, {Samuel}, \& {Faucher-Gigu{\`e}re}}]{santistevan_origin_2021}
{Santistevan}, I.~B., {Wetzel}, A., {Sanderson}, R.~E., {et~al.} 2021, \mnras, 505, 921

\bibitem[{Saunders {et~al.}(2004)Saunders, Bridges, Gillingham, Haynes, Smith, Whittard, Churilov, Lankshear, Croom, Jones, \& Boshuizen}]{saunders_aaomega_2004}
Saunders, W., Bridges, T., Gillingham, P., {et~al.} 2004, in Ground-based {Instrumentation} for {Astronomy}, Vol. 5492 (SPIE), 389--400

\bibitem[{Scannapieco {et~al.}(2003)Scannapieco, Schneider, \& Ferrara}]{scannapieco_detectability_2003}
Scannapieco, E., Schneider, R., \& Ferrara, A. 2003, \apj, 589, 35

\bibitem[{Schlegel {et~al.}(1998)Schlegel, Finkbeiner, \& Davis}]{schlegel_maps_1998}
Schlegel, D.~J., Finkbeiner, D.~P., \& Davis, M. 1998, \apj, 500, 525

\bibitem[{{Schneider} {et~al.}(2012){Schneider}, {Omukai}, {Limongi}, {Ferrara}, {Salvaterra}, {Chieffi}, \& {Bianchi}}]{schneider_formation_2012}
{Schneider}, R., {Omukai}, K., {Limongi}, M., {et~al.} 2012, \mnras, 423, L60

\bibitem[{Schönrich \& Binney(2009)}]{schonrich_chemical_2009}
Schönrich, R., \& Binney, J. 2009, Monthly Notices of the Royal Astronomical Society, 396, 203

\bibitem[{Schönrich {et~al.}(2010)Schönrich, Binney, \& Dehnen}]{schonrich_local_2010}
Schönrich, R., Binney, J., \& Dehnen, W. 2010, \mnras, 403, 1829

\bibitem[{Schörck {et~al.}(2009)Schörck, Christlieb, Cohen, Beers, Shectman, Thompson, McWilliam, Bessell, Norris, Meléndez, Ramírez, Haynes, Cass, Hartley, Russell, Watson, Zickgraf, Behnke, Fechner, Fuhrmeister, Barklem, Edvardsson, Frebel, Wisotzki, \& Reimers}]{schorck_stellar_2009}
Schörck, T., Christlieb, N., Cohen, J.~G., {et~al.} 2009, \aap, 507, 817

\bibitem[{Sestito {et~al.}(2019)Sestito, Longeard, Martin, Starkenburg, Fouesneau, González Hernández, Arentsen, Ibata, Aguado, Carlberg, Jablonka, Navarro, Tolstoy, \& Venn}]{sestito_tracing_2019}
Sestito, F., Longeard, N., Martin, N.~F., {et~al.} 2019, \mnras, 484, 2166

\bibitem[{Sestito {et~al.}(2020)Sestito, Martin, Starkenburg, Arentsen, Ibata, Longeard, Kielty, Youakim, Venn, Aguado, Carlberg, González Hernández, Hill, Jablonka, Kordopatis, Malhan, Navarro, Sánchez-Janssen, Thomas, Tolstoy, Wilson, Palicio, Bialek, Garcia-Dias, Lucchesi, North, Osorio, Patrick, \& Peralta de Arriba}]{sestito_pristine_2020}
Sestito, F., Martin, N.~F., Starkenburg, E., {et~al.} 2020, \mnras: Letters, 497, L7

\bibitem[{Sestito {et~al.}(2021)Sestito, Buck, Starkenburg, Martin, Navarro, Venn, Obreja, Jablonka, \& Macciò}]{sestito_exploring_2021}
Sestito, F., Buck, T., Starkenburg, E., {et~al.} 2021, \mnras, 500, 3750

\bibitem[{Sharp {et~al.}(2006)Sharp, Saunders, Smith, Churilov, Correll, Dawson, Farrel, Frost, Haynes, Heald, Lankshear, Mayfield, Waller, \& Whittard}]{sharp_performance_2006}
Sharp, R., Saunders, W., Smith, G., {et~al.} 2006, 6269, 62690G

\bibitem[{Singh {et~al.}(2020)Singh, Hansen, Byrgesen, Reichert, \& Reggiani}]{singh_empirical_2020}
Singh, D., Hansen, C.~J., Byrgesen, J.~S., Reichert, M., \& Reggiani, H.~M. 2020, \aap, 634, A72

\bibitem[{Starkenburg {et~al.}(2010)Starkenburg, Hill, Tolstoy, González~Hernández, Irwin, Helmi, Battaglia, Jablonka, Tafelmeyer, Shetrone, Venn, \& de~Boer}]{starkenburg_nir_2010}
Starkenburg, E., Hill, V., Tolstoy, E., {et~al.} 2010, \aap, 513, A34

\bibitem[{Starkenburg {et~al.}(2017)Starkenburg, Martin, Youakim, Aguado, Allende~Prieto, Arentsen, Bernard, Bonifacio, Caffau, Carlberg, Côté, Fouesneau, François, Franke, González~Hernández, Gwyn, Hill, Ibata, Jablonka, Longeard, McConnachie, Navarro, Sánchez-Janssen, Tolstoy, \& Venn}]{starkenburg_pristine_2017}
Starkenburg, E., Martin, N., Youakim, K., {et~al.} 2017, \mnras, 471, 2587

\bibitem[{Umeda \& Nomoto(2002)}]{umeda_nucleosynthesis_2002}
Umeda, H., \& Nomoto, K. 2002, \apj, 565, 385

\bibitem[{{Viswanathan} {et~al.}(2024){Viswanathan}, {Yuan}, {Ardern-Arentsen}, {Starkenburg}, {Martin}, {Youakim}, {Ibata}, {Sestito}, {Matsuno}, {Allende Prieto}, {Barwell}, {Bayer}, {Doliva-Dolinsky}, {Fernandez-Alvar}, {Galan-de Anta}, {Jhass}, {Longeard}, {Arroyo-Polonio}, {Massana}, {Montelius}, {Rusterucci}, {Santos}, {Thomas}, {Vitali}, {Wu}, {Yarker}, {Ye}, {Aguado}, {Gran}, \& {Navarro}}]{viswanathan_pristine_2024}
{Viswanathan}, A., {Yuan}, Z., {Ardern-Arentsen}, A., {et~al.} 2024, arXiv e-prints, arXiv:2405.13124

\bibitem[{Wehrhahn {et~al.}(2023)Wehrhahn, Piskunov, \& Ryabchikova}]{wehrhahn_pysme_2023}
Wehrhahn, A., Piskunov, N., \& Ryabchikova, T. 2023, \aap, 671, A171

\bibitem[{Whitehouse {et~al.}(2018)Whitehouse, Farihi, Green, Wilson, \& Subasavage}]{whitehouse_dwarf_2018}
Whitehouse, L.~J., Farihi, J., Green, P.~J., Wilson, T.~G., \& Subasavage, J.~P. 2018, \mnras, 479, 3873

\bibitem[{{Xylakis-Dornbusch} {et~al.}(2024){Xylakis-Dornbusch}, {Christlieb}, {Hansen}, {Nordlander}, {Webber}, \& {Marshall}}]{xylakis-dornbusch_metallicities_2024}
{Xylakis-Dornbusch}, T., {Christlieb}, N., {Hansen}, T.~T., {et~al.} 2024, \aap, 687, A177

\bibitem[{{Yao} {et~al.}(2024){Yao}, {Ji}, {Koposov}, \& {Limberg}}]{yao_188000_2023}
{Yao}, Y., {Ji}, A.~P., {Koposov}, S.~E., \& {Limberg}, G. 2024, \mnras, 527, 10937

\bibitem[{Yong {et~al.}(2021)Yong, Da Costa, Bessell, Chiti, {A Frebel}, Gao, Lind, Mackey, Marino, Murphy, Nordlander, Asplund, Casey, Kobayashi, Norris, \& Schmidt}]{yong_high-resolution_2021}
Yong, D., Da Costa, G.~S., Bessell, M.~S., {et~al.} 2021, \mnras, 507, 4102

\bibitem[{Youakim {et~al.}(2020)Youakim, Starkenburg, Martin, Matijevič, Aguado, Allende~Prieto, Arentsen, Bonifacio, Carlberg, González~Hernández, Hill, Kordopatis, Lardo, Navarro, Jablonka, Sánchez~Janssen, Sestito, Thomas, \& Venn}]{youakim_pristine_2020}
Youakim, K., Starkenburg, E., Martin, N.~F., {et~al.} 2020, \mnras, 492, 4986

\bibitem[{Zackrisson {et~al.}(2011)Zackrisson, Rydberg, Schaerer, Östlin, \& Tuli}]{zackrisson_spectral_2011}
Zackrisson, E., Rydberg, C.-E., Schaerer, D., Östlin, G., \& Tuli, M. 2011, \apj, 740, 13

\bibitem[{Zackrisson {et~al.}(2012)Zackrisson, Zitrin, Trenti, Rydberg, Guaita, Schaerer, Broadhurst, Östlin, \& Ström}]{zackrisson_detecting_2012}
Zackrisson, E., Zitrin, A., Trenti, M., {et~al.} 2012, \mnras, 427, 2212

\bibitem[{Zackrisson {et~al.}(2024)Zackrisson, Hultquist, Kordt, Diego, Nabizadeh, Vikaeus, Meena, Zitrin, Volpato, Lundqvist, Welch, Costa, \& Windhorst}]{zackrisson_detection_2024}
Zackrisson, E., Hultquist, A., Kordt, A., {et~al.} 2024, \mnras, 533, 2727

\bibitem[{{Zhang} {et~al.}(2023){Zhang}, {Green}, \& {Rix}}]{zhang_parameters_2023}
{Zhang}, X., {Green}, G.~M., \& {Rix}, H.-W. 2023, \mnras, 524, 1855

\bibitem[{Zwitter {et~al.}(2008)Zwitter, Siebert, Munari, Freeman, Siviero, Watson, Fulbright, Wyse, Campbell, Seabroke, Williams, Steinmetz, Bienaymé, Gilmore, Grebel, Helmi, Navarro, Anguiano, Boeche, Burton, Cass, Dawe, Fiegert, Hartley, Russell, Veltz, Bailin, Binney, Bland-Hawthorn, Brown, Dehnen, Evans, Fiorentin, Fiorucci, Gerhard, Gibson, Kelz, Kujken, Matijevič, Minchev, Parker, Peñarrubia, Quillen, Read, Reid, Roeser, Ruchti, Scholz, Smith, Sordo, Tolstoi, Tomasella, Vidrih, \& Boer}]{zwitter_radial_2008}
Zwitter, T., Siebert, A., Munari, U., {et~al.} 2008, \aj, 136, 421

\end{thebibliography}

\appendix

\section{Comparisons to Literature}
\label{sec:gaia xp compare}
Here, we compare our stellar parameters for the stars in common with those from the following surveys that make use of \textit{Gaia} XP data: \citet{zhang_parameters_2023}, \citet{andrae_gaia_2023}, and \citet{li_aspgap_2023}, noting the comparison with \citet{martin_pristine_2023} was discussed in Section \ref{subsec:xp validation}). We also compare our data with that from the spectroscopic PIGS survey \citet{ardern-arentsen_pristine_2024}, which used 2dF+AAOmega data like our sample.

\subsection{Zhang et al. (2023)}
\begin{figure*}[tp]
    \centering
    \includegraphics[width=1\linewidth]{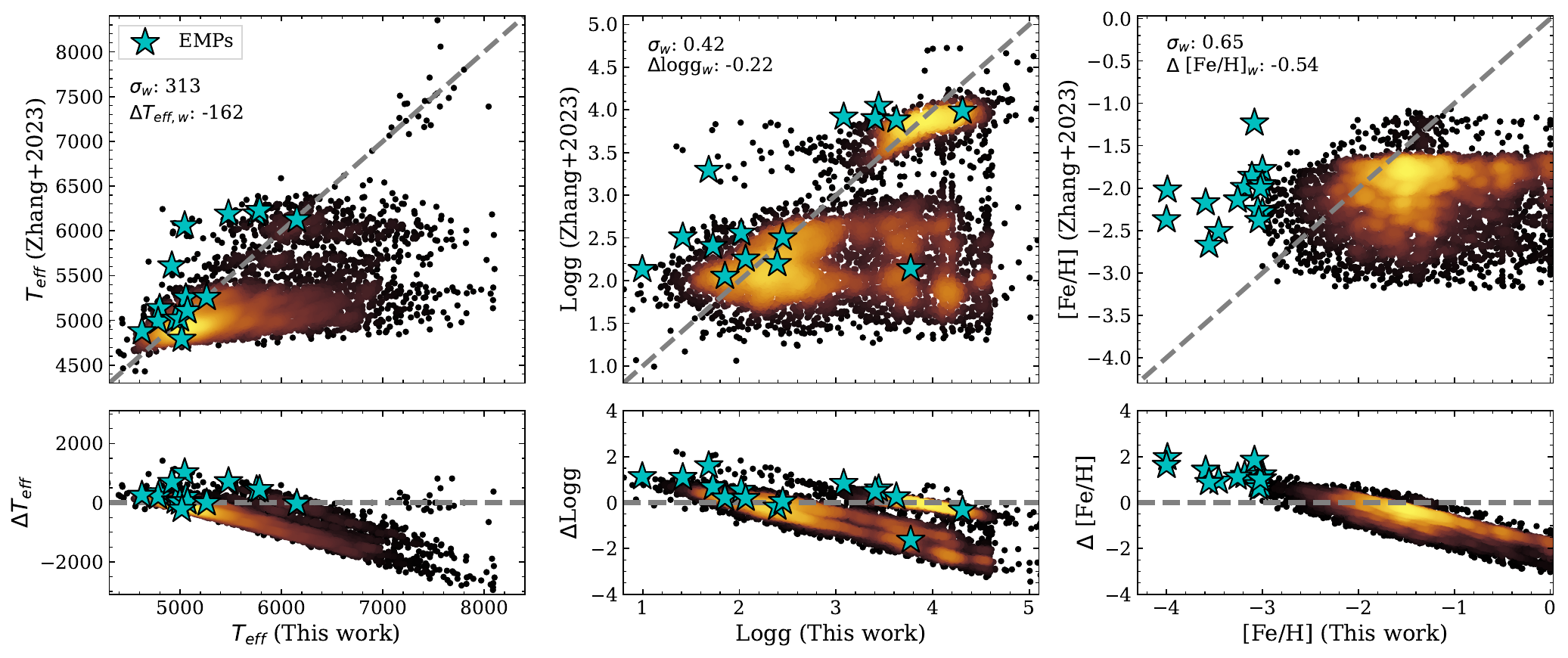}
    \caption{Stellar parameter comparison between us and \citet{zhang_parameters_2023} for our sample of 5783 stars. The weighted standard deviations and mean differences are shown directly on each subplot. The EMPs (blue star symbols) are found to be more metal-rich than our values suggest.}
    \label{fig:aat zhang}
\end{figure*}

We compare our spectroscopic stellar parameter estimates to those from the  \citet{zhang_parameters_2023} catalogue in Fig.\,\ref{fig:aat zhang}. Since \citet{zhang_parameters_2023} provide errors, we also show the weighted standard deviations and mean differences, which are written directly on the subplots. Note that the sample shown includes stars above the selection cut of $\FeH_{\rm{Zhang}} \leq -1.5$\,dex. These were selected for the calibration field ra\_0103-7050 containing globular cluster NGC\,362. 

The $\Teff$ correlation between the two datasets ($\sigma_w = 317$\,K, $\Delta T_{\rm{eff}, w}=-164$\,K) is reasonable for stars within the over-density region at $T_{\textrm{eff}} = 5000$\,K, and a smaller group of stars at $T_{\textrm{eff}} = 6000$\,K, but the majority are below the 1:1 line, showing that our $\Teff$'s are hotter than what \citet{zhang_parameters_2023} suggests. This is the consequence of reddening, as we can demonstrate that stars further to the right along the x-axis have higher reddening than those closer to the 1:1 line. 

Similarly, the $\logg$ relationship ($\sigma_w = 0.42$\,dex, $\Delta \logg_w=-0.23$\,dex) has two main over-density regions that agree with \citet{zhang_parameters_2023}: largest at log\,$g=4.0$\,dex and at $\logg = 2.5$\,dex, but again, our analysis suggests that majority of our stars have greater $\logg$'s. The authors mentioned that a strong degeneracy between their inferred $\logg$'s and \textit{Gaia} parallax was present, due to the inability to measure line widths in \textit{Gaia} XP spectra as a result of the poor spectral resolution. The issue with parallax also increased uncertainty in our sample. Therefore, stars that are dwarfs in both datasets are generally in agreement with each other. For giants, the less precise parallax measurements result in larger differences in our results. We also see that between $2.0 \leq \logg_{\textrm{Zhang}} < 2.5$, the range in surface gravities from our approach is vast, covering every evolutionary stage with $2.0 \leq \logg_{\textrm{This work}} < 4.5$. 

The $\FeH$ correlation ($\sigma_w = 0.66$\,dex, $\Delta \FeH_w=-0.61$\,dex), again, has agreement in the over density region at $\FeH =-1.5$\,dex, but no correlation is shown outside of this. The hard cut at $\FeH = -1.5$ in \citet{zhang_parameters_2023} is due to us selecting stars with metallicities below this. The stars between $-1.0$ and $-1.5$ result from filling out fibres for calibration field ra\_0103-7050 containing globular cluster NGC\,362. There may be a dependence on reddening, with stars with high $E(B-V)$ values typically showing larger discrepancies than those less reddened, but it is not the main driving factor, making the $\Teff -- \FeH$ correlation seen in Fig.\,\ref{fig:aat params} unclear. As with $\logg$, the reason is again most likely due to the low spectral resolution of the \textit{Gaia} XP spectra. 

\subsection{Andrae et al. (2023)}
\begin{figure*}[tp]
    \centering
    \includegraphics[width=1\linewidth]{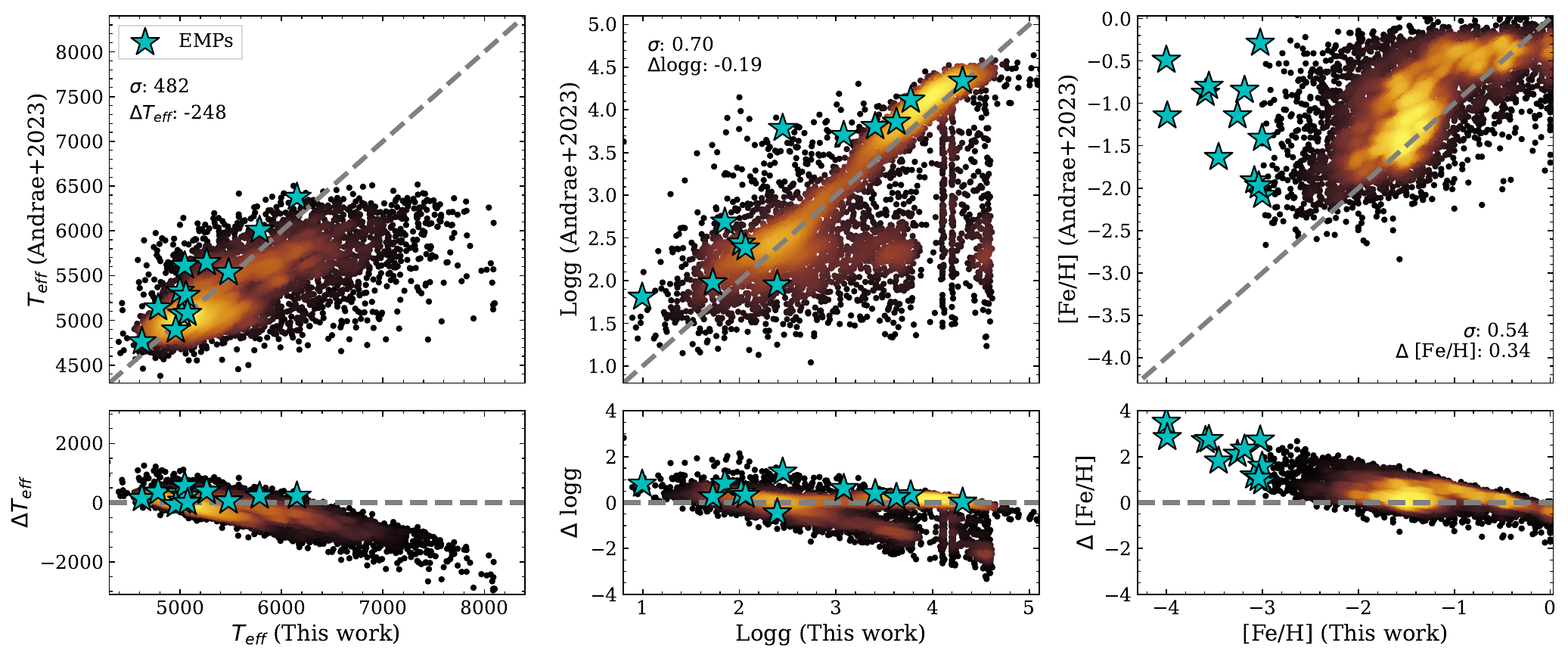}
    \caption{Stellar parameter comparison between us and \citet{andrae_robust_2023} for 7659 stars. The comparison includes 13 of the 15 stars we identified as EMPs. Since errors were not provided, only the unweighted standard deviations and mean differences for each parameter are shown.}
    \label{fig:aat andrae}
\end{figure*}

Using the XGBoost algorithm trained on APOGEE stellar parameters, \citet{andrae_robust_2023} derived stellar parameters for $\sim 175$ million \textit{Gaia} XP stars. This catalogue includes 5655 stars in common with our work (with variables removed), including 13 of the 15 stars we identify as EMPs. The stellar parameter comparison between for the stars in common is shown in Fig.\,\ref{fig:aat andrae}.

The correlation between $\Teff$ for the two datasets ($\sigma = 493$\,K, $\Delta T_{\rm{eff}} = -250$\,K) again shows the similar behaviour towards our $\Teff$'s seen before with \citet{zhang_parameters_2023}, where for a certain $T_{\textrm{eff, XGBoost}}$, the range of $T_{\textrm{eff, This work}}$ is large. Again, this is due to reddening. The 13 EMPs are close to the 1:1 line, showing that the agreement for these stars is good.

For surface gravity ($\sigma = 0.70$\,dex, $\Delta \logg = -0.19$\,dex), we see that a large portion of stars follow the 1:1 line, but a distinct population that has gathered below the line. Both analyses place these stars on the RGB (rather than on the horizontal branch). For this, reddening was again the main reason, with stars in the concentration of points below the 1:1 line ($\logg_{\textrm{XGBoost}} \approx 2.3$ and $\logg_{\textrm{This work}} \approx 3.3$) overall having greater reddening (E(B-V)\,$\geq 0.8$) than those closer to the 1:1 line. Stars in the vertical streaks (the vertical lines at $\logg_{\textrm{This work}} \approx 4.1$ and 4.2) are also highly reddened (E(B-V)\,$\geq 1.0$). The reason for these streaks is likely due to high $\Teff$ estimates placing stars at the turn-off point. Because of this, we are selecting two distinct $\logg$ values depending on their $\FeH$ values or on the age of the isochrones, causing the vertical bands seen in the data.

The metallicity comparison ($\sigma = 0.54$\,dex, $\Delta \FeH = 0.34$\,dex) shows that from $\FeH_{\rm{This work}} = 0$ to $-1.5$, the metallicities from XGBoost remains relatively consistent at $\FeH_{\rm{XGBoost}} = -0.5$, before they drop to lower values and closer to our values. Of note is very few of our EMPs have metallicities below $\FeH_{\rm{XGBoost}} = -2.0$, with several even having values above $\FeH_{\rm{XGBoost}} = -1.0$.

\subsection{Li et al. (2023)}
\begin{figure*}[tp]
    \centering
    \includegraphics[width=1\linewidth]{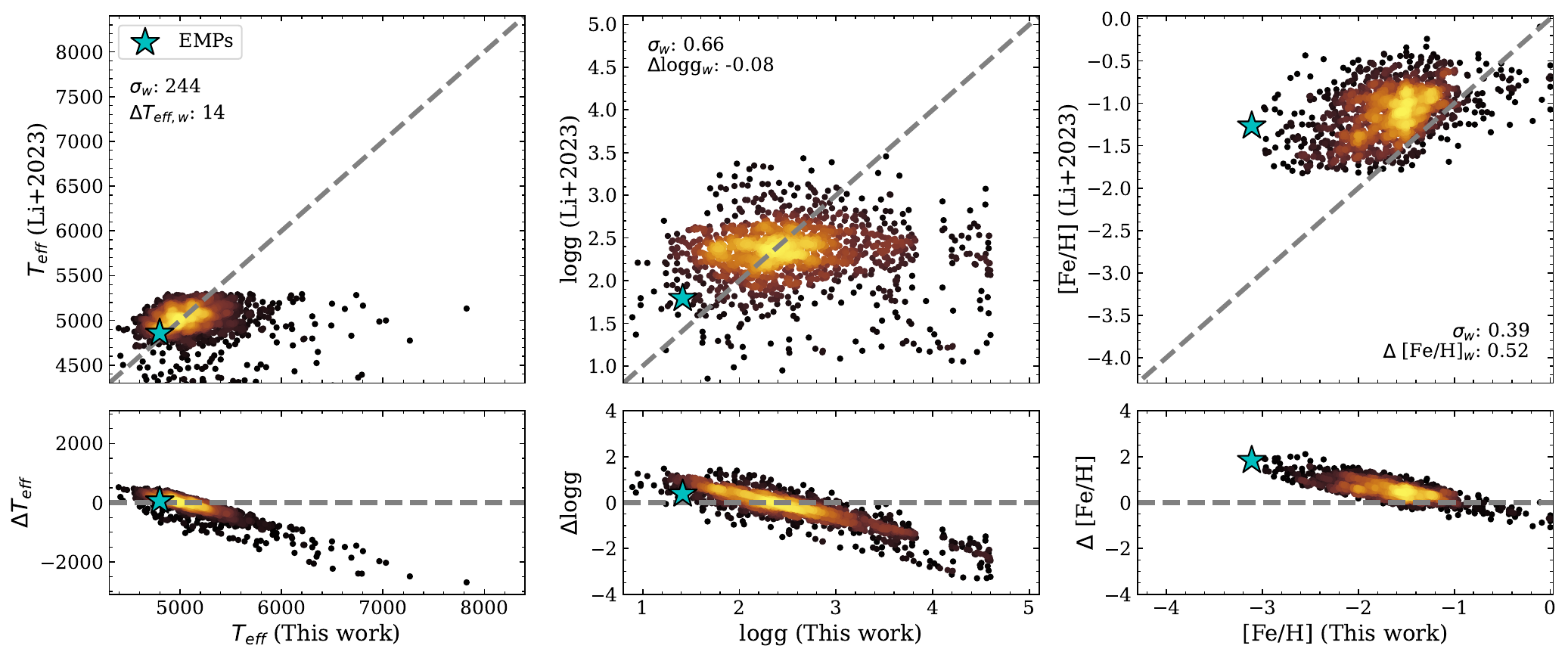}
    \caption{Stellar parameter comparison between us and \citet{li_aspgap_2023} for 1454 stars, with one of them identified as an EMP. Weighted standard deviations and mean differences are also shown.}
    \label{fig:aat li}
\end{figure*}

With the focus on RGB stars, \citet{li_aspgap_2023} estimated the metallicities of 23 million RGBs  from \textit{Gaia} XP as part of \textit{AspGap}, a neural-network based regression system trained on high resolution spectra taken with APOGEE to infer stellar parameters from \textit{Gaia} XP. Their catalogue includes 1454 of our stars, including one EMP star. We show comparison of our stellar parameters to \textit{AspGap} in Fig.\,\ref{fig:aat li}.

For all stars in common, we find mean differences and standard deviations of ($\sigma_w = 250$\,K, $\Delta T_{\rm{eff}, w}=15$\,K) for $\Teff$, ($\sigma_w = 0.65$\,dex, $\Delta \logg_w=-0.070$\,dex) for $\logg$, and ($\sigma_w = 0.40$\,dex, $\Delta \FeH_w=0.53$\,dex) for $\FeH$. No obvious correlations are present.

For $\Teff$, we see a compact grouping at 5000\,K on both axes, consistent with typical $\Teff$'s for stars on the RGB. We also see a smattering of stars stretching towards high $T_{\rm{eff, This work}}$, again the result of reddening. For $\logg$, we see a similar grouping at 2.5\,dex on both axes, but the distribution of $\logg_{\rm{This work}}$ is wider than $\logg_{\rm{AspGap}}$. This is the result of the selection function used in \textit{AspGap} from APOGEE to ensure their stars are RGB, which holds mostly true for our stars. For $\FeH$, we again see a compact grouping, but interestingly, this one does not pass through the 1:1 line. For all our stars in common, the metallicities predicted by \citet{li_aspgap_2023} are all above the 1:1 line by $\Delta \FeH_w=0.53$\,dex, showing that overall, they derive higher metallicities than what we find. This is particularly true for the one EMP star in common, having a difference of 1.8\,dex in metallicity between our two analyses.

\subsection{Ardern-Arentsen et al. (2024)}
\begin{figure*}[tp]
    \centering
    \includegraphics[width=1\linewidth]{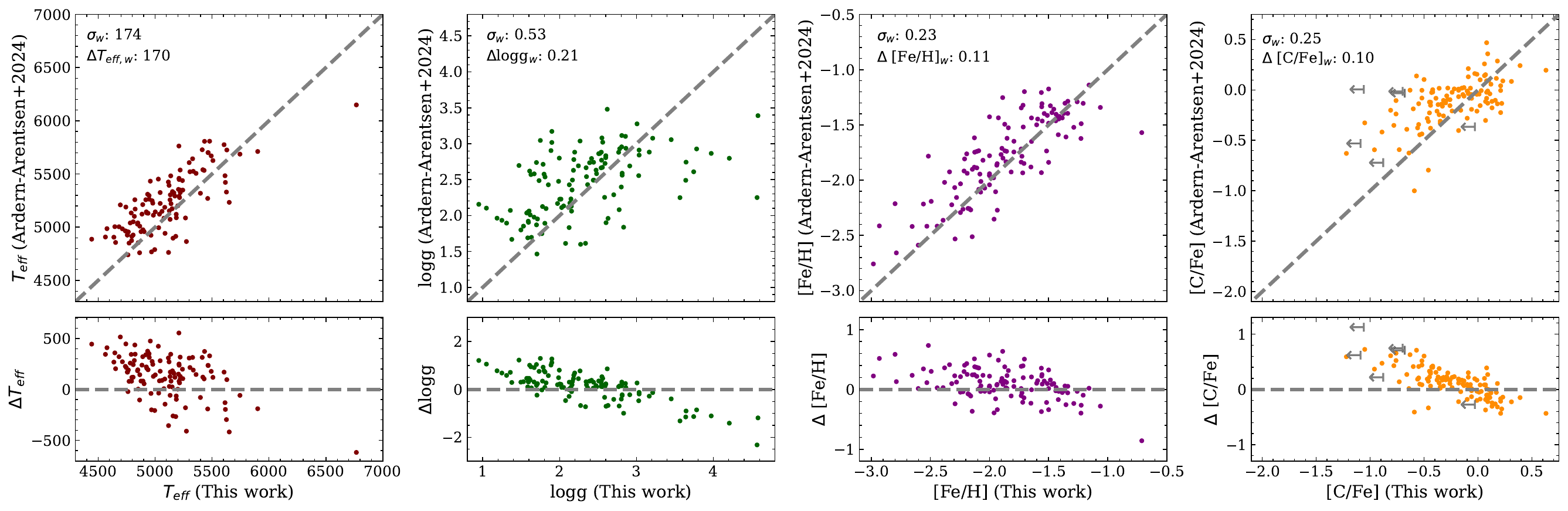}
    \caption{Stellar parameter comparison between us and \citet{ardern-arentsen_pristine_2024} survey for 121 stars in common, with $\XFe{C}$ shown as an additional comparison. Stars with non-detections in the $\XFe{C}$ comparison are shown by leftward-facing arrows representing their upper limits. The weighted standard deviations and mean differences are shown directly on the plots.}
    \label{fig:aat pigs}
\end{figure*}

The Pristine Inner Galaxy Survey (PIGS) \citep{starkenburg_pristine_2017, arentsen_pristine_2020} uses Ca H\&K photometry from MegaCam on the 3.6m Canada-France-Hawaii Telescope (CFHT) to identify metal-poor stars within the inner Galactic bulge region. Candidate metal-poor stars are followed up on the AAT with 2dF+AAOmega, using a very similar instrumental setup to us. \citet{ardern-arentsen_pristine_2024} derived spectroscopic stellar parameters (including $\XFe{C}$) for 13235 stars, with 117 of our stars in common. We show the comparison of stellar parameters in Fig.\,\ref{fig:aat pigs}.

For all four stellar parameters, we see excellent consistency between our values and those of \citet{ardern-arentsen_pristine_2024}. For $\Teff$ ($\sigma_w = 173$\,K, $\Delta T_{\rm{eff}, w}=168$\,K), the scatter is minimal between us, with their temperatures on average slightly hotter. One star (the only one over $\Teff{} = 6000$\,K) has a discrepancy of 600\,K, with it being hotter in our work. It shows significant differences in other stellar parameters, which mostly vanish if we adopted their $\Teff$ value.

The comparison with $\logg$ ($\sigma_w = 0.50$, $\Delta \logg_{w}=0.24$) is good, with majority of the stars having higher values in \citet{ardern-arentsen_pristine_2024}. Their surface gravities were derived using a mix of spectroscopic and parallax information, so some temperature dependence is present. Given that their $\Teff$ values are higher overall, the higher $\logg$'s may be the result of this. 

For $\FeH$ ($\sigma_w = 0.24$, $\Delta \FeH{}_{w}=0.11$), the comparison is excellent, showing the lowest scatter for all datasets we have compared to. With the WiFeS comparison in Fig.\,\ref{fig:wifes vs aat}, we had $\sigma_w = 0.29$, and for our globular cluster analysis in Fig.\,\ref{fig:gc results}, the weighted standard deviations varied from $\sigma_w = 0.13$ to 0.23 dex. The $\sigma_w$ values being similar to those from our WiFeS and globular cluster comparisons means our measurements and errors are reliable.

For the $\XFe{C}$ comparison ($\sigma_w = 0.25$, $\Delta \XFe{C}_{w}=0.10$), we see that overall, \citet{ardern-arentsen_pristine_2024} has higher $\XFe{C}$ abundances than us, but the trend is acceptable. The majority of the stars compared are carbon-depleted ($\XFe{C} < 0.0$), with a handful being carbon-normal ($0.0 \leq \XFe{C} < 0.7$). 

\section{CH G-Band EMP Fits}
\label{append:gband fits}
The CH G-band fits for the 15 EMP stars across wavelength region $4150 \leq \lambda \leq 4450$ \AA{} are shown in Fig.\,\ref{fig:emp carb fits 1} and Fig.\,\ref{fig:emp carb fits 2}. Note, the two figures are zoomed in over the regions $4260 \leq \lambda \leq 4370$ \AA{} to highlight the CH G-band region. Stars with detections (red) have their best $\XFe{C}$ value fitted, alongside their statistical fitting error represented by the shaded region. The $\XFe{C}$ systematic abundance error estimates are shown in the legend. Stars with non-detections (blue) have their upper limit values fitted, which is then quoted in the legend. These values are shown in Table\,\ref{tab:emp cfe}.

\begin{table}[tp]
    \centering
    \caption{$\XFe{C}$ abundance values alongside their systematic abundance error estimates and $1\sigma$ statistical fitting error for our EMP stars. Those with non-detections have their upper limits shown, with the fitting error left empty.}
    \begin{tabular}{lcc}
        \hline
        Star ID & $\XFe{C}$ & $\sigma_{\rm stat}$ \\
        \hline
        ra\_0103-7050\_s236 & $-$0.32 $\pm$ 0.02 & 0.12 \\
        ra\_0752-5047\_s47 & 0.09 $\pm$ 0.07 & 0.17 \\
        ra\_1604-2712\_s188 & $<$ 1.39 & \\
        ra\_1604-2712\_s292 & 1.3 $\pm$ 0.1 & 0.22 \\
        ra\_1633-2814\_s130 & 0.4 $\pm$ 0.2 & 0.17 \\
        ra\_1633-2814\_s284 & $<$ 1.33 & \\
        ra\_1639-2632\_s419 & 0.0 $\pm$ 0.1 & 0.080 \\
        ra\_1648-2642\_s91 & 0.47 $\pm$ 0.08 & 0.095 \\
        ra\_1659-2154\_s114 & $<$ 0.93 & \\
        ra\_1659-2154\_s261 & 0.30 $\pm$ 0.05 & 0.080 \\
        ra\_1659-2154\_s347 & 0.6 $\pm$ 0.2 & 0.15 \\
        ra\_1752-4300\_s155 & 0.05 $\pm$ 0.06 & 0.075 \\
        ra\_1752-4300\_s269 & 0.17 $\pm$ 0.05 & 0.070 \\
        ra\_1832-3457\_s438 & $<$ 1.14 & \\
        ra\_1853-3255\_s51 & 0.79 $\pm$ 0.03 & 0.020 \\
        \hline
    \end{tabular}
    \label{tab:emp cfe}
\end{table}

\begin{figure*}[tp]  
    \centering
     \begin{subfigure}{0.49\textwidth}
        \centering
        \includegraphics[width=0.85\linewidth]{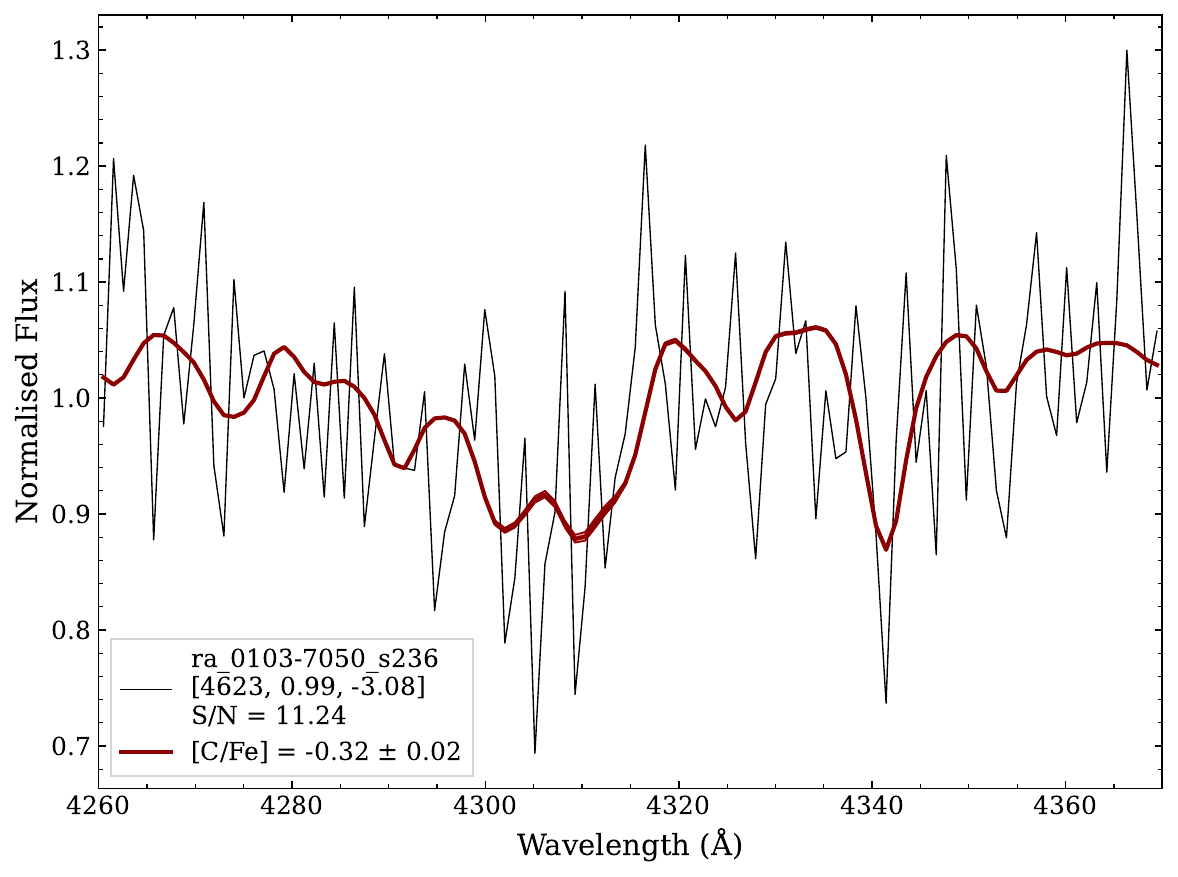}
    \end{subfigure}
    \hfill
    \begin{subfigure}{0.49\textwidth}
        \centering
        \includegraphics[width=0.85\linewidth]{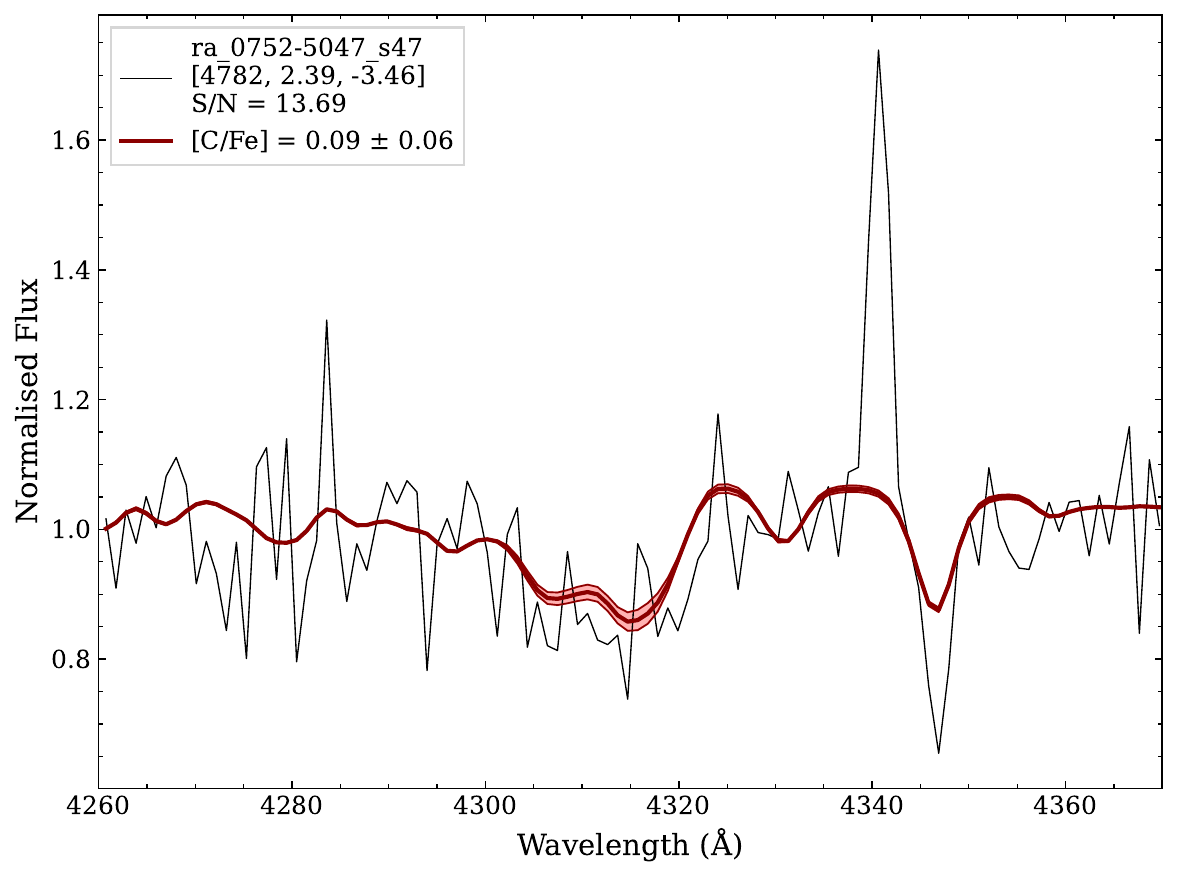}
    \end{subfigure}

    \begin{subfigure}{0.49\textwidth}
        \centering
        \includegraphics[width=0.85\linewidth]{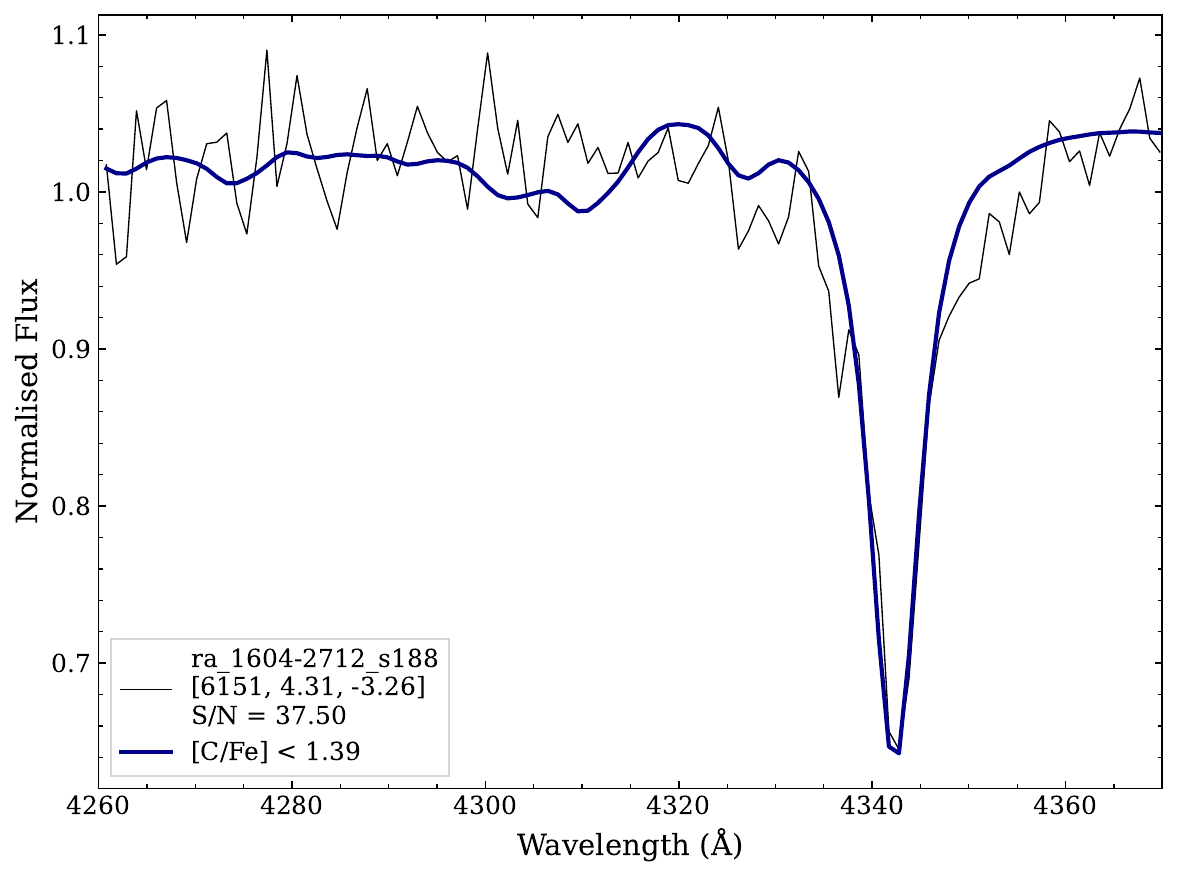}
    \end{subfigure}
    \hfill
    \begin{subfigure}{0.49\textwidth}
        \centering
        \includegraphics[width=0.85\linewidth]{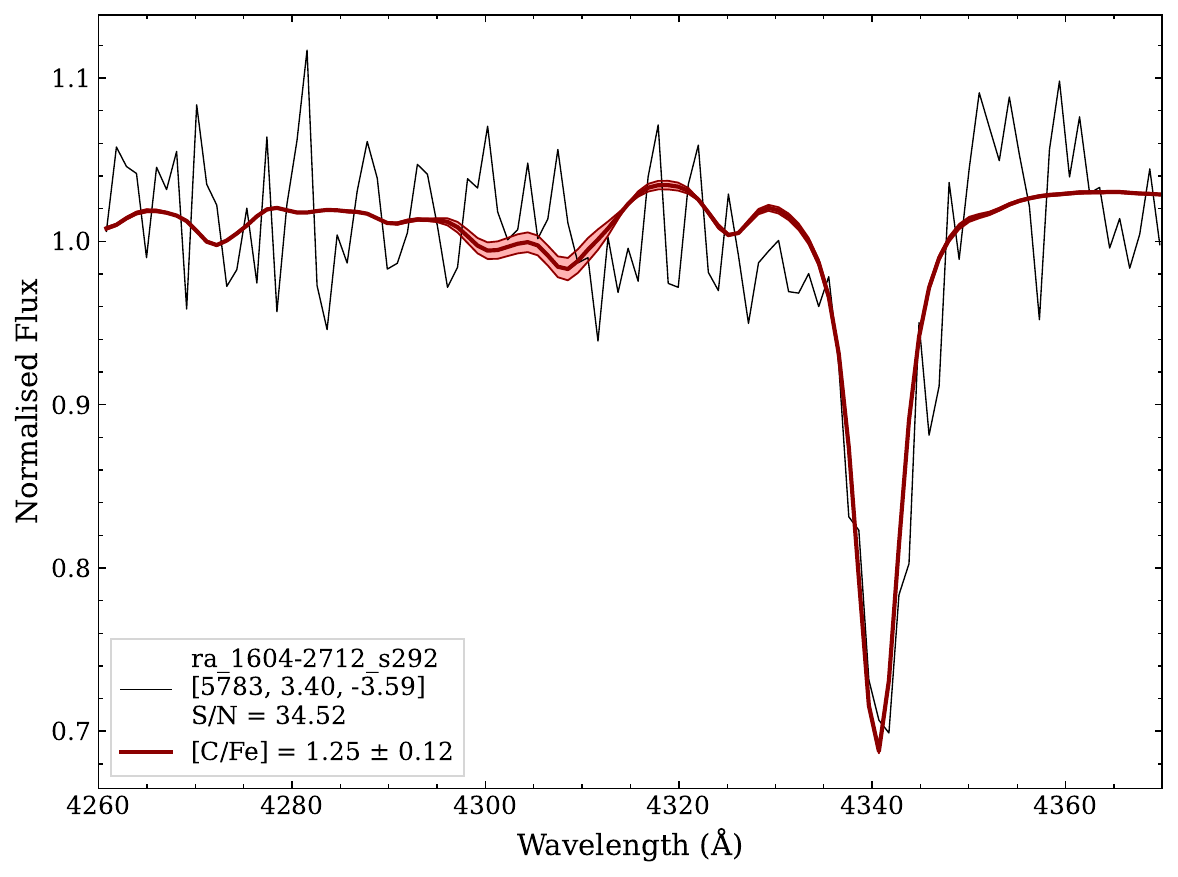}
    \end{subfigure}

    \begin{subfigure}{0.49\textwidth}
        \centering
        \includegraphics[width=0.85\linewidth]{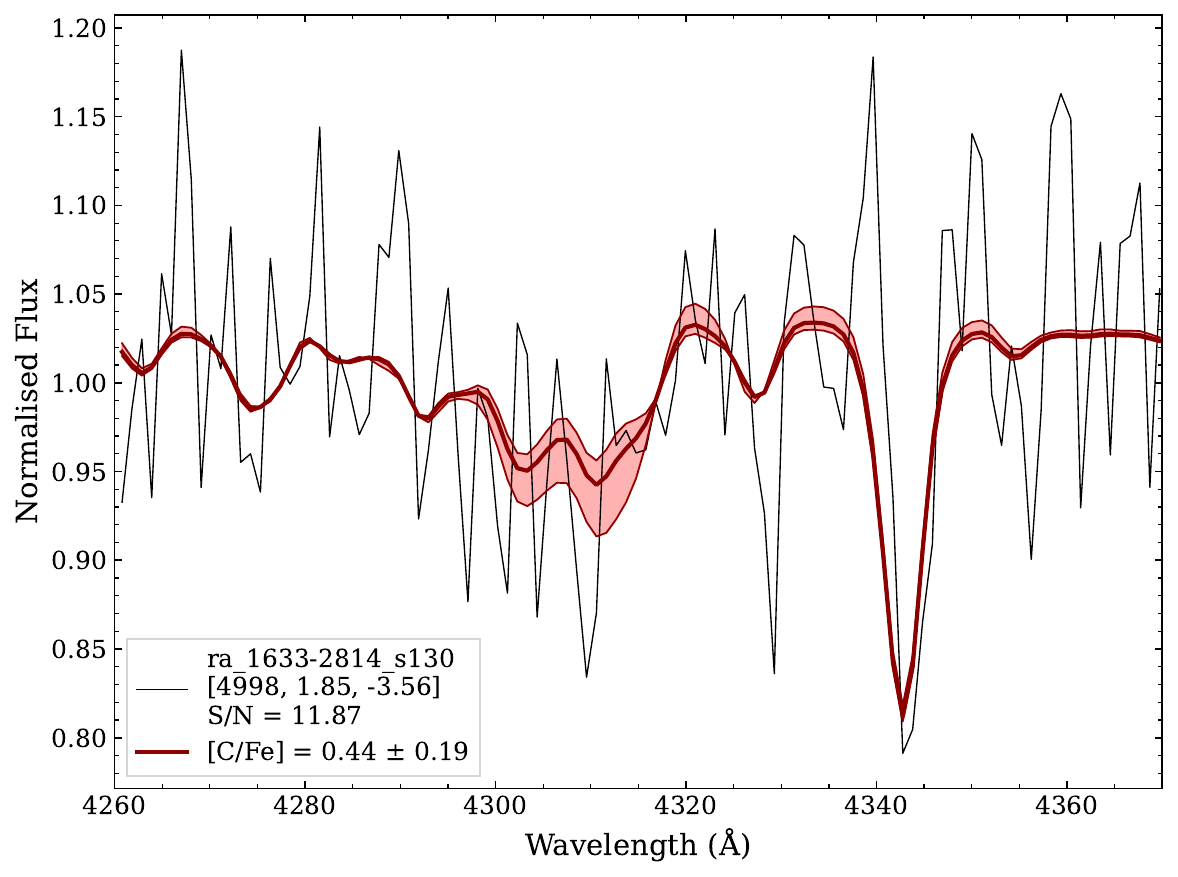}
    \end{subfigure}
    \hfill
    \begin{subfigure}{0.49\textwidth}
        \centering
        \includegraphics[width=0.85\linewidth]{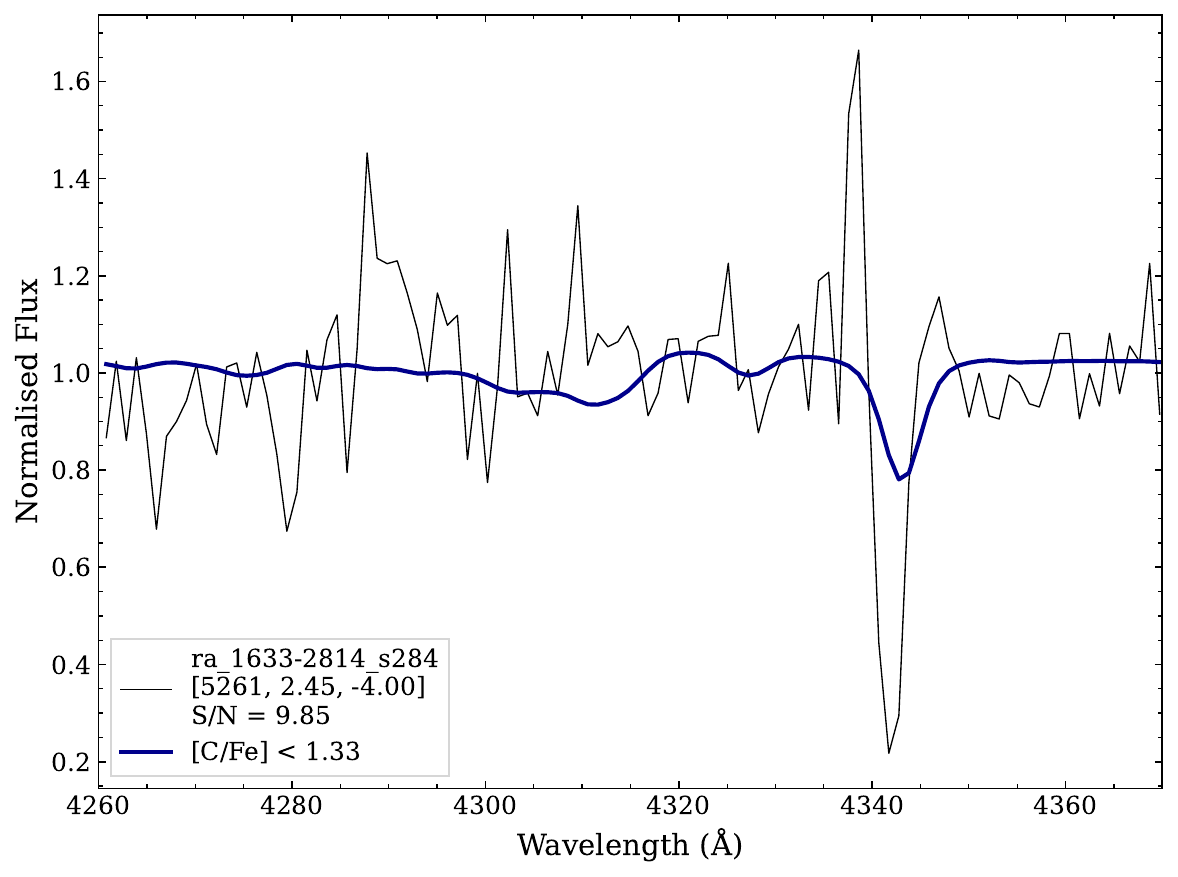}
    \end{subfigure}
    
    \begin{subfigure}{0.49\textwidth}
        \centering
        \includegraphics[width=0.85\linewidth]{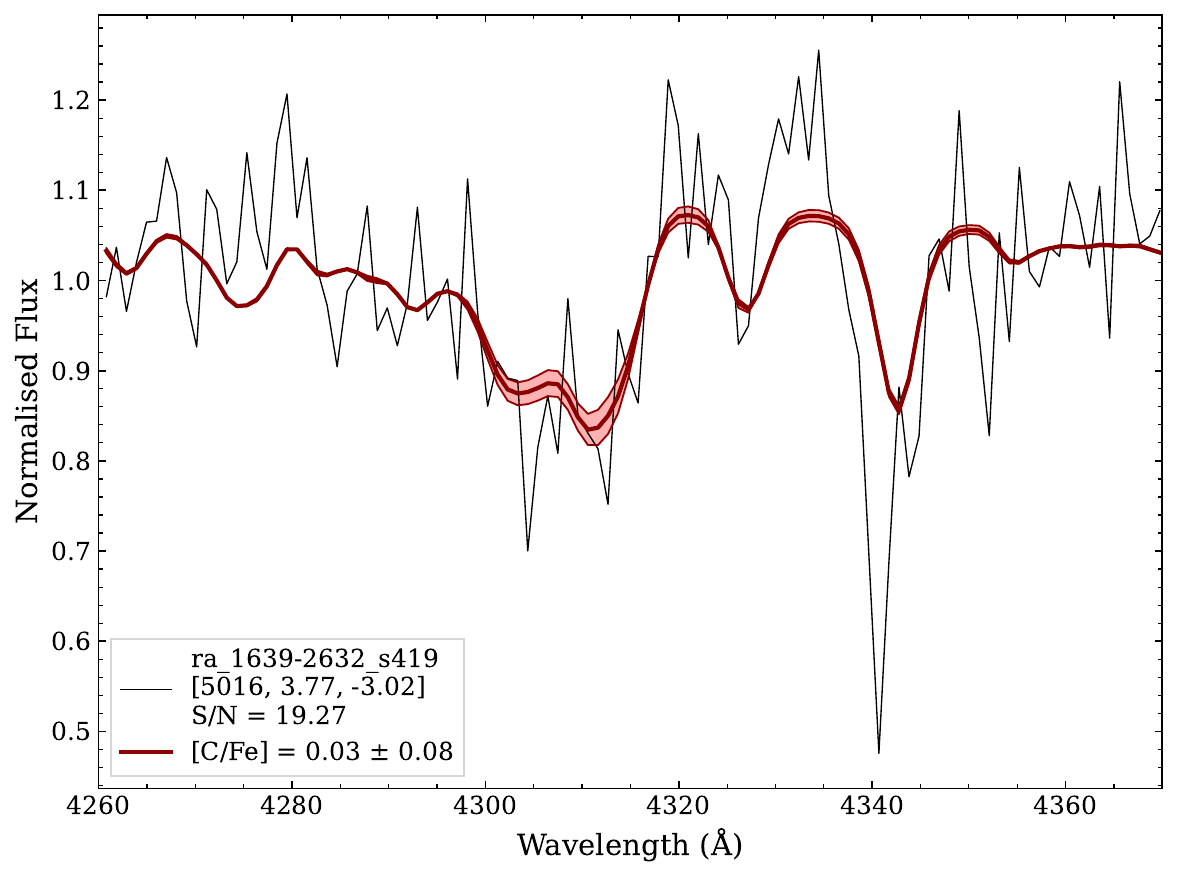}
    \end{subfigure}
    \hfill
    \begin{subfigure}{0.49\textwidth}
        \centering
        \includegraphics[width=0.85\linewidth]{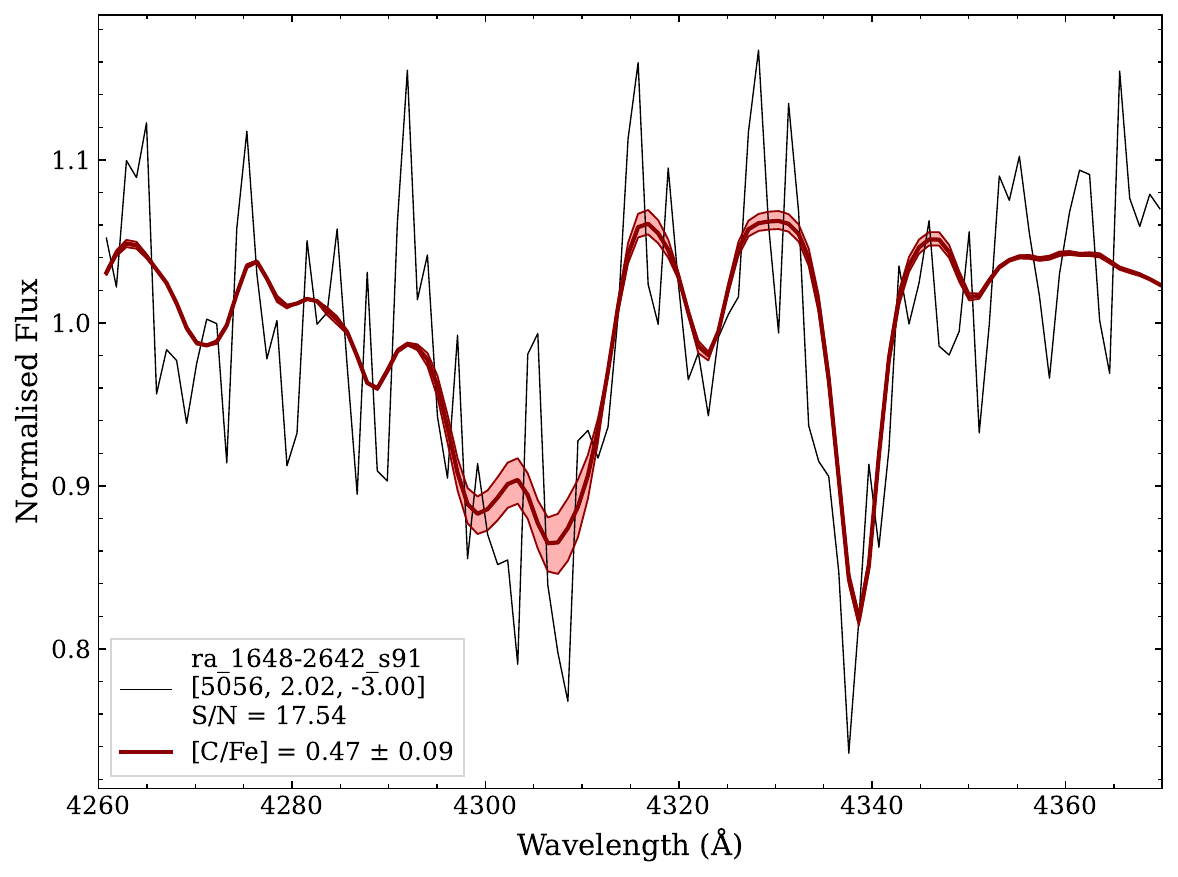}
    \end{subfigure}
    \caption{CH G-band fits for the 15 EMP stars across the wavelength region $4260 \leq \lambda \leq 4370$\,\AA{} (using continuum regions $4150-4200$\,\AA{} and $4400-4450$\,\AA{}). Black line is the observed data, and for stars with detections: the thick red line is the best-fitted [C/Fe] value, and the shaded region is the $1\sigma$ statistical error (values shown in Table\,\ref{tab:emp cfe}). Stars with non-detections have a thick blue line showing their upper limits. Stellar parameters $\Teff$, $\logg$ and $\FeH$ are in the legend, alongside the S/N measured in the red spectra. The total errors taking into account uncertainties in stellar parameters are quoted in the legend with the best-fitting $\XFe{C}$ value. Stars ra\_1604-2712\_18 and ra\_1633-2814\_284 have non-detectable [C/Fe] measurements, with upper-limits quoted in legend. Continues onto \ref{fig:emp carb fits 2}.}
    \label{fig:emp carb fits 1}
\end{figure*}

\begin{figure*}[tp] 
    \centering
    \begin{subfigure}{0.49\textwidth}
        \centering
        \includegraphics[width=0.85\linewidth]{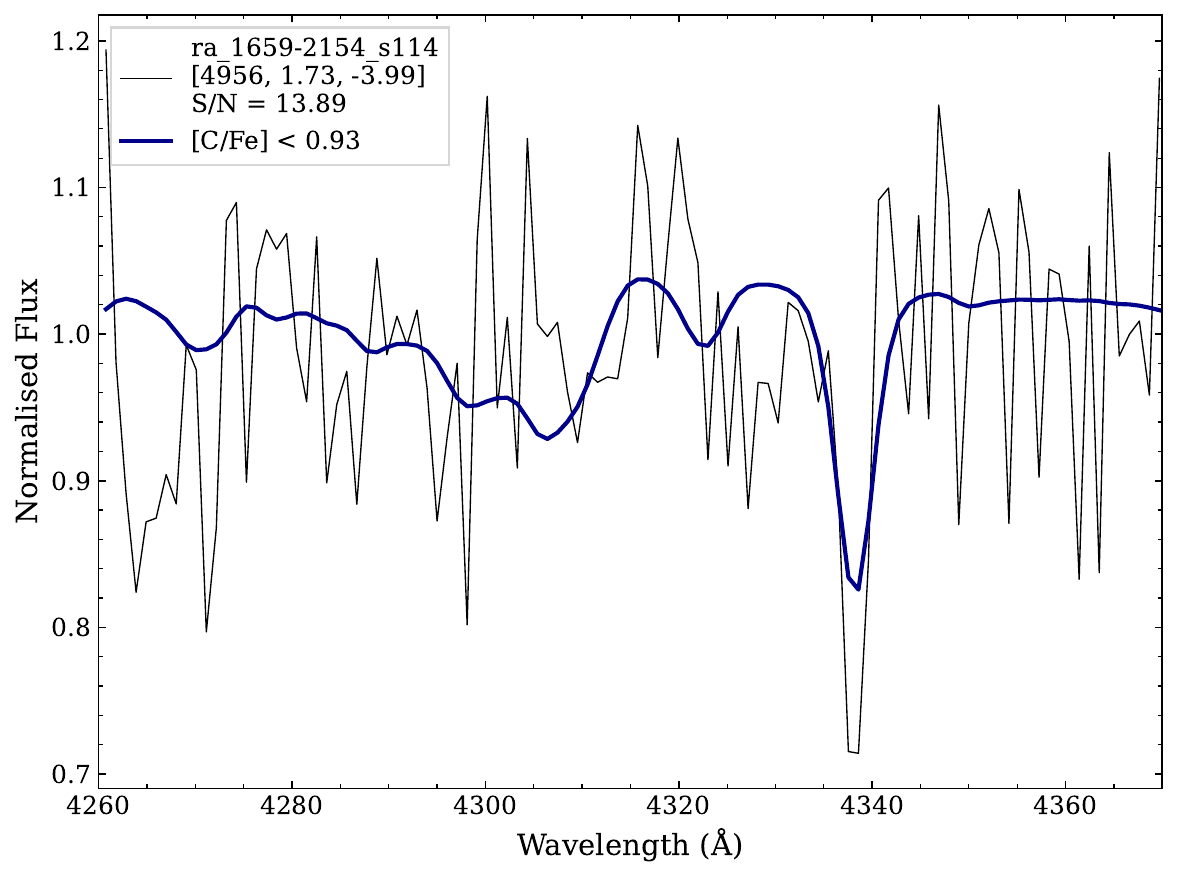}
    \end{subfigure}
    \hfill
    \begin{subfigure}{0.49\textwidth}
        \centering
        \includegraphics[width=0.85\linewidth]{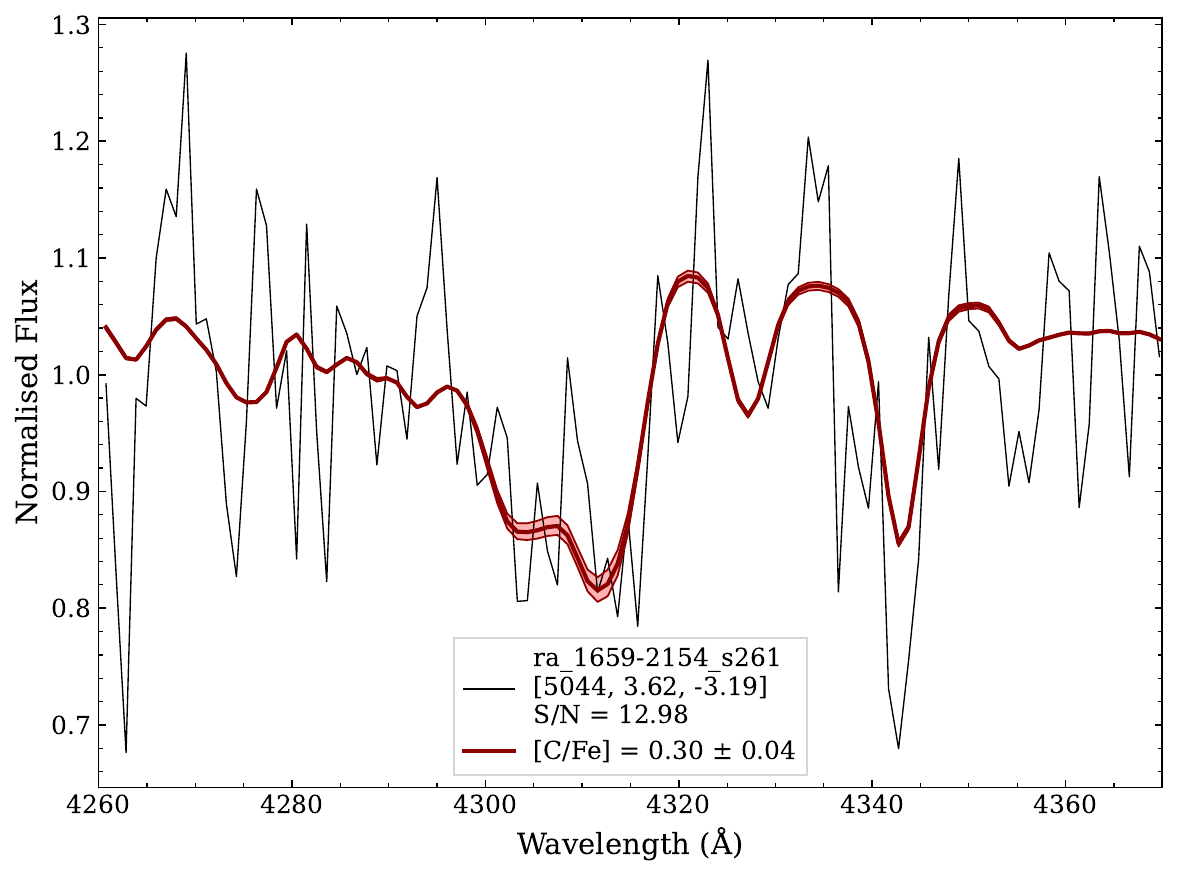}
    \end{subfigure}

    \begin{subfigure}{0.49\textwidth}
        \centering
        \includegraphics[width=0.85\linewidth]{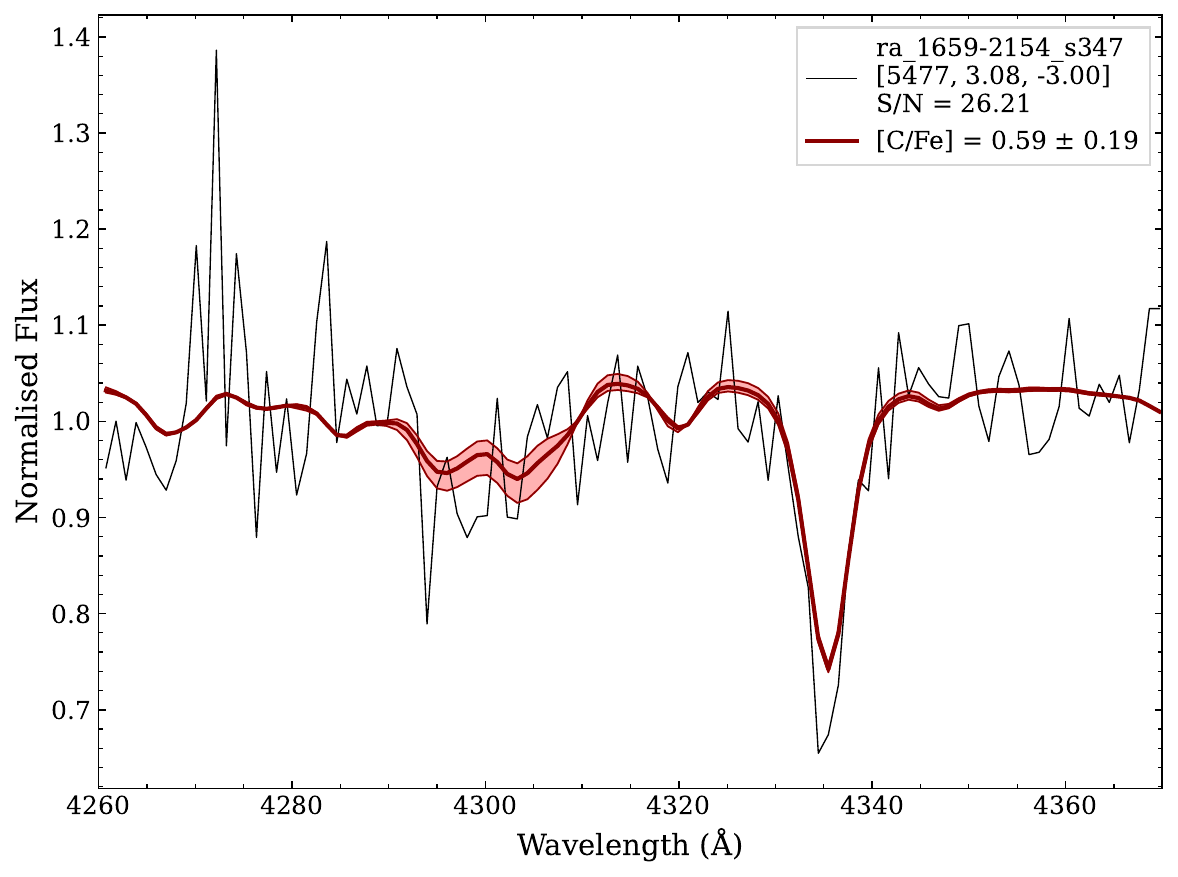}
    \end{subfigure}
    \hfill
    \begin{subfigure}{0.49\textwidth}
        \centering
        \includegraphics[width=0.85\linewidth]{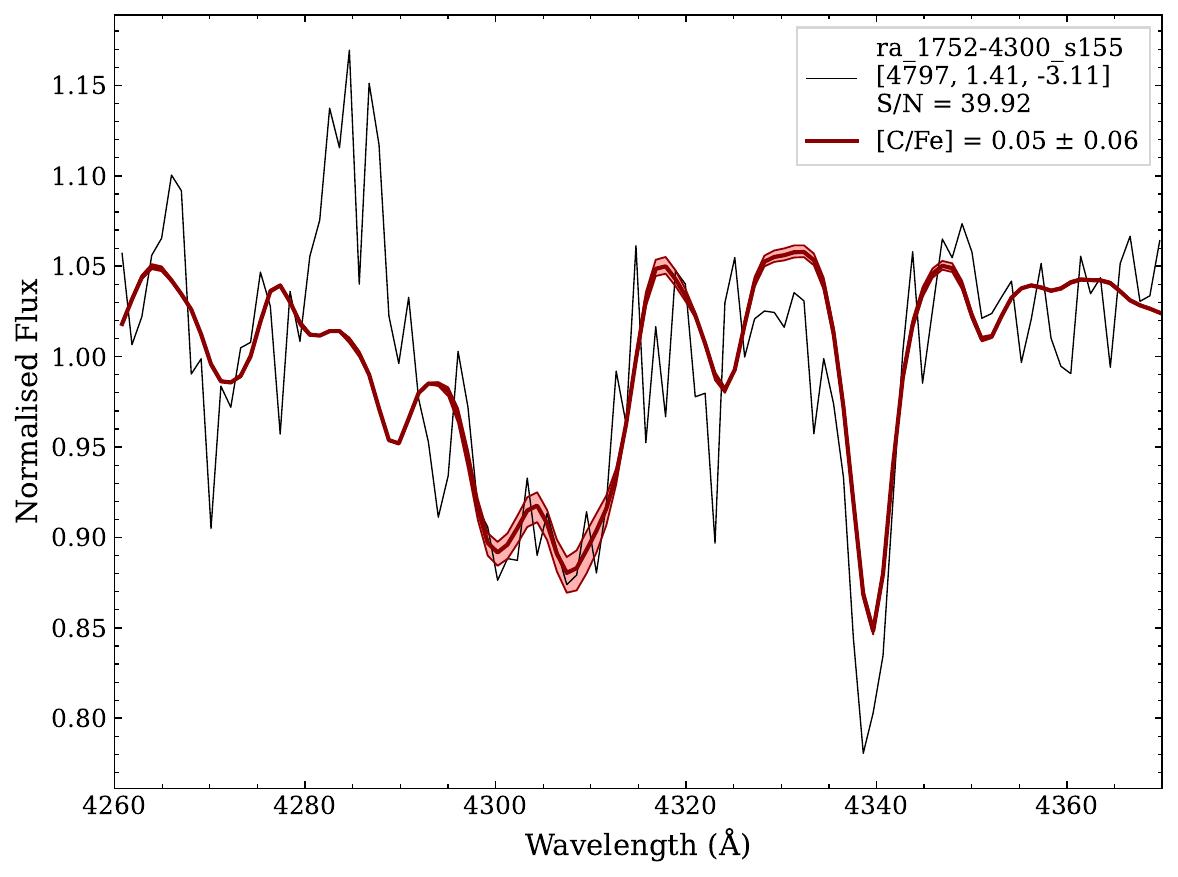}
    \end{subfigure}

    \begin{subfigure}{0.49\textwidth}
        \centering
        \includegraphics[width=0.85\linewidth]{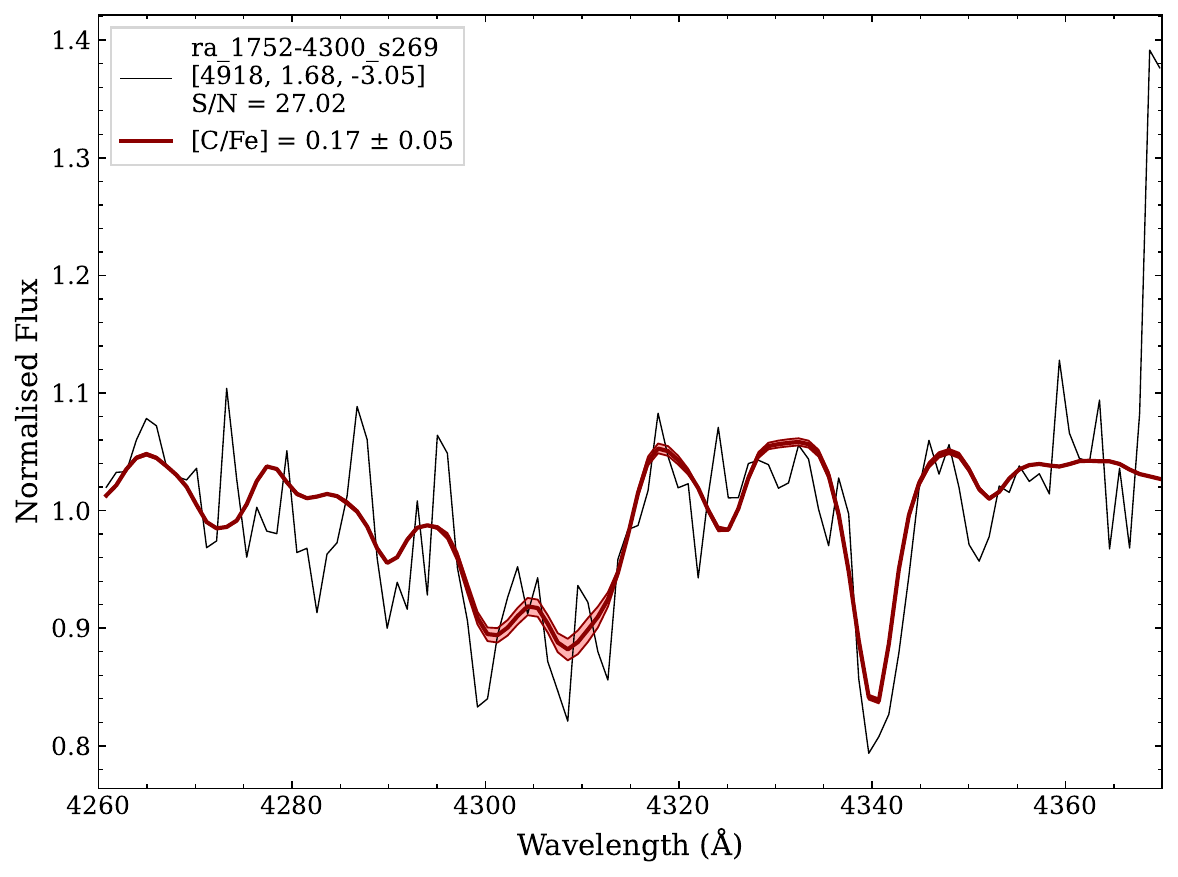}
    \end{subfigure}
    \hfill
    \begin{subfigure}{0.49\textwidth}
        \centering
        \includegraphics[width=0.85\linewidth]{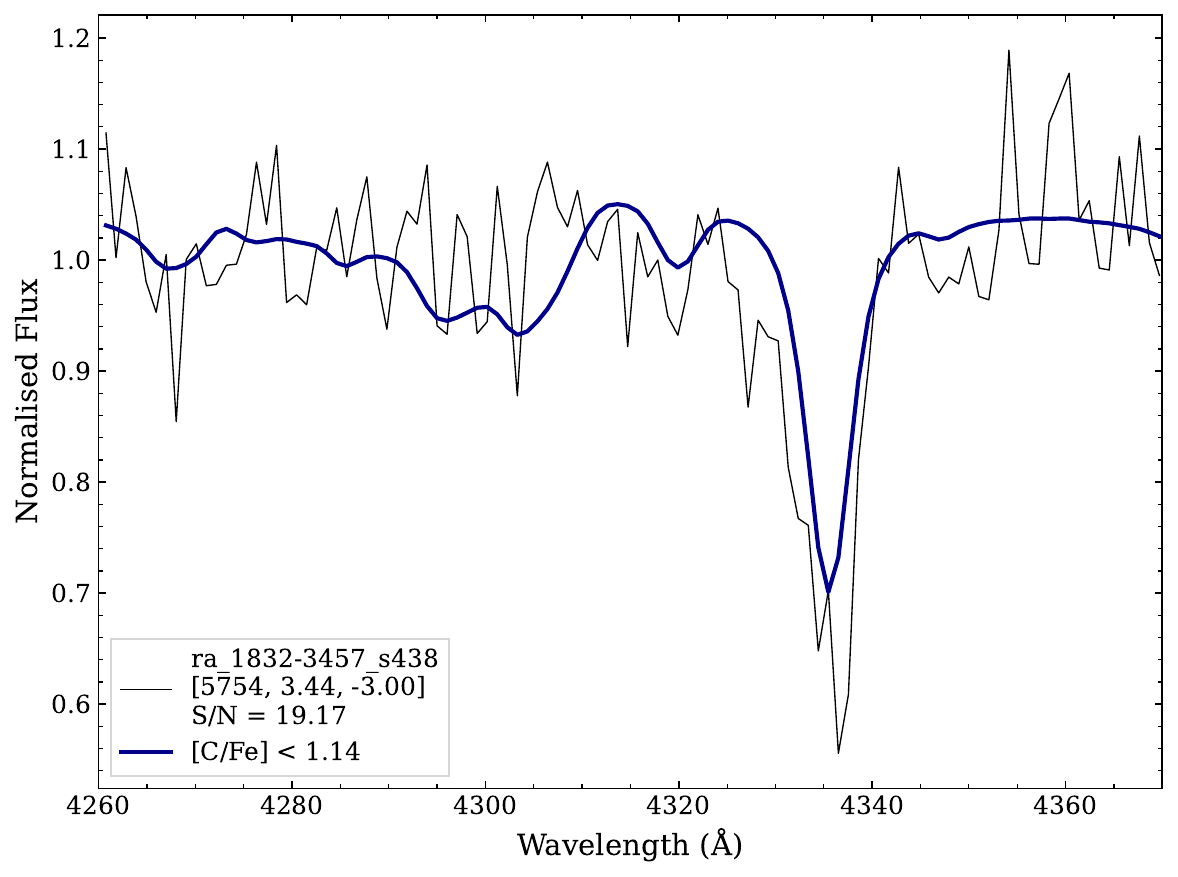}
    \end{subfigure}

    \begin{subfigure}{0.49\textwidth}
        \centering
        \includegraphics[width=0.85\linewidth]{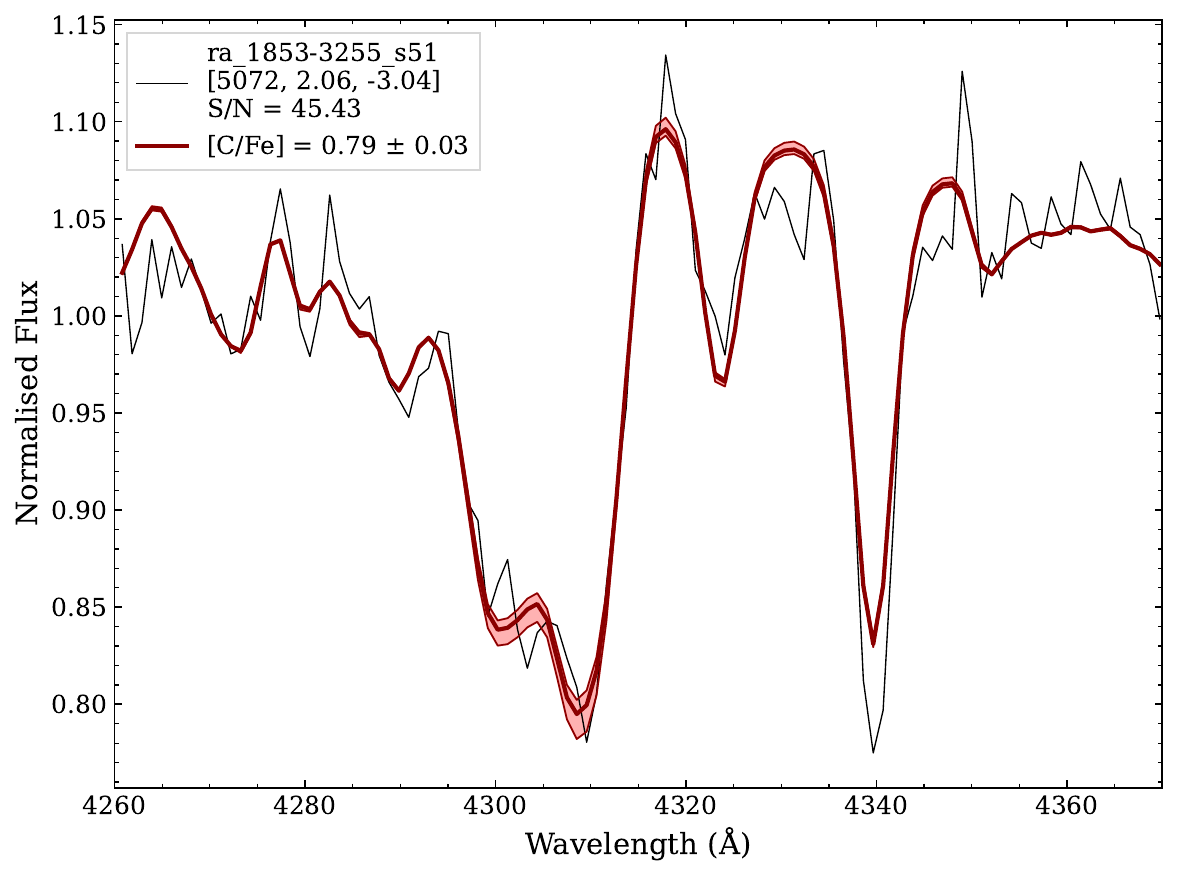}
    \end{subfigure}
    \caption{Continuation of Fig.\,\ref{fig:emp carb fits 1}. Stars ra\_1659-2154\_114 and ra\_1832-3457\_438 also have non-detectable [C/Fe] measurements.}
    \label{fig:emp carb fits 2}
\end{figure*}

\end{document}